\documentclass{article}   

\usepackage{graphicx}
\usepackage{hyperref}
\hypersetup{
    colorlinks=true,
    linkcolor=blue,
    urlcolor=blue,
    citecolor=blue,
}

\usepackage{subcaption}

\usepackage{float}
\usepackage{amssymb}
\usepackage{xfrac}
\usepackage{miller}

\usepackage{siunitx}
\usepackage{mathtools}
\usepackage{booktabs} 	% Formatting for tables
\usepackage[title]{appendix}
\newcommand{\figref}[1]{Figure~\ref{#1}}

\newcommand{\eqnref}[1]{Eq.~(\ref{#1})}

\newcommand{\tabref}[1]{Table~\ref{#1}}
\newcommand{\secref}[1]{Section~\ref{#1}}

\DeclareMathAlphabet{\mathsfsl}{OT1}{cmss}{m}{sl}

\newcommand{\mechmet}{{\bfseries{\slshape{MechMet}}}}

\newcommand{\neper}{{\bfseries{\slshape{Neper}}}}

\newcommand{\vctr}[1]{\boldsymbol{#1}}

\newcommand{\tnsr}[1]{\mathsfsl{#1 }}

\begin{document}

\title{A Virtual Diffractometer For Creating Synthetic HEDM Images of Tessellated and Meshed Finite Element Polycrystals}

\author{Paul R. Dawson \and Matthew P. Miller}

\date{Sibley School of Mechanical and Aerospace Engineering\\ Cornell University\\  Ithaca, NY 14850, USA \\ \today}

\maketitle

\begin{abstract}
To assist in the planning of in situ loading, HEDM experiments by generating synthetic diffraction images of virtual samples in loaded and unloaded states.   The user designates a target grain in the virtual sample and specifies the set of reflections for which images are to be generated.  The code generates several intermediate images, including: (1) the points of intersection between the diffracted beam direction vector the detector plane; (2) reflection-dependent frequency distributions of the diffraction angle, $\omega$; and (3) plots of the diffraction volume-weighted intensity distributions for the specified set of reflections.  The final sets of plots are facsimiles of pixelated detector images which take into account characteristics of specific detectors, including its pixel size and point spread behavior. 
%\keywords{discrete harmonics \and finite element \and  reduced-order descriptions \and polycrystal data fields}
\end{abstract}

\section{Introduction}
\label{sec:intro}
High Energy Diffraction Microscopy (HEDM), also known as 3D X-ray Diffraction (3DXRD), refers to a branch of high-energy x-ray diffraction that is devoted to measuring the internal state of polycrystalline solids, often while under mechanical loading~\cite{poulsen_book, suter_jemt_2008, lienert_jom_2010, lienert_titanium_2009, Oddershede2010a, lienert_jom_2010, Bernier2011}. In particular, it is possible to conduct experiments involving mechanical loading while using HEDM to interrogate the orientations and distortion of the crystallographic lattice at the sub-grain scale~\cite{20202708900950}.   Measuring lattice orientations is important because: (1) crystals comprising a solid usually display anisotropic properties associated with the crystal structure, making orientation a need-to-know quantity and (2) internal deviations in orientations from a nominal value result from deformation and thermal processing and can have substantial effects on a material's properties. 
The latter lattice distortions are important because it can be possible with the precision needed to quantify the elastic strains, and critically for engineering applications, the changes in strain with changes in load which allow for evaluation of stress. 

While HEDM experiments have potential to deliver data of great value,  challenges persist in designing and conducting experiments that achieve research goals.  
Because high energy X-ray facilities are costly, access to them is limited and awarded on a merit basis.  
Consequently, researchers are motivated to be as productive as possible in using the precious beam time.  
As HEDM experiments all are unique to some extent, customization of the experimental set-up is typical making it difficult to 'get it right' the first time.  
For example, it is difficult to assess the extent to which detectors images will reveal behaviors of interest.  
As a result,  many times experiments fail to achieve goals articulated in the proposal and either (1) the experiments must be redone after modifying the set-up or (2) the experimental goals must be  changed on-the-fly to reflect how experimental limitations influence the character of the data.

For these reasons, tools that can provide {\it a priori } estimation of the experimental observations can provide insight into an experiment in advance of conducting the experiment.  Such tools could benefit a researcher's efforts in a number of ways: providing insight into the mechanical environment present within loaded samples during the experiment;  assessing if the diffracted beams will generate signals that can be detected by the instrumentation; and, determining if the diffraction data in the end will be definitive for purposes of the experimental goals. 
For example, Obstalecki, Wong, Dawson and Miller~\cite{obs_won_daw_mil_14}
investigated the microstructural origins of peak broadening in a copper alloy subjected to cyclic loading.
Bertin and Cai~\cite{bertin2018computation} have proposed a method to compute virtual 
diffraction patterns from discrete dislocation structures in single crystals.  Intensity distributions are computed for the detector images using a previously published ray-tracing method.
Pagan, Jones, Bernier and Than~\cite{Pagan2020} 
published a framework for finite energy bandwidth-based diffraction that can be used for simulating x-ray diffraction patterns gathered during {\it in situ} laser-melting processes.
Ribart, King, Ludwig, Bertoldo, Joao and Proudhon~\cite{Ribart:vl5004}  examined the influence of using local versus grain-averaged lattice orientations in  synthetic diffraction images of virtual samples reconstructed from of {\it in situ} diffraction contrast tomography (DCT) data.

This article describes one of a suite of simulations tools that enables  researchers to perform virtual mechanical loading experiments in advance to estimate the ability of the experimental set-up and procedure to expose the sought-after behaviors.  
It builds on a previous contribution reported by Wong, Park, Miller and Dawson~\cite{won_par_mil_daw_13}.   
Specifically, it performs the analog of a diffraction measurement by creating a set of diffraction spot images corresponding to specified crystallographic reflections at designated points in the loading regime.  
Here, diffraction spot or peak is used to mean the detector reading associated with the diffracted beam associated with a small part of the sample volume having a shared lattice orientation  (typically a grain or grain subvolume). 

The article is organized as follows. 
\secref{sec:background}  provides background useful for understanding the functioning of a virtual diffractometer in conjunction with {\it in situ} loading HEDM experiments.   
\secref{sec:comp_tasks} then details the methodologies associated with the distinct steps employed in creating a synthetic 
diffraction spot and its corresponding detector image.
\secref{sec:demonstration} illustrates the functionality of the virtual diffractometer via the tensile loading of a stainless steel sample.
\secref{sec:summary} summarizes the capabilities of the virtual diffractometer.

\section{Background and Goals for the Virtual Diffractometer}
\label{sec:background}

\subsection{Using diffraction measurements with {\it in situ} loading experiments}
\label{sec:insitudiffraction}
X-ray diffraction provides a powerful method for measuring internal structural changes during in mechanical testing of crystalline solids.
Succinctly stated, high-energy x-rays are used to measure the evolving microstructural state within samples of the material that are subjected to mechanical loads that induce deformation.  
High-energy x-rays are well suited for such measurements because (1) the beams are  sufficiently bright to pass through samples, thereby facilitating interrogation of interior grains, and (2) the diffraction events may be sufficiently resolved in space and time to detect both lattice orientation and lattice distortion with accuracies needed for engineering applications.

\begin{figure}[htbp]
   \centering
   \includegraphics[width=0.5\textwidth]{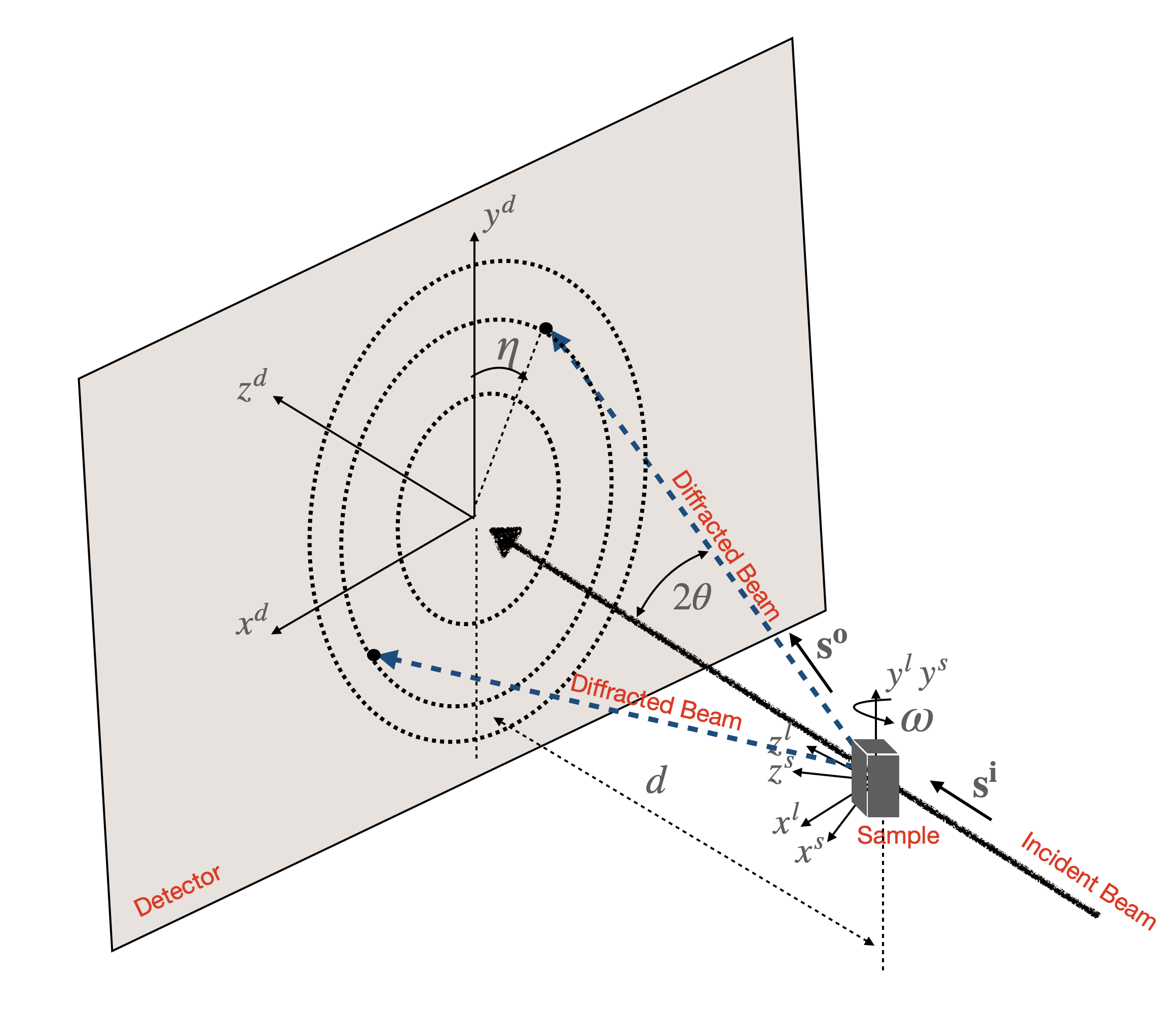} 
   \caption{Physical set-up being simulated by the virtual diffractometer.  Modified version of figure from \cite{won_par_mil_daw_13}.   }
   \label{fig:expsetup}
\end{figure}

Basic diffraction elements of an  {\it in situ} loading HEDM experiment are illustrated in \figref{fig:expsetup}.  A load frame applies loads to the sample concurrently with the diffraction measurements, but is not shown as the focus here is on the diffraction measurements.  
The incident x-ray beam is shown to penetrate the loaded sample and to spawn two diffracted beams  from two different crystallographic planes within a single diffraction volume.   
Typically at any given time, there are many such diffracted beams originating from suitably oriented crystallographic planes within a diffraction volume.  
For now, we concentrate on a few  diffracted beams originating from within a single grain that are associated with chosen families of crystallographic planes.  

The theory underpinning the basic diffraction geometry shown in  \figref{fig:expsetup} can be understood with basic principles laid out originally by Bragg and  subsequently refined by others~\cite{Cullity2001}.
Bragg's law expresses the relationship between the wavelength of the incident X-ray beam, $\lambda$, the lattice  interplanar spacing, $d$, within a diffraction volume, and the angle  $\theta$ between the incident and diffracted beams as:
\begin{equation}
\lambda = 2 d \rm{sin}(\theta)
\label{eqn:braggslaw}
\end{equation}
To fully define the diffraction geometry, the spatial orientation of the diffracting planes enters the mathematical description.
For simplicity here, the orientation is denoted by its normal vector, $\vctr{n}$; its connection to families of crystallographic planes will be made explicit in \secref{sec:diffraction_from_elemental_volumes}.
Let $\vctr{s}^i$ and $\vctr{s}^o$ denote unit vectors in the direction of the incident and diffracted beams, respectively,  as also shown in \figref{fig:expsetup}.
Assuming elastic scattering, the scattering vector,  $\vctr{q}$, is defined as a bisector of $\vctr{s}^i$ and $\vctr{s}^o$ as:
\begin{equation}
\vctr{q} \equiv \frac{\vctr{s}^o - \vctr{s}^i}{\lambda} 
\label{eqn:scatteringvector}
\end{equation}
Constructive interference during diffraction, leading to an observable diffracted beam, occurs 
when the lattice plane normal and the scattering vector align.  Mathematically, $\vctr{n} \| \vctr{q}$.
The diffracted beam forms a spot on the detector with finite size owing to the finite volume from which it originates.  The larger the diffracting volume, the larger the spot, in general.  
The spot embodies much more information that just the diffraction volume. The shape of the spot not only relates to the shape of the diffraction volume, but also the distortions of the crystal lattice within the volume, the variations in lattice orientation over the diffraction volume, and more.  Moreover, as the diffraction volume deforms under load, the spots evolve in manners that depict the changes to orientation and distortion over the diffraction volume.  

When applied as the instrumentation for mechanical test, the direction change of an outgoing beam can be used to extract the lattice strain in the direction normal to the scattering planes.  Further, changes to the conditions for diffraction, as described in the next section indicate changes to the lattice orientation within the diffraction volume.  Thus, using diffraction measurements together with {\it in situ } loading, it is possible to produce spatial maps of the elastic strain and lattice orientation over a polycrystalline sample with sub-grain definition.

\subsection{Overview of the  virtual diffractometer}
\label{sec:diffractometer_overview}
 
It is the intent of the virtual diffractometer described herein to simulate the evolution of diffraction spots that could be observed in a physical experiment and to relate these to the deformations induced by the mechanical loading. 
To relate the simulated diffraction images to actual ones measured, the simulations must 
replicate the experiments in a number of critical ways.  
Namely, the  experimental facility information required must include: (1) the beam energy, spread and input direction vector and (2) the detector size, orientation, and center position.
For the virtual sample, required data include:
(1) the sample material  unstrained lattice spacing and crystallographic reflections of interest, (2) the virtual sample definition (tessellation defining grains and their lattice orientations plus the associated finite element mesh), and (3) lattice orientation and elastic strain tensors for every finite element with the grains.
With these input data, the virtual diffractometer output should include: (1) basic diffraction parameters on an element-by-element basis over a tessellated and meshed virtual sample; (2) a map of the intersections of diffracted beams with the plane of the detector for all elements within a designated target grain; (3) relative intensity distributions for spots created on the detector for designated scattering vectors; and, (4) strains associated with each spot.

The virtual diffractometer reported here follows the same overall framework used  to implement the methodology documented in  \cite{won_par_mil_daw_13}. 
That methodology  divided the complete set of tasks in creating a detector image into a number of distinct steps (referred to as Steps 1 to 4):
\begin{enumerate}
\item Establish the diffraction conditions for each finite element within a target grain (diffracted beam direction and sample rotation angle).
\item Compute the intersections of the diffracted beam with the detector plane for sets of points within every element of the target grain).
\item Combine contributions from Step 2 to form an intensity distribution over a patch lying on the diffraction plane.
\item Process the intensity distribution to form a detector image taking into account the detector resolution.
\end{enumerate}
There are  several differences between the method reported here and that in \cite{won_par_mil_daw_13}:
\begin{itemize}
\item In creating the diffraction spot for the target grain, the present code combines the contributions to the diffracted beam intensity for each reflection regardless of the sample rotation angle, $\omega$ (rather than for discrete intervals in  $\omega$). Supplementary plots are created that give distributions of rotation angles for each reflection from the target grain. 
\item Spatial resolution within every finite element of the target grain is generated using the elemental quadrature points of the target grain elements,  as is commonly employed in finite element practice for performing volume integrations.  
Quadrature weights become the intensity weights, thus giving a diffraction-volume weighted interpretation to each diffracted beam to intersect the detector plane.
\item To determine the relative intensity distribution over the plane of the detector, a local region in the vicinity of a diffraction spot is discretized with a 2D finite element mesh.  Simple bi-linear quadrilateral elements are employed to facilitate subsequent mapping to a detector pixel array. 
\item  Diffracted beams  are projected onto the 2D detector mesh where they are treated as the centers of Gaussian distributions.  An intensity field is determined for the entire set of diffracted beams using a finite-element $\rm{L}_2$ inner product formalism, as detailed in the subsection devoted to Step 3.. 
\item The average normal strain for a single spot is computed over all diffraction (elemental) volumes that contribute to that  reflection.  Each contributing strain value is  computed from the Cauchy formula using the reciprocal lattice vector in sample coordinates.
\item Finally, a detector image is created as a derivative of the intensity distribution.   The detector image incorporates  pixel-based binning of the relative intensity distribution completed in Step 3.  In the methodology reported here, a finite element mesh is defined that directly coincides with detector pixels.  The intensity distribution computed in Step 3 is mapped onto this mesh to provide the detector image.
\end{itemize}
Details of each task for the present code are laid out in the following subsections.

\subsection{Coordinate systems and related notation}
\label{sec:coordsystems}

Several right-handed Cartesian coordinate systems displayed in \figref{fig:expsetup} are used during the computations performed in the virtual diffractometer.  
These are defined in \tabref{tab:coordsys}.
Note the following points:
\begin{itemize}
\item The crystal coordinate systems (one for each crystal) have base vectors coincident with the lattice directions.
(See Figure 2-10 of \cite{Cullity2001}.) Currently, cubic and hexagonal crystal types are available.  
The crystal coordinate system orientations are defined relative to the sample coordinate system via the lattice orientations provided as part of the virtual sample definition.
\item The sample coordinate system and the laboratory coordinate system share the same origin and the $\vctr{e}^s_2$ and $\vctr{e}^l_2$ base vectors are aligned.  This conforms to any experimental station where HEDM is performed. 
\item The sample coordinate system is attached to the sample with the following provisions: the loading direction coincides with $\vctr{e}^s_2$; the sample is rotated during  an experiment by the angle $\omega$ about $\vctr{e}^s_2$ to facilitate placing diffraction spots on the detector (this could be generalized to permit tilting of the load frame).  
\item The incident beam has a fixed direction in the laboratory coordinate system.  This is readily changed. 
\item The detector coordinate system is defined relative to the laboratory coordinate system by a vector offset of the origins of the laboratory and detector coordinate systems.  The detector plane normal vector is $ \vctr{e}^d_3 $.  $ \vctr{e}^d_1 $ and  $ \vctr{e}^d_2 $   lie in the detector plane and may be oriented as the user chooses.  Note $ \vctr{e}^d_3 = 1 $ corresponds to the orientation shown in  \figref{fig:expsetup}. The detector plots generated by the code will appear as viewed looking toward the detector in the direction opposite the direction of the incident beam.  If $ \vctr{e}^d_3 = -1 $ is designated the plots will be a mirror image about the $ \vctr{e}^d_2 $ axis and appear as viewed looking toward the detector in same direction as the incident beam.
\item The detector patch coordinate systems are two-dimensional, lying in the detector plane and displaced from the detector coordinate system to position their origins in the vicinity of particular spots.   
\end{itemize}

The convention for a position vector within space given as $\vctr{x}$.  The vector has components depending on the coordinate system as indicated in \tabref{tab:coordsys}.  
Vectors are written as  $ \vctr{t}  = t^c_i \vctr{e}^c_i = t^s_i \vctr{e}^s_i$ using Einstein summation convection on component indices (subscripts). 
A change of basis between systems isngiven by a transformation of the form: $ t^s_i = Q^{sc}_{ij} t^c_j$, where $Q^{sc}_{ij} = \vctr{e}^s_i \cdot \vctr{e}^c_j $ for two systems sharing the same origin.
Spatial dependence for a field vector is given as $\vctr{t}(\vctr{x})$ or $\vctr{t}(x^s_i)$ (for example).

\begin{table}[ht]
	\centering
	\caption{Coordinate system notation.}
	\begin{tabular} {| c | c | c | c |}	\hline	
		Name & Dimensions & Base vectors & Components of $\vctr{x}$  \\ \hline
		crystal & 3 & $ \vctr{e}^c_i $ & $x^c_i $ \\ 
		sample & 3 & $ \vctr{e}^s_i $ & $x^s_i $ \\ 
		laboratory & 3 & $ \vctr{e}^l_i $ & $x^l_i $ \\ 
		detector & 3 & $ \vctr{e}^d_i $ & $x^d_i $ \\ 
		detector patch  & 2 & $ \vctr{e}^p_i $ & $x^p_i $ \\ \hline	
	\end{tabular}	
	\label{tab:coordsys}
\end{table} 

\section{Computational Tasks of a Virtual Diffractometer}
\label{sec:comp_tasks}

\subsection{Step 1: Determining the diffraction conditions of elemental volumes.}
\label{sec:diffraction_from_elemental_volumes}

Using the elemental lattice orientations and lattice (elastic) strains previously determined from a finite element simulation
(and entered here as data),
the objectives of Step 1 are to compute the basic diffraction conditions illustrated in \figref{fig:expsetup}.  
This consists of computing the angles, $\omega$, $\eta$, and $\theta$, and the diffracted beam directions (one for each reflection) for each diffraction center in the virtual sample (each quadrature point of every element of the target grain.)
The methodology used here is similar in its overall structure to what is reported by Wong, Park, Miller and Dawson~\cite{won_par_mil_daw_13}
although the notation has been revised. 
An alternative approach is described in \cite{Pagan2020}.

We first define reference lattice vectors\footnote{Also called lattice translation vectors by Cullity and Stock~\cite{Cullity2001} .}.
For  cubic crystals, the reference lattice vectors are: 
\begin{equation}
\vctr{a}^0_i = a \vctr{e}^c_i
\label{eqn:cubicreferencelatticevector}
\end{equation}
 which are obtained simply by scaling the coordinate system base vectors by the lattice parameter. 
The reference lattice  vectors are aligned with the edges of the simple cubic Bravais lattice of an undeformed unit cell.  
For hexagonal crystals, the reference lattice vectors are:
\begin{align*}
\vctr{a}^0_1 &= a \vctr{e}^c_1 \\
\vctr{a}^0_2 &= -\frac{1}{2} a \vctr{e}^c_1 + \frac{\sqrt{3}}{2}a \vctr{e}^c_2 \\
%\vctr{a}^0_3 &= -\frac{1}{2} a \vctr{e}^c_1 - \frac{\sqrt{3}}{2}a \vctr{e}^c_2 \\
\vctr{a}^0_3 &= c \vctr{e}^c_3 
\label{eqn:hexreferencelatticevector}
\end{align*}
These reference lattice  vectors are aligned with the edges of the hexagonal Bravais lattice of an undeformed unit cell.  
The lattice  vectors, $\vctr{a}$, in the current configuration are computed in  the deformed configuration using the pure elastic stretch, $\tnsr{V}^e$, as defined in the kinematic decomposition of an 
an elastoplastic deformation~\cite{Dawson_FEpX_2015}:
\begin{equation}
\vctr{a}_{i} = \tnsr{V}^e \vctr{a}^0_{i} 
\label{eqn:adeformed}
\end{equation}
For a linear elastic simulation, $\tnsr{V}^e$ simplifies to the stretch tensor computed from the polar decomposition of the deformation gradient.

Reciprocal lattice  vectors, $\vctr{b}_i$, are defined from the lattice vectors~\cite{Krawitz:2001} as:
\begin{gather}
\vctr{b}_{1} = \frac{\vctr{a}_{2} \times \vctr{a}_{3}}{\vctr{a}_{1} \cdot \vctr{a}_{2} \times \vctr{a}_{3}} \qquad \qquad
\vctr{b}_{2} = \frac{\vctr{a}_{3} \times \vctr{a}_{1}}{\vctr{a}_{1} \cdot \vctr{a}_{2} \times \vctr{a}_{3}} \qquad \qquad
\vctr{b}_{3} = \frac{\vctr{a}_{1} \times \vctr{a}_{2}}{\vctr{a}_{1} \cdot \vctr{a}_{2} \times \vctr{a}_{3}}
\end{gather}
The reciprocal lattice vectors often are used as base vectors in reciprocal space, but here we 
use them to define  the normals to the crystallographic planes mentioned in \secref{sec:insitudiffraction}.
To emphasize their role in the diffraction events, we refer to these as reflection vectors, $\vctr{r}^{hkl}$,
where the superscript $hkl$ denotes the Miller indices of a family of crystallographic planes
and write them as: 
\begin{equation}
\vctr{r}^{hkl}(x^c_k) = h \vctr{b}_{1} + k \vctr{b}_{2} + l \vctr{b}_{3}
\label{eq:reflectionvector}
\end{equation}
for each crystallographic plane of interest.  The reflection vectors are referred to also as the reciprocal vectors in the literature.
Here, the parameters $ h$, $ k $ and $l$ are the Miller indices for crystallographic planes $\{ h k l \}$ for cubic crystal types and
$h=u-t$, $k=v-t$ and $l=w$ for crystallographic planes $\{ u v t w\}$ for hexagonal crystal types using the 4 index notation.\footnote{See \cite{Cullity2001} Section 2-8.}
Note that in using the $hkl$ parameters explicitly to define the reflection vectors, these vectors are implicitly written with components in the crystal coordinate system as shown in \eqnref{eq:reflectionvector}.  Applying the Laue condition, the reflection vectors equate to the scattering vectors for the $hkl$ reflections to define the condition for constructive interference in diffraction.  That is, $\vctr{r}^{hkl}=  \vctr{q}^{hkl} $.

Continuing this step to identify the diffraction conditions, $\vctr{r}^{hkl}$ is written in the sample coordinate system indicated in \figref{fig:expsetup} using the transformation matrix $Q^{sc}_{ij}$ that converts a vector in the crystal coordinate system to one in the sample coordinate system: 
\begin{equation}
{r}^{hkl}_i (x^s_k)= Q^{sc}_{ij} \, {r}^{hkl}_j(x^c_k)
\label{eqn:Hsam}
\end{equation}
Likewise, this vector is written in the laboratory coordinate systems (again as shown in \figref{fig:expsetup}),
with the aid of a transformation matrix, $Q^{ls}_{ij}$ 
 \begin{equation}
{r}^{hkl}_i (x^l_k)= Q^{ls}_{ij} \, {r}^{hkl}_j(x^s_k)
\label{eqn:Hlab}
\end{equation}
Applying Bragg's law (\eqnref{eqn:braggslaw}), the reflection vector in laboratory coordinates can be written after normalization in terms of 
the azimuthal angle, $\eta$, and the Bragg angle, $\theta$, of the diffraction spot are depicted in \figref{fig:expsetup}:
\begin{equation}
\vctr{\bar r}^{hkl}(x^l_k) = \frac{\vctr{r}^{hkl}(x^l_k)}{\parallel \vctr{r}^{hkl} \parallel} = 
\left\{
\begin{array}{c}
- \sin \eta \cos \theta \\
\cos \eta \cos \theta \\
- \sin \theta
\end{array}
\right\}
\label{eqn:normrl}
\end{equation}
For the the particular case of the relation between the sample frame and laboratory frame consisting solely of
a rotation of the sample frame about the its $\vctr{e}^s_2$ axis 
by the angle $\omega$,  $Q^{ls}_{ij}$ is given in matrix form as
\begin{equation}
[ Q^{ls} ]= \left[
\begin{array}{ccc}
\cos \omega & 0 & \sin \omega \\
0 & 1 & 0 \\
-\sin \omega & 0 & \cos \omega
\end{array}
\right]
\label{eqn:Rls}
\end{equation}
Finally,  we designate the components of the normalized reflection vector in the sample coordinate system as $(\bar r^{s}_{1}, \bar r^{s}_{2}, \bar r^{s}_{3})$. 
Combining Equations \ref{eqn:Hlab}, \ref{eqn:normrl} and \ref{eqn:Rls}, provides the matrix equation,
\begin{equation}
\left[
\begin{array}{ccc}
\cos \omega & 0 & \sin \omega \\
0 & 1 & 0 \\
-\sin \omega & 0 & \cos \omega
\end{array}
\right]
\left\{
\begin{array}{c}
\bar r^{s}_{1}\\
\bar r^{s}_{2}\\
\bar r^{s}_{3}
\end{array}
\right\}
=
\left\{
\begin{array}{c}
- \sin \eta \cos \theta \\
\cos \eta \cos \theta \\
- \sin \theta
\end{array}
\right\}
\label{eq:matrix_equation}
\end{equation}
from which the $\omega$-position at which the diffraction condition is fulfilled for a specific $\vctr{r}^{hkl}$ and a specific x-ray wavelength $\lambda$ can be determined. 
The third equation in \eqnref{eq:matrix_equation} provides a nonlinear equation for $\omega$ given that $\theta$ is known.  $\omega$ is evaluated from this equation using an optimization routine with the constraint that the solution is in the interval of $0 \le \omega \le \pi$.  
Note that this equation for determining $\omega$ admits  the possibility of multiple solutions
(here, 0,1 or 2 solutions are possible).
Within the code we exclude imaginary solutions.
If two solutions are possible, the code chooses the
solution that provides the lesser value of the objective function.
The first two equations of \eqnref{eq:matrix_equation} are then used to solve for $\eta$, again using an optimization routine.  For $\eta$, the solution interval is constrained to lie within $-\pi  \le \eta \le \pi$.  

With the components of the lattice reciprocal vectors known in laboratory coordinates by enforcing the Laue condition,  the direction of the diffracted beams, $\vctr{s}^o$, for each reflection is readily computed from \eqnref{eqn:scatteringvector}.
Further, the normal strain component in the direction of the normal to the diffracting plane also
can be computed using the now known scattering vector:
\begin{equation}
\epsilon^{hkl}_{qq} =\vctr{r}^{hkl} \cdot \tnsr{\epsilon} \cdot \vctr{r}^{hkl}
\label{eqn:normalstrain}
\end{equation}

\subsection{Step 2: Determining intersections of diffracted beams with the detector plane.}
\label{sec:beam_plane_intersection} 

\begin{figure}[htbp]
	\centering		
		\includegraphics[width=1\linewidth]{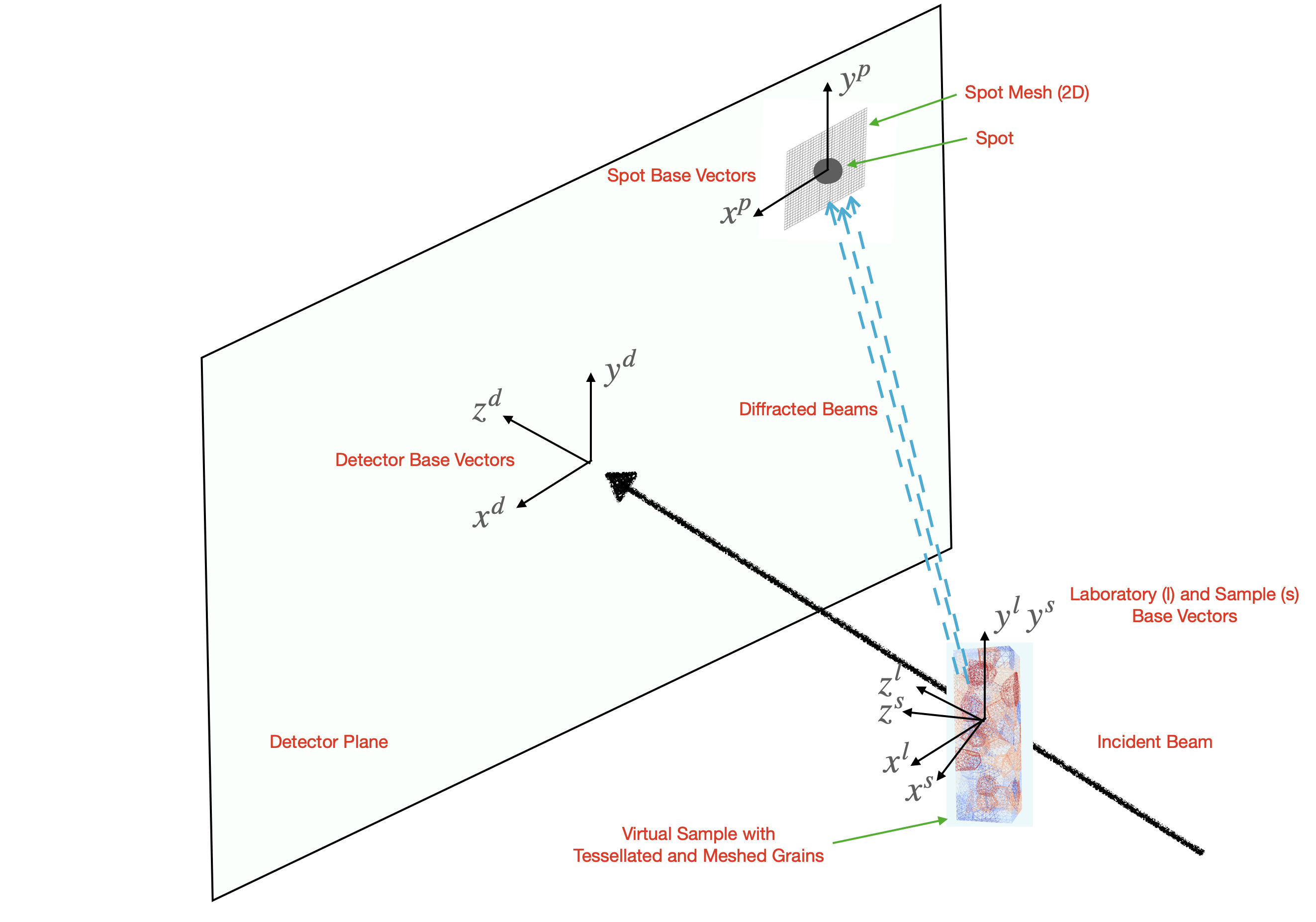}
		\caption{Schematic of the projection of diffracted beams onto the plane of the detector.}
		\label{fig:projection2detector}
\end{figure}

Using the diffraction conditions computed in Step 1,
the objective of Step 2 is to compute the intersection points between the diffracted beams and the plane of the detector
for every quadrature point of every element of the target grain.
The computation is depicted schematically in \figref{fig:projection2detector} and is performed in the following manner.

\begin{figure}[htbp]
	\centering		
		\includegraphics[width=0.5\linewidth]{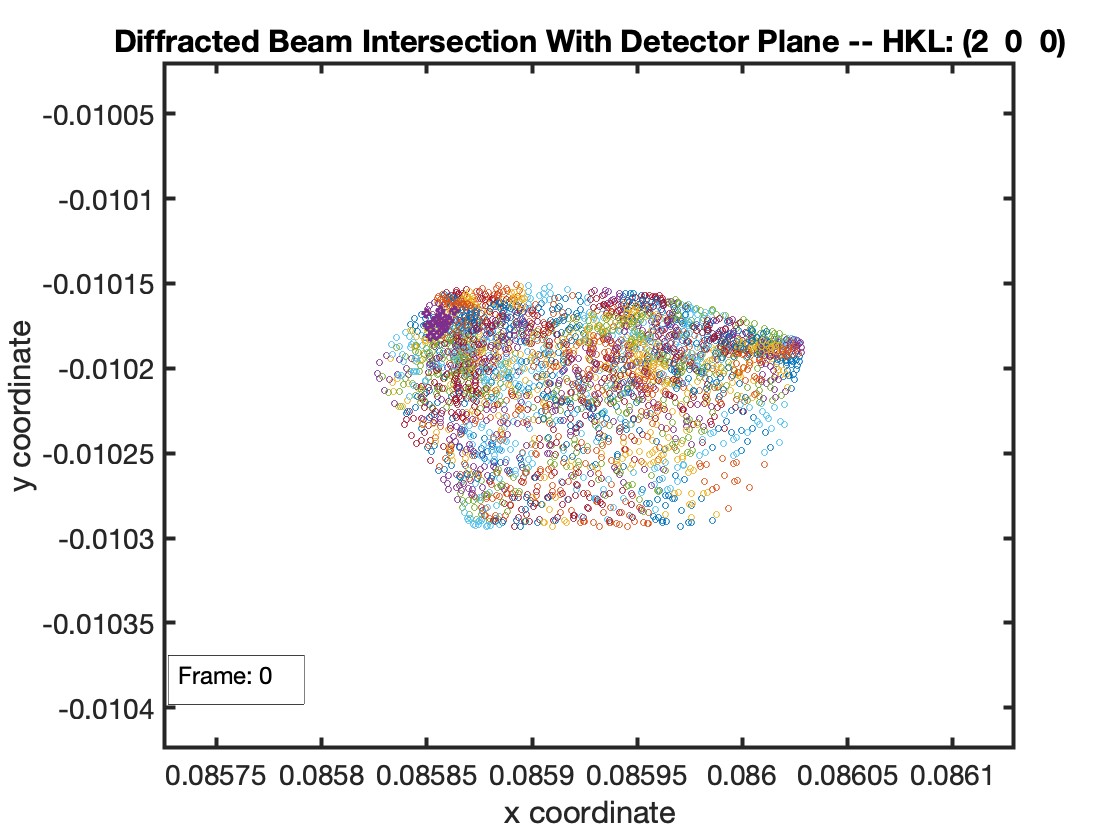}
		\caption{Projected points on the detector plane for one crystallographic reflection from within a typical grain. Each point represents a the diffraction volume associated with a quadrature point from within a finite element of the virtual sample.  }
		\label{fig:pointsondetector}
\end{figure}

 A line is defined that passes through the position of the quadrature point (defined as $\vctr{p}^q$) in the direction of the diffracted beam. 
The point of intersection of this line and the detector plane  is defined as $\vctr{p}^d$ and is given by: 
\begin{equation}
\vctr{p}^d = \vctr{p}^{q} + \alpha_d*\vctr{s}^o 
\label{eqn:pt_on_plane}
\end{equation}
where the distance along the line, $\alpha_d$, is defined by:
\begin{equation}
\alpha_d = \frac{\vctr{c}^d \cdot \vctr{n}^d}{\vctr{p}^d \cdot \vctr{n}^d}
\label{eqn:distance2plane}
\end{equation}
Here, $\vctr{c}^d$ is the line segment connecting the detector center to the origin of the laboratory coordinate system.  $\vctr{n}^d$ is the unit normal vector to the detector plane (also serving as the third base vector of the detector coordinate system, $\vctr{e}^d_3$.)
Recall that $\vctr{s}^o $ is a unit vector in the direction of the diffracted beam.
Finally, the point is written in the detector coordinate system, which consists of projecting the point onto the detector plane
as shown in \figref{fig:pointsondetector}.  
We denote the two-dimensional,  detector plane coordinate system using the $\vctr{e}^p$ base vectors.

\begin{figure}[htbp]
	\centering		
		\includegraphics[width=0.4\linewidth]{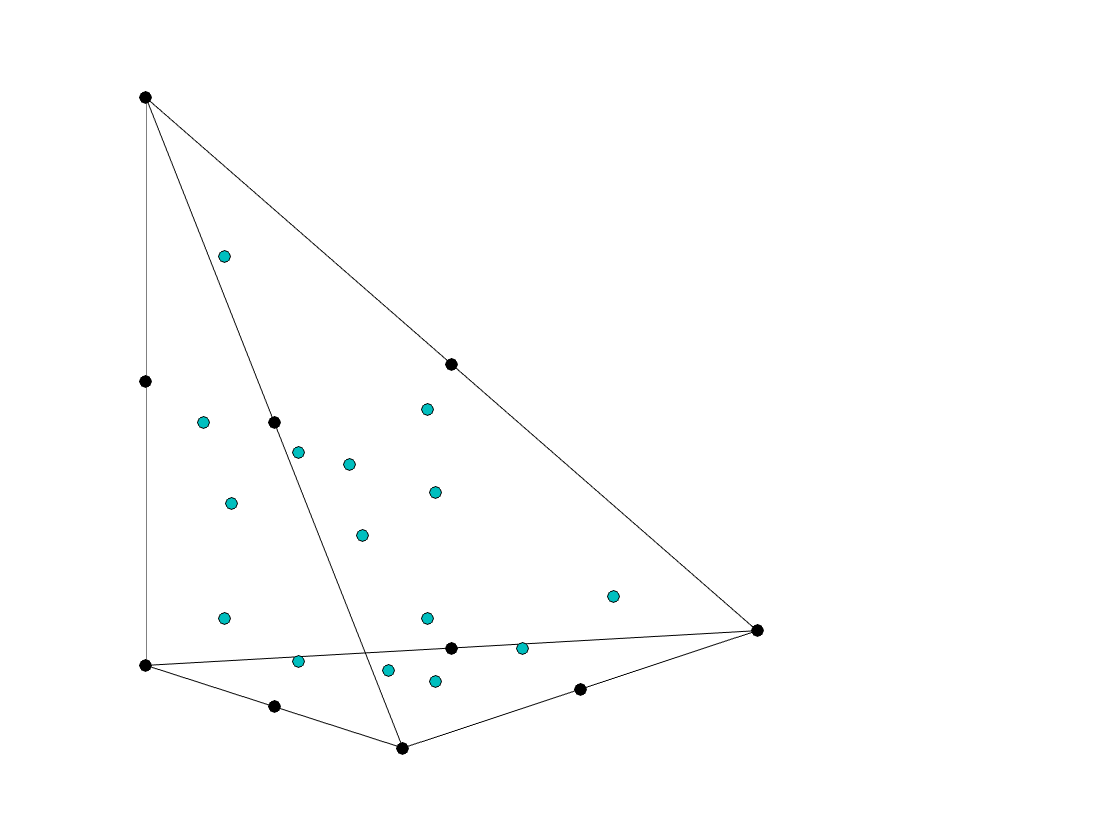}
		\caption{Ten-node tetrahedral element showing nodes (black) and quadrature points (green).}
		\label{fig:tennodetet}
\end{figure}

The quadrature point positions within an element are shown in \figref{fig:tennodetet}.  Each point carries with it an intensity, which is the product of the quadrature point weight and the elemental volume.  
Thus the weight represents the volume within the sample that is associated with the point. Note that the sum of the weights over an element is the element volume and the sum of this sum taken over all elements in the grain is the grain volume. 
Also computed at this point in the code are the distances traveled by diffracted beams within the sample -- that is, the distance between a quadrature point and the point at which a diffracted beam exits the sample for every quadrature point and reflection. 
The methodology follows the same approach as used to compute the distances to the detector plane. 
The intra-sample travel distances are used together with absorption properties stemming from the X-ray's interaction with the sample material to compute the attenuation of the diffracted beams in Step 4.

\subsection{Step 3: Determining an intensity field over the detector plane.}
\label{sec:intensity_fields}

The objective of Step 3 is to create a smooth (continuous) intensity field from the point-wise intensity data computed in Task 2 using the following procedure: 
\begin{enumerate}
\item Define a patch on the detector plane from the spans of the detector coordinates, $x^d_i$. (Individual spot typically cover a small portion of the detector, which is why they are call spots).  Position a reference mesh over the patch using a parametric mapping of the nodal point coordinates of the reference detector mesh.
\item Compute an array of values defined by the distances between the projected point positions and  quadrature point positions for elements of the re-positioned detector mesh. 
These distances are used to distribute the relative intensity for each projected point spatially in the vicinity of the projected point.  Specifically, the relative intensity of the diffracted beam derived from a quadrature point in one element of the target grain, $i^q(\vctr{x}^p)$, is spread over an area surrounding its point of intersection on the detector according to a Gaussian spread function:
\begin{equation}
i^q(\vctr{x}^p) = i^q_0 \left(\frac{1}{\sqrt{2 \pi} \tilde \xi} \right) \exp{\left(- \frac{1}{2}(\frac{\xi}{\tilde \xi})^2\right)}
\label{eqn:gaussianspreadfunction}
\end{equation}
where $\xi = \| \vctr{x}^p-\vctr{p}^p \|$, $\vctr{p}^p $ is the point position on the detector defined in \eqnref{eqn:pt_on_plane}, and $\tilde \xi$ is the square root of the variance of the intensity distribution.  The value of $i^q_0$ is the attenuated weight associated with the diffracted beam from quadrature point $q$. 
\item Compute nodal point values for the intensity distribution resulting from all projected points using an $\rm L_2$ inner product procedure described below.
\item Making use of the piecewise interpolation functions, display a continuous relative intensity distribution to represent  an image of the diffraction spot.
\end{enumerate}

The benefit is using an $\rm L_2$ approach for determining the field representation for the total relative intensity, $ i(\vctr{x}^p) $,  
that it provides a best fit to the net sum of the distributions of the quadrature point relative intensities given in \eqnref{eqn:gaussianspreadfunction}.
The total relative intensity is represented  over a patch on the detector that encompasses a spot using a piecewise polynomial approximation commonly employed in finite element interpolation: 
\begin{equation}
 i(\vctr{x}^p) \,=\,  \left[ N (\vctr{x}^p) \right] \{I \}
\label{interpolation}
\end{equation}
where $ \{I \}$ are the nodal point values of the approximation and  the interpolation functions, $\left[ N (\vctr{x}^p) \right] $.  This spatial approximation possess $C^0$ continuity (continuity of the field across element boundaries, but not of the field's spatial derivatives.)
To determine the nodal values, $ \{I \}$,
a weighted residual, $R$, is formed over detector patch, $A$, using \eqnref{interpolation}:  
\begin{equation}
\label{eqn:eqres}
R\,=\, \int_{A} \Psi(\vctr{x}^p) \left[ i(\vctr{x}^p)  - [N(\vctr{x}^p)] \{ I \} \right] \, \rm{d}A
\end{equation}
where $\Psi(\vctr{x}^p)$ is the weighting function and $A$ is the patch area.
Standard finite element procedures are followed to develop a matrix equation for the nodal point intensities from \eqnref{eqn:eqres}:
\begin{equation}
\left[ A \right] \{I \} \,=\, \{ B \} 
\label{eqmatrix}
\end{equation}
where the elemental contributions for $\left[ A \right]$ and $\{ B \}  $, respectively, are:
\begin{equation}
\left[ A^e \right] \,=\, \int_{A^e}  [ N(\vctr{x}^p)]^T [ N(\vctr{x}^p)] \rm{d}A 
\label{Amatrix}
\end{equation}
and 
\begin{equation}
\{ B^e \}  \,=\, \int_{A^e}  [ N(\vctr{x}^p)]^T i(\vctr{x}^p) \rm{d}A 
\label{Bmatrix}
\end{equation}
The projected point intensities enter the residual through the evaluation of $\{ B^e \} $.
Because the integral that defines $\{ B^e \} $ is evaluated by numerical quadrature, 
pointwise evaluation of  $\{ i(\vctr{x}^p) \}$ is all that is required:
\begin{equation}
\{ B^e \}  \,=\, \sum_{i=1}^{n^{qp}} [ N(\vctr{x}^p |^{qp})]^T i(\vctr{x}^p|^{qp})  w^{qp} \Delta A 
%\{ B^e \}  \,=\, \sum_{i=1}^{nqp} [ N(\vctr{x}^p |^{qp}|^{qp})]^T i(\vctr{x}^p|^{qp}) w  \DeltaA 
\label{eq:BMatrix_quadrature}
\end{equation}
where $n^{qp}$ is the number of quadrature points used with the detector mesh elements and $w^{qp} $ is the associated
quadrature point weight.
The pointwise values of the total relative intensity are evaluated by summing the individual contributions from
all of the projected beams:
\begin{equation}
 i(\vctr{x}^p|^{qp}) \,=\, \sum_{i=1}^{n^{pp}} i^q(\vctr{x}^p|^{qp}) 
\label{eq:qpvalue}
\end{equation}
where $n^{pp}$ is the total of all  beams projected onto the detector from the virtual sample target grain (product of the number of elements in the target grain times the number of quadrature points within each of those elements).  
Solving \eqnref{eqmatrix} for the nodal point values of the total relative intensity, $ \{I \}$,  completes the information needed to construct 
the piecewise approximation for $i(\vctr{x}^p)$ using \eqnref{interpolation} thereby giving a spatial map of the spot intensity in its vicinity.

\subsection{Step 4: Creating a detector image.}
\label{sec:detector_image}

The objective of Step 4 is to create a facsimile of the detector image using the intensity distribution computed in Step 3.  The detector image is constrained by the detector's construction.  Namely, it is an array of pixels with fixed size and shape.  Each pixel displays a single value of intensity determined principally by the total flux from diffracted beam that shines upon it. However, this value can be altered to some extent by neighboring pixels, referred to as point spread.  The procedure for constructing a detector images includes the steps summarized below.
\begin{enumerate}
\item  For each reflection, a mesh with one-to-one element-to-pixel construction is defined that overlays the detector patch discretized in Step 3.  
The elements are superparametric: geometry is mapped with bi-linear interpolation; the field variable (intensity) is mapping with piecewise constant interpolation.  We recognize that using parametric elements for this purpose appears to introduce unnecessary complexity, but defend the choice as it opens the door to including greater detail regarding the physical behavior of detector pixels later.  
\item  There is a relation between the intensity distribution (Step 3) mesh and the detector(Step 4, pixel) mesh.  Namely, the intensity distribution mesh is a refinement of the pixel mesh.  The intensity distribution mesh subdivides the detector mesh elements into one or more subdivisions in both detector in-plane directions. In the demonstration example shown in \secref{sec:demonstration}, for instance, there are 256 (16x16)  intensity distribution mesh elements with each detector mesh element.  The two meshes are made in coordination: in Step 3, the patch around a spot is sized to be an even integer multiple of the pixel size.  The intensity distribution mesh is then defined to have its elements lying within a single detector mesh element.  
\item  Mapping of a relative intensity distribution computed in Step 3 to its corresponding detector mesh is done simply by averaging the integrated intensities of the elemental distributions over the intensity distribution elements with a detector element.  The integrations over individual elements is performed by quadrature at the end of Step 3 for every element in the intensity distribution mesh.  
\item  A point spread function is then applied to adjust the elemental (pixel) values over the detector mesh. The point spread function reported in \cite{won_par_mil_daw_13} has been used here.  However, this function is detector-dependent, so it is anticipated users will modify the code as appropriate at the point indicated within the code.  Before applying the point spread function, the detector mesh is enlarged by a factor of two in each direction (expanded about its center) to accommodate the larger numbers of pixels being activated as a consequence of the spread.
\item Visualization of the image completes the step. This image is a patchwork of constant value squares with each square coinciding with a pixel. 
\end{enumerate}

Step 4 above is carried out in discretized manner using a template to modify elemental (pixel) values according to the spread function. 
The spread function $f(s)$ is a decreasing function of the distance, $s$, measured from the center of a pixel.  The energy reaching the detector above a chosen pixel is re-distributed across it and its neighbors according to the proportions specified by this function. 
For the spread function implemented, the effect of the spread extends only about 2 pixels in each coordinate direction.  
In this case, the template for implementing the spread function for a single pixel is given by:
\begin{equation}
\begin{bmatrix}
 &  & \tilde f(2s^p) &  & \\
 & \tilde f({\sqrt{2}}s^p) & \tilde f(s^p) & \tilde f({\sqrt{2}}s^p) & \\
\tilde f(2s^p) & \tilde f(s^p) & \tilde f(0) & \tilde f(s^p) & \tilde f(2s^p)\\
 & \tilde f({\sqrt{2}}s^p) & \tilde f(s^p) & \tilde f({\sqrt{2}}s^p) & \\
 &  & \tilde f(2s^p) &  & \\
\end{bmatrix}
\label{eq:ps_template}
\end{equation}
where $s^p$ is the center-to-center distance between neighboring pixels (equivalently, the pixel size)
and $\tilde f$ indicates a rescaling of $f$ values in the template so that the sum equals unity.  The rescaling is done to preserve the total energy over the spot during the application of the point spread function.

The template values for each pixel are re-written as a vector with values appearing in the rows for the corresponding elements of the detector mesh.  The set of vectors are concatenated to form a point spread matrix, $[ PS ] $,
which is used to map the pixel values prior to application of the point spread function to the final detector values:
\begin{equation}
\{I^d\} = [ PS ] \{I^{d_0}\} 
\label{eq:ps_matrix}
\end{equation}
where $\{I^d\}$ and $\{I^{d_0}\}$ are the elemental (pixel) values over the detector mesh and before application of the point spread function, respectively.  The sum of the pixel intensities over the detector mesh subsequent to application of the point spread function equals the sum prior to its application. 

\section{Demonstration Example}
\label{sec:demonstration}

\subsection{Problem definition}
\label{sec:demo_definition}

\begin{figure}[h]
   \centering
   \includegraphics[width=3.0in]{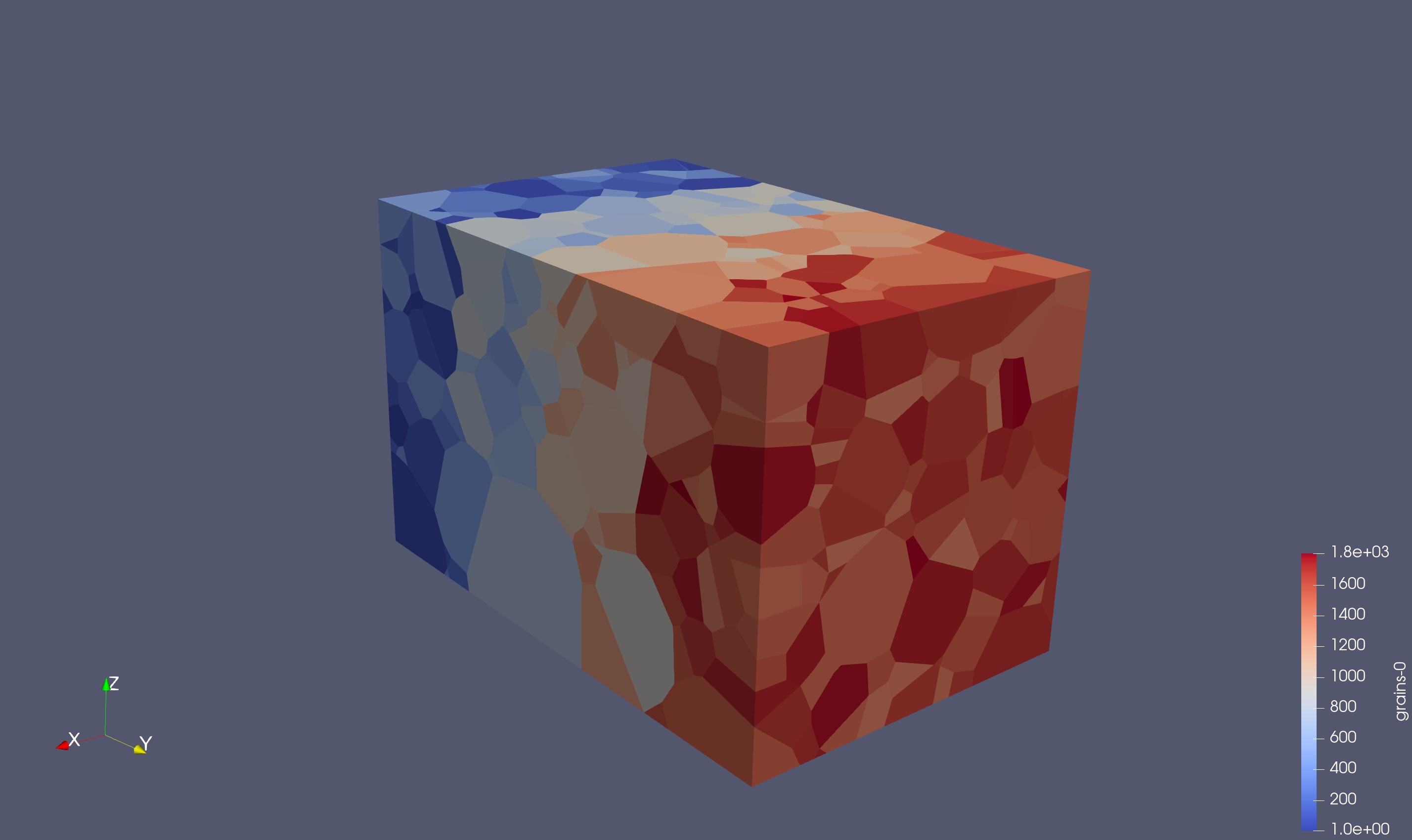}
   \caption{\neper\, tesselation of complete stainless steel sample. Color indicates the grain number.}
   \label{fig:aps_ss_full_grains}
\end{figure}

To demonstrate the functionality of the virtual diffractometer we show results for the tensile loading of a
stainless steel sample.  
The HEDM data needed to define the virtual sample were provided by Dr.~Jun-Sang Park.\footnote{
This research used resources of the Advanced Photon Source, a U.S. Department of Energy (DOE) Office of Science user facility at Argonne National Laboratory and is based on research supported by the U.S. DOE Office of Science-Basic Energy Sciences, under Contract No. DE-AC02-06CH11357.}
The full experimental records include comprehensive sets of data at numerous points in a loading program.  
For this demonstration, we use only the grain center-of-mass and lattice orientation information at the initial (unloaded) state to define a virtual sample using \neper. 
This is an anticipated starting point for using the virtual diffractometer as a planning tool for HEDM experiments.
The full virtual sample has $\approx 1800 $ grains  and is shown in \figref{fig:aps_ss_full_grains}.  

\begin{figure}[htbp]
	\centering		
	\begin{subfigure}{.3\textwidth}
		\centering
		\includegraphics[width=1\linewidth]{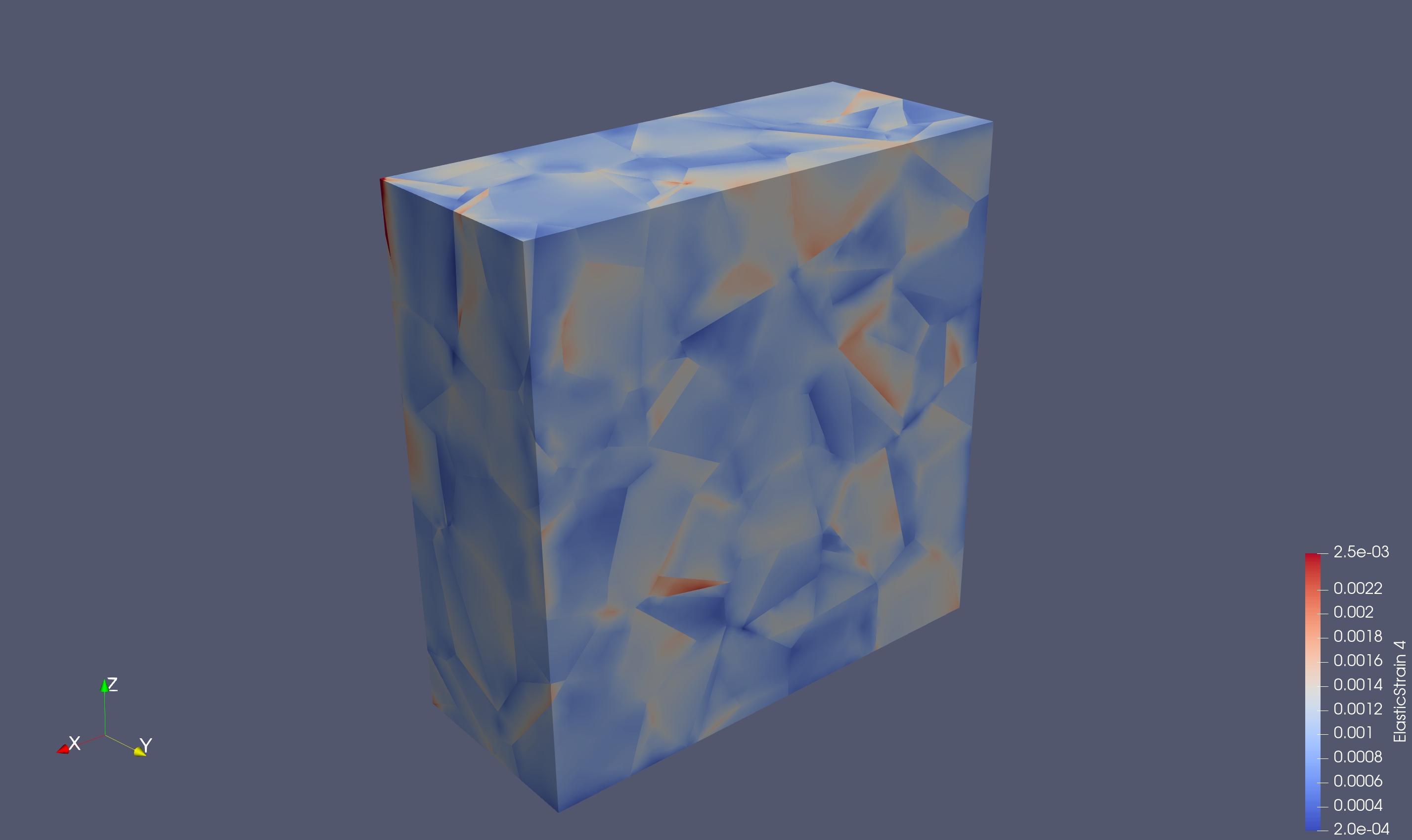}
		\caption{ }
		\label{fig:epsyy_allgrain}
	\end{subfigure}%
	\quad
	\begin{subfigure}{.3\textwidth}
		\centering
		\includegraphics[width=1\linewidth]{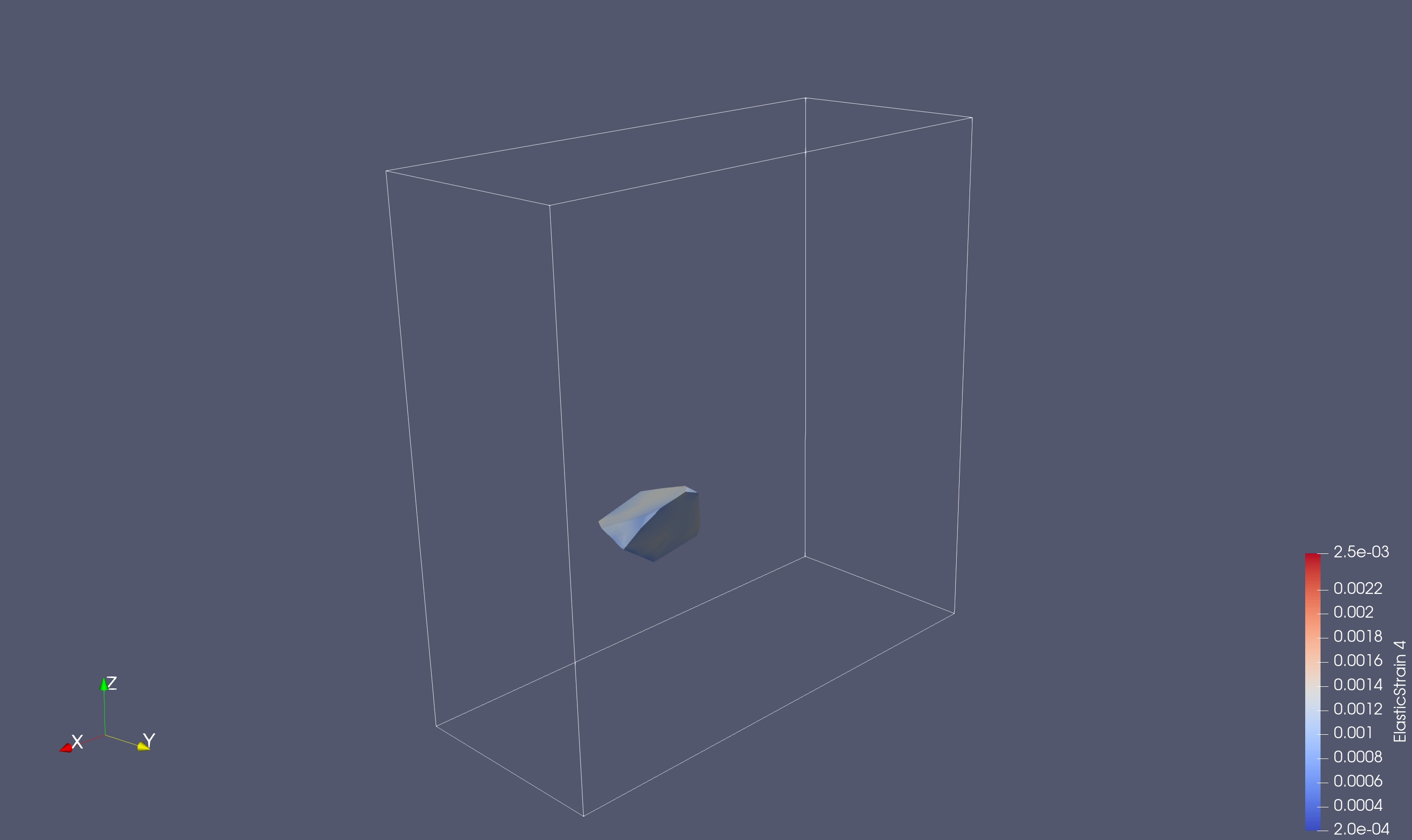}
		\caption{ }
		\label{fig:epsyy_grain80}
	\end{subfigure}	
	\quad
	\begin{subfigure}{.3\textwidth}
		\centering
		\includegraphics[width=1\linewidth]{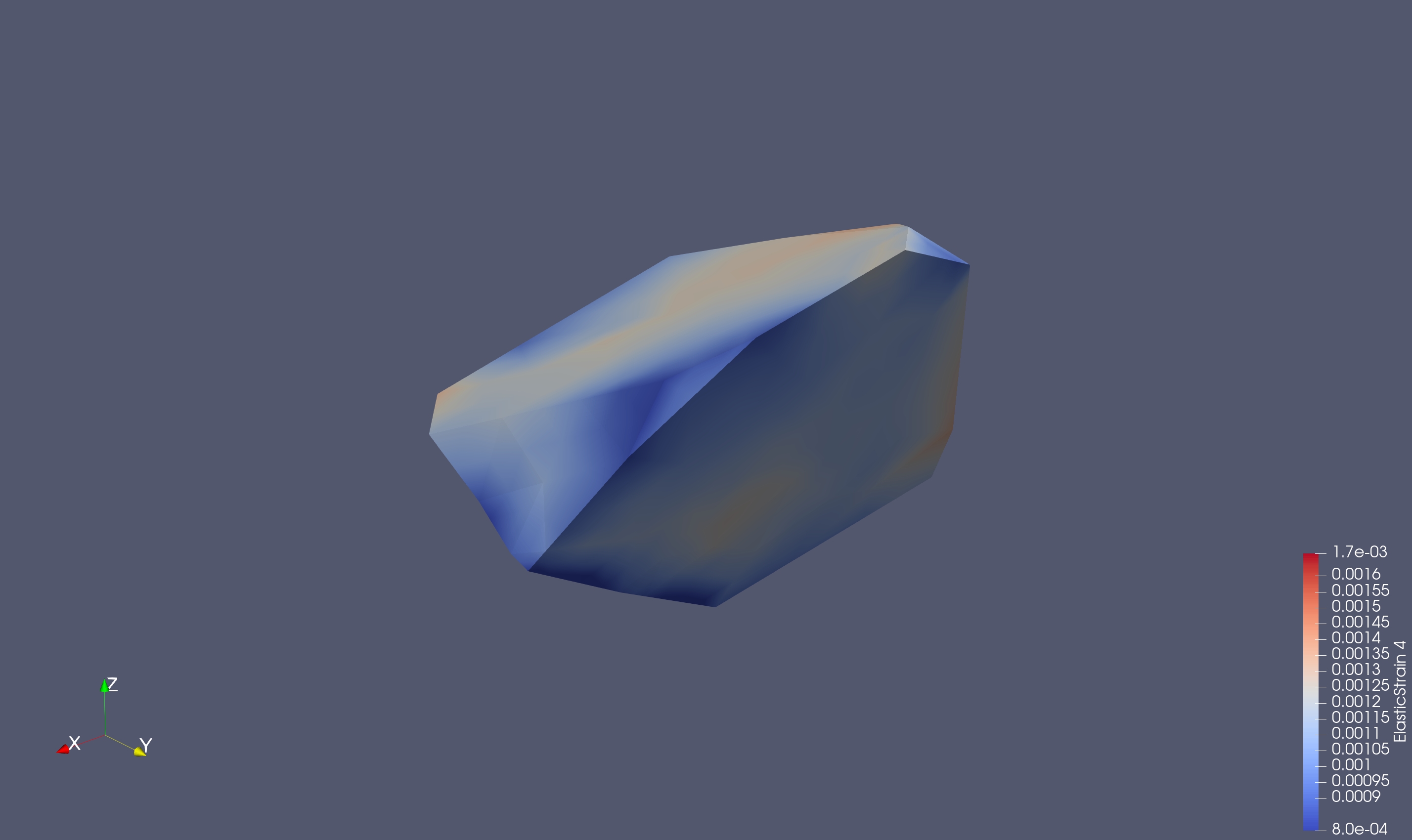}
		\caption{ }
		\label{fig:epsyy_grain80_zaxis}
	\end{subfigure}
		\caption{\mechmet\,  model of Layer 2 showing (a) the elastic strain distribution (axial component) over the entire layer (b) the elastic strain distribution  (axial component) over target grain (80) and (c) a close-up of the target grain (80), again with elastic strain distribution  (axial component) plotted. }
		\label{fig:elementalstrains}
\end{figure}

The experimental data were collected in four separate layers from which the complete sample was reconstructed. 
For the virtual diffractometer demonstration, we employ only one of the four subsets of data (Layer 2).
Layer 2 has $\approx 440$ grains and the \neper-generated virtual sample is shown in \figref{fig:elementalstrains}.
The tessellated sample was discretized with \neper\, with 78,696 10-node tetrahedral elements.  This is a relatively
coarse mesh, but is adequate for the purpose of demonstrating the virtual diffractometer capabilities.
Using this mesh, the elastic strain distribution was computed using  \mechmet. 
For this purpose, the single crystal elastic moduli for stainless steel were taken from literature and are given in \tabref{tab:ElasticConstants}.

In the \mechmet\,  simulation,  the sample was extended along its $y-$axis to induce a nominal
axial strain of  0.1\%.  
The elastic strain distribution, shown in  \figref{fig:elementalstrains}a, shows the spatial heterogeneity in the strain field
typical of polycrystals with moderate levels of elastic anisotropy (as is the case with stainless steel).
Note that the strain fields are smooth over each grain, but display discontinuities at the grain boundaries.
This reflects a post-processing step performed in \mechmet\, as the raw results generated using $C^0$ elements 
do not possess intra-element continuity.
The virtual diffractometer currently assumes that the strain data from the mechanical simulation will 
consist of a value for each component of the strain tensor for each element of the mesh and that this value is the average value for the element. 
\begin{table}[ht]
	\centering
	\caption{Single crystal elastic constants using the strength of materials convention ($\tau_{44} = c_{44} \gamma_{44}$).}
	\begin{tabular} {c c c c c}		
		phase & $C_{11}$ & $C_{12}$ & $C_{44}$&   source \\
		& (GPa) & (GPa) & (GPa)  & - \\ \hline
		FCC & 205 & 138 & 126& \cite{Ledbetter01a} \\ 
	\end{tabular}	
	\label{tab:ElasticConstants}
\end{table}

\begin{figure}[ht]
   \centering
   \includegraphics[width=2.5in]{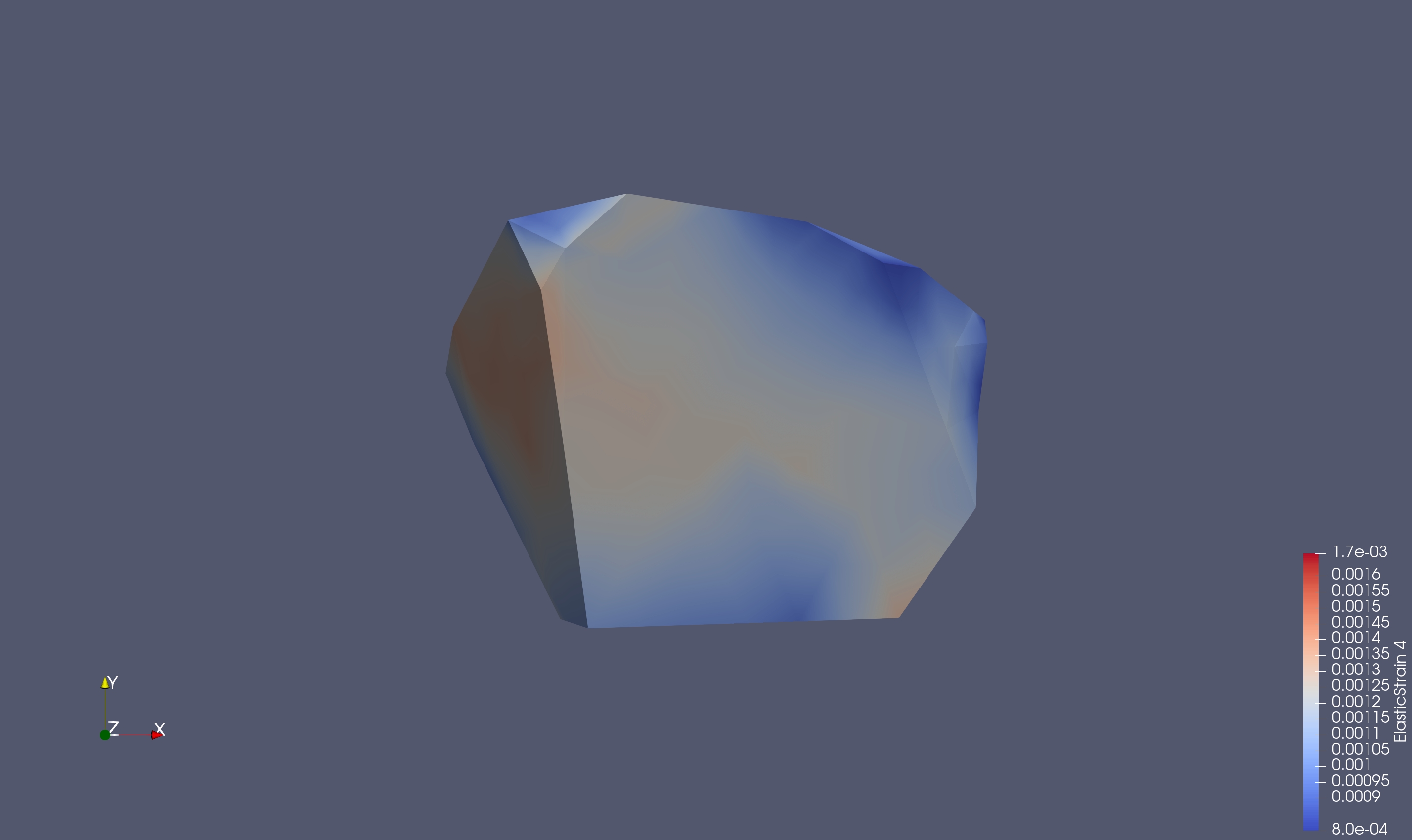}
   \caption{Close-up of target grain (80) viewed along an axis parallel to, but opposite, the incident beam direction vector  and showing axial strain associated with an 0.1\% extension in the $y$ direction.}
   \label{fig:targetgrainzaxis}
\end{figure}

Grain 80 is the designated target grain and also is shown in  \figref{fig:elementalstrains}a and \figref{fig:elementalstrains}b.
Grain 80 was selected because it is an interior grain of relatively average size.  
There are 249 tetrahedral elements discretizing this grain in the finite element mesh.  
A second view of Grain 80 appears in  \figref{fig:targetgrainzaxis}.
Here, the viewing axis of the plot is parallel to, but opposite, the incident beam direction. This view is helpful later when 
examining images generated by the virtual diffractometer.  
This stainless steel has a face-centered cubic crystal structure. 
The plane families of interest were specified to be $\{ 111 \}$, $\{ 200 \}$, and $\{ 220 \}$.
%\clearpage

\subsection{Results from Step 1: Elemental diffraction conditions }
\label{sec:demo_task1}

If requested by the user, the diffractometer will create plots and images 
associated with each step summarized in \secref{sec:comp_tasks}.
This subsection and the two that follow present these results step-by-step, beginning with Step 1.
Note that the detector coordinate system is as shown 
in \figref{fig:projection2detector}, which means that the detector is viewed from a direction opposite the direction of the incident beam.

The virtual diffractometer prompts the user for a scattering vector of interest.  
The code then identifies the particular reflection for each family of reflections having its reflection (reciprocal) vector  closest to the scattering vector designated by the user.   
Only the angle between the loading direction and the scattering vector is considered in choosing the particular reflection. 
In this demonstration example, two cases are examined with the virtual diffractometer, one for spots associated with near-axial strains and the other with spots associated with near-transverse strains.  
Thus, angles of 0 and $\pi/2$ were designated.  
None of the chosen reflections exactly match these angles, but rather are the reflections that are closest to those angles.  
For convenience, these are called the 'near-axial' and 'near-traverse' cases for 0 and $\pi/2$ angles, respectively.
The identified scattering vector directions depend on the grain's lattice orientation and thus will differ from grain-to-grain.

\begin{figure}[htbp]
	\centering		
	\begin{subfigure}{.4\textwidth}
		\centering
		\includegraphics[width=1\linewidth]{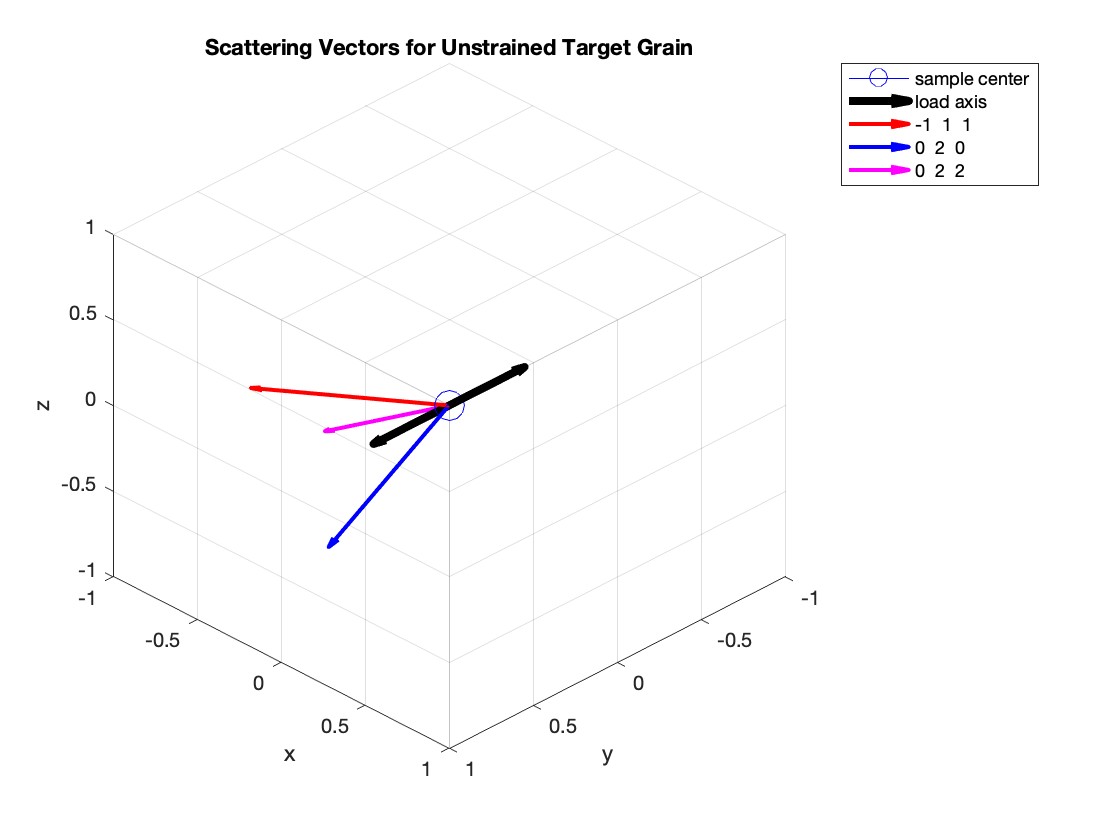}
		\caption{ }
		\label{fig:scattering_vectors_0deg}
	\end{subfigure}%
	\quad
	\begin{subfigure}{.4\textwidth}
		\centering
		\includegraphics[width=1\linewidth]{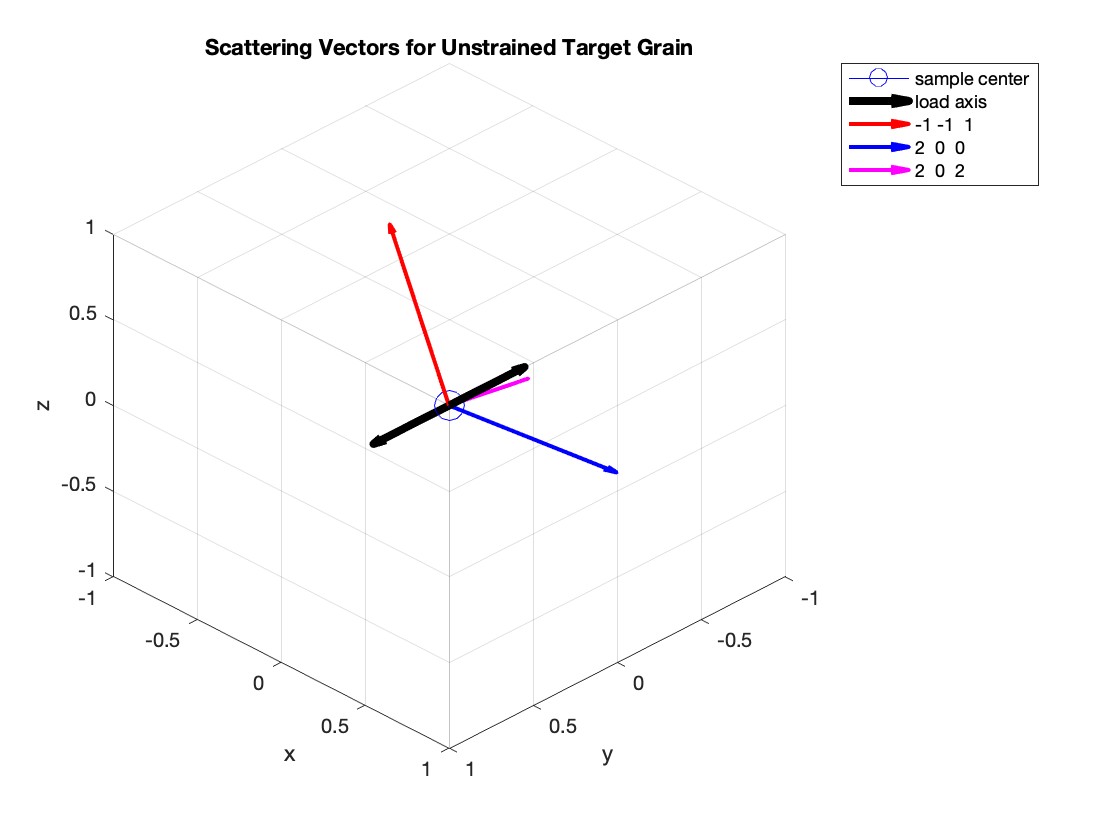}
		\caption{ }
		\label{fig:scattering_vectors_90deg}
	\end{subfigure}	
		\caption{Scattering vectors chosen by the virtual diffractometer to best match axial and transverse directions: (a) ones most closely aligned with the axial (loading) axis and (b) ones most closely transverse to the loading axis. }
		\label{fig:scatteringvectors}
\end{figure}

The scattering vectors are shown in \figref{fig:scatteringvectors} for the 'near-axial' and 'near-traverse' cases.
The legends indicate the particular reflection ($hkl$) chosen for each case. 
A table of scattering vectors with associated normal strain values is given in \tabref{tab:diffractionconditions}.   
Note that the 'near axial' normal strains are positive and the 'near traverse' normal strains are negative.  
We see that the $\bar111$ axial strains are smaller than the $020$ axial strains.  
We expect this trend but have to caution that the scattering vectors are not closely aligned with the loading direction nor each other, so this trend is not definitive.  
The detector positions will be discussed in \secref{sec:demo_task3}.
\begin{table}[ht]
	\centering
	\caption{Approximate detector position and associated average normal strains for near-axial and near-transverse scattering vectors.}
	\begin{tabular} {c c  c c }		
		$hkl$ & $x^p_1$ (m) & $x^p_2$  (m) & $\epsilon^{hkl}_{qq}$\\ \hline
	        $\bar111$ & 0.042 & 0.062 &   $0.35\times 10^{-3}$ \\
	        $020$ & -0.037 & 0.078 &   $1.24\times 10^{-3}$ \\
	        $022$ &  0.044 & 0.114 & $0.70\times 10^{-3}$ \\ \hline

	        $\bar1\bar11$ & 0.073 & -0.016 &  $-0.01\times 10^{-3}$ \\
	        $200$ &  0.086 & -0.010 & $-0.44\times 10^{-3}$ \\ 
	        $202$ & 0.120 & 0.025 &   $-0.41\times 10^{-3}$ \\ \hline
	\end{tabular}	
	\label{tab:diffractionconditions}
\end{table}

The  sample rotation angle, $\omega$, is a critical diffraction parameter in the experimental set-up.  
In an undeformed grain with spatially uniform lattice orientation, there is a single rotation angle for the entire grain 
for each  $hkl$ plane. 
However, if either of these conditions is not met, the sample rotation angle differs from point-to-point within the grain
(and thus from element-to-element within the mesh).
The virtual diffractometer computes this angle for all designated reflection planes for every element of the target grain and generates frequency distribution plots for the loaded states.
The code loads orientation and strain data for each load step.  
The load steps are referred to as 'frames' and number from 1 to the maximum number of load steps.
Frame 0 refers to the initial unloaded state.  Orientations for this frame are taken from the \neper\, file and
the strain is assumed to be zero. 
Note that distributions for the diffraction angle $\eta$ and the Bragg angle $\theta$ also are computed, but are not
shown here. 

\begin{figure}[htbp]
	\centering		
	\begin{subfigure}{.3\textwidth}
		\centering
		\includegraphics[width=1\linewidth]{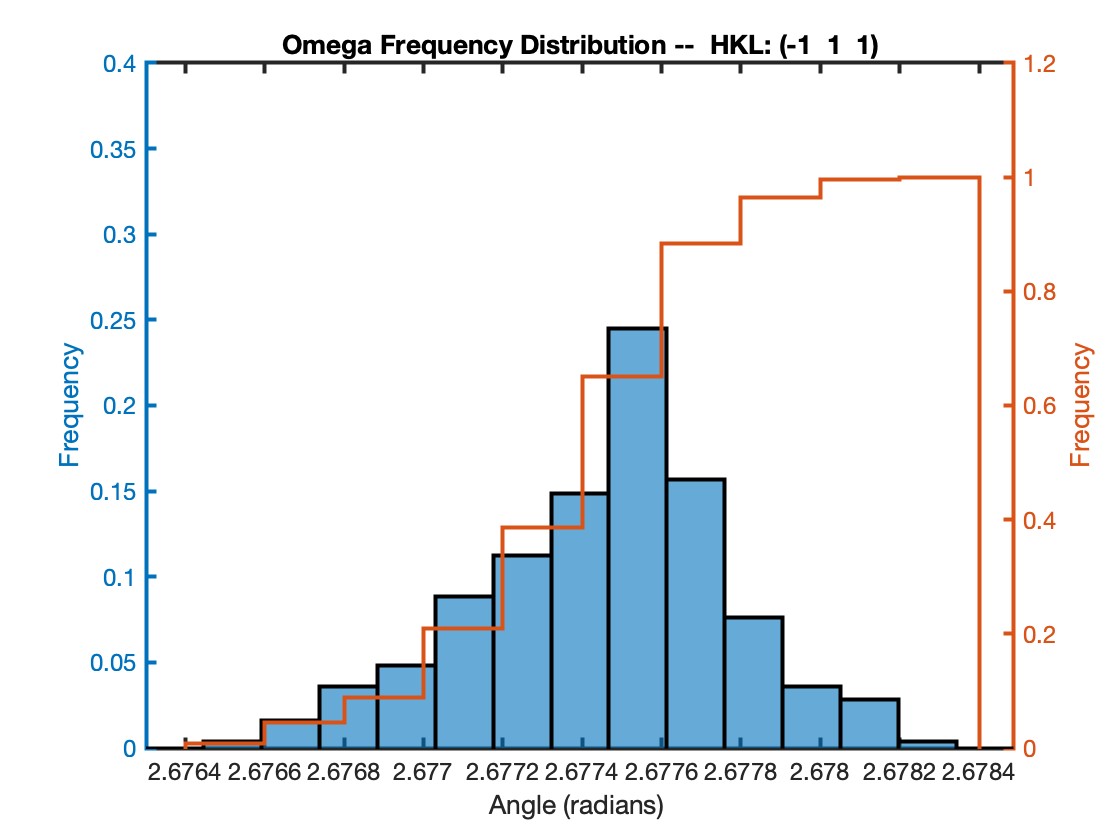}
		\caption{ }
		\label{fig:omega_frequency_bar111l}
	\end{subfigure}%
	\quad
	\begin{subfigure}{.3\textwidth}
		\centering
		\includegraphics[width=1\linewidth]{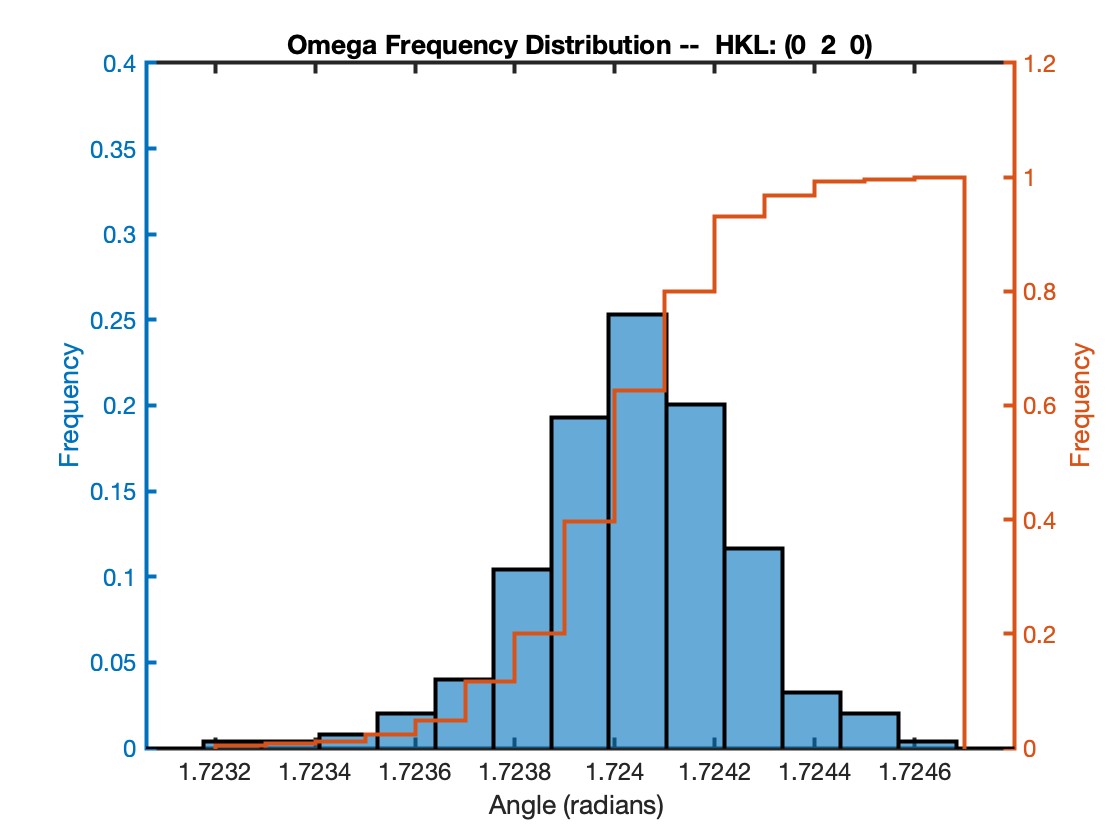}
		\caption{ }
		\label{fig:omega_frequency_020l}
	\end{subfigure}	
	\quad
	\begin{subfigure}{.3\textwidth}
		\centering
		\includegraphics[width=1\linewidth]{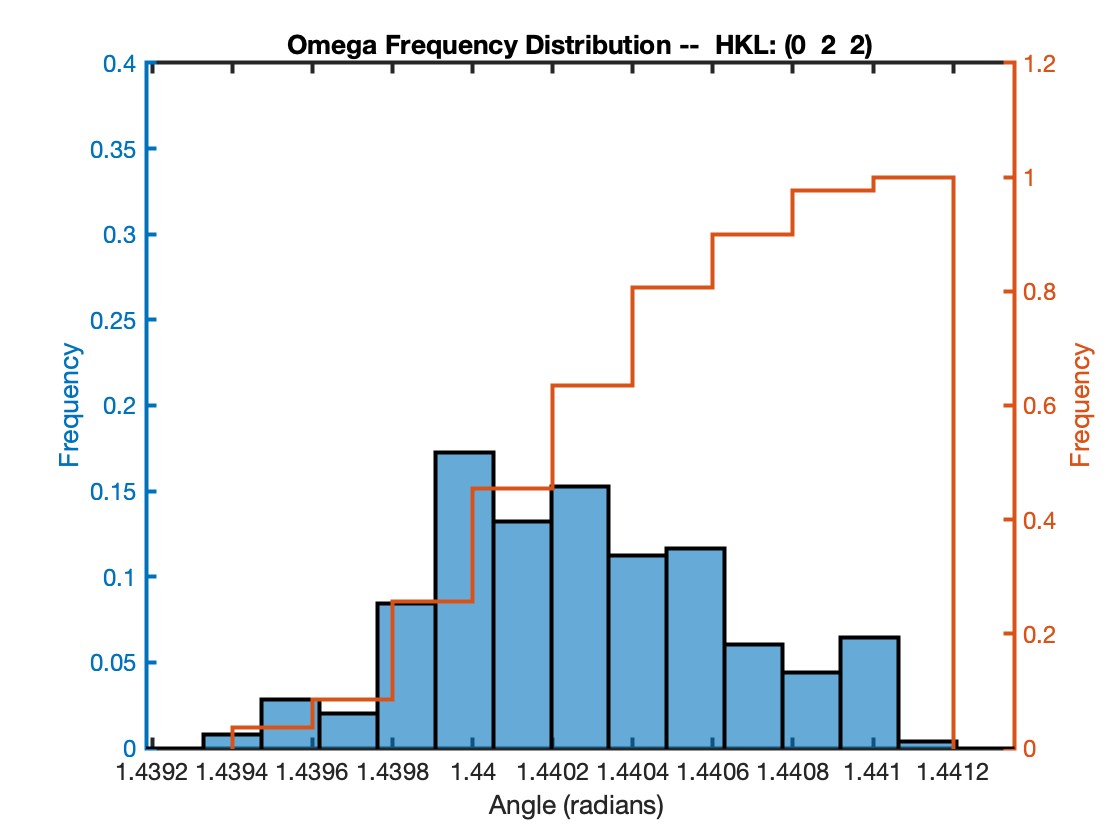}
		\caption{ }
		\label{fig:omega_frequency_022l}
	\end{subfigure}
		\caption{Frequency distributions for sample rotation angle, $\omega$,   for the most axial scattering vectors: (a) $\bar111$ reflection, (b) 020 reflection, and (c) 022 reflection. }
		\label{fig:omega_frequency_loaded_0deg}
\end{figure}

\begin{figure}[htbp]
	\centering		
	\begin{subfigure}{.3\textwidth}
		\centering
		\includegraphics[width=1\linewidth]{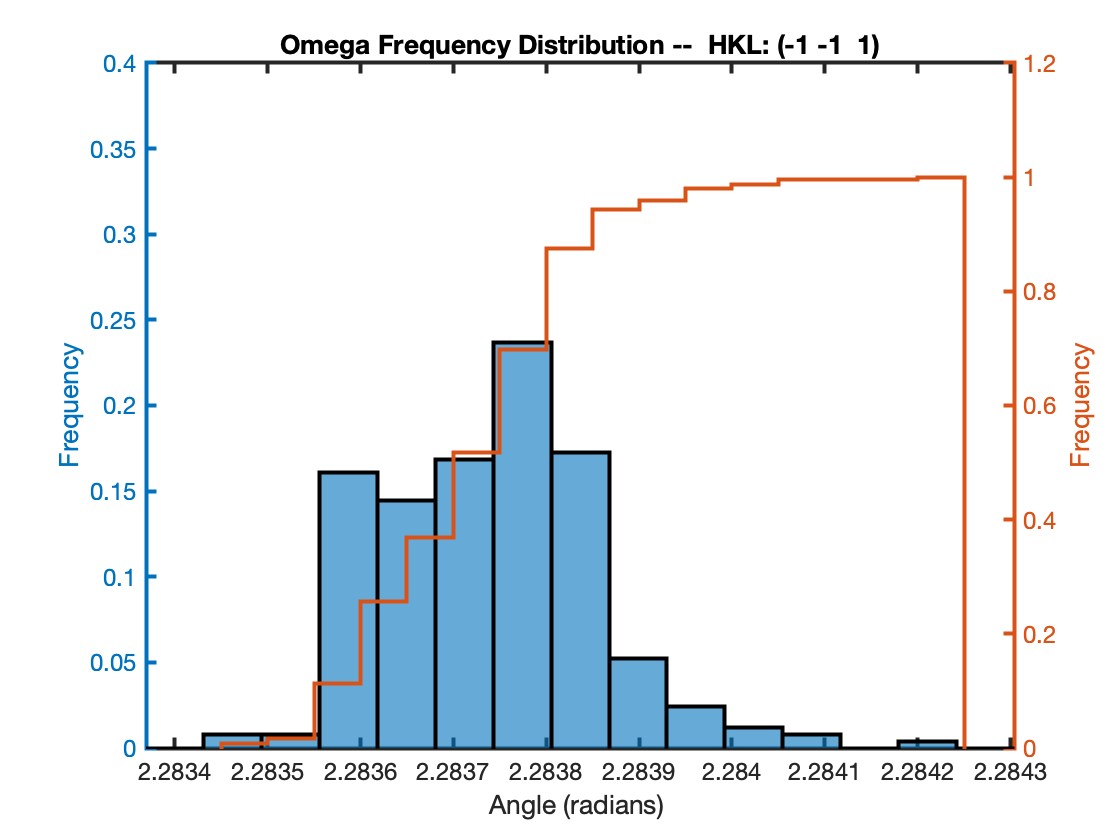}
		\caption{ }
		\label{fig:omega_frequency_bar1bar11l}
	\end{subfigure}%
	\quad
	\begin{subfigure}{.3\textwidth}
		\centering
		\includegraphics[width=1\linewidth]{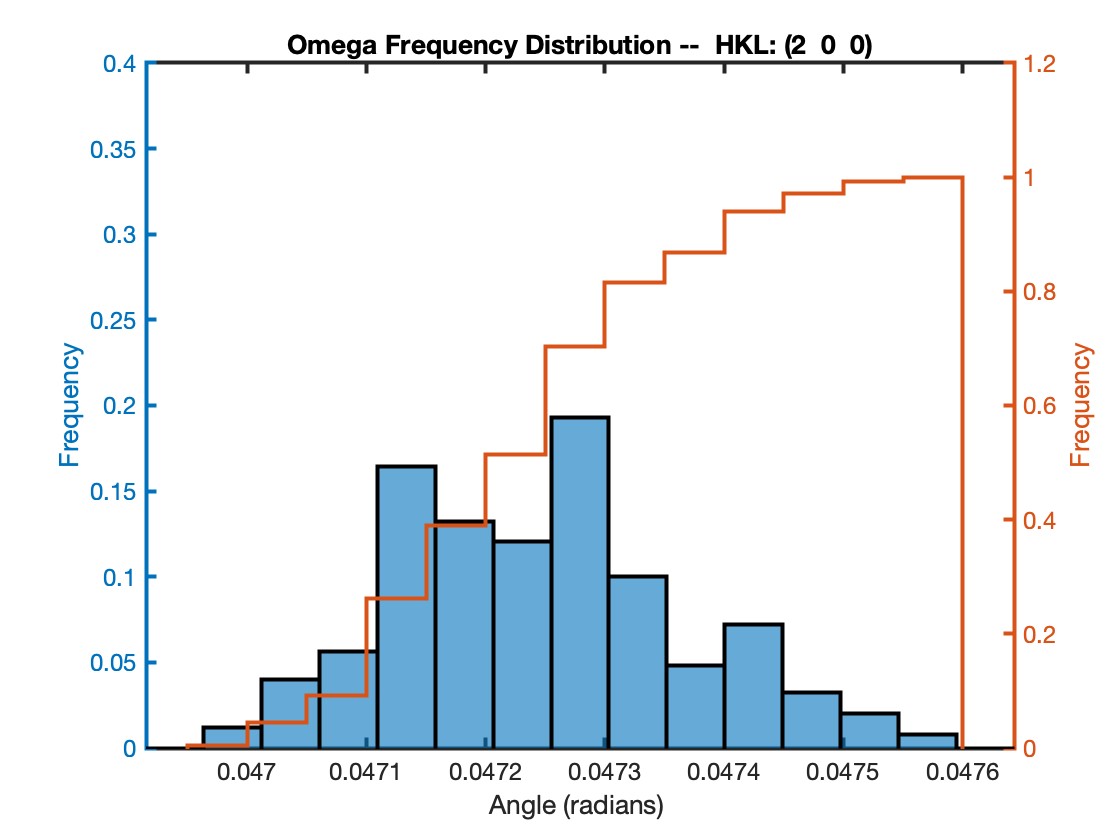}
		\caption{ }
		\label{fig:omega_frequency_200l}
	\end{subfigure}	
	\quad
	\begin{subfigure}{.3\textwidth}
		\centering
		\includegraphics[width=1\linewidth]{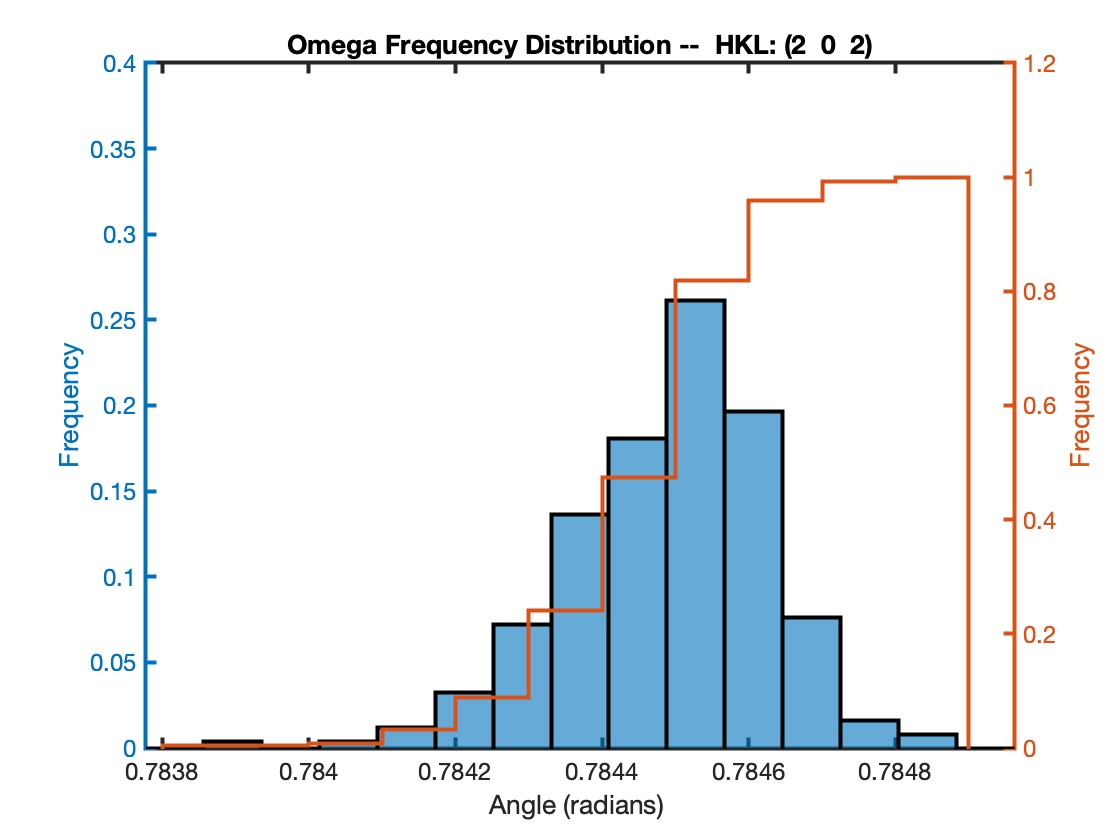}
		\caption{ }
		\label{fig:omega_frequency_202l}
	\end{subfigure}
		\caption{Frequency distributions for sample rotation angle, $\omega$, for the most transverse scattering vectors: (a) $\bar1\bar11$ reflection, (b) 200 reflection, and (c) 202 reflection.    }
		\label{fig:omega_frequency_loaded_90deg}
\end{figure}

The $\omega$\, frequency distributions for the 'near-axial' scattering vectors are shown in \figref{fig:omega_frequency_loaded_0deg}.   
The spread in $\omega$\, is a consequence of the elastic strain altering the lattice vectors as 
indicated by \eqnref{eqn:adeformed}.
For all reflections, the widths of the distributions are small, but are consistent with the typically small spot size of
a crystal that has not undergone plastic deformation.   
 The distributions differ among the three reflections owing to the spatial heterogeneity in the strain field.
 Qualitatively similar $\omega$\, frequency distributions arise for the 'near-transverse' scattering vector
 as seen in \figref{fig:omega_frequency_loaded_90deg}.
 Comparing across all of the frequency distributions, one can see that the strain heterogeneity 
 affects the distributions for the same reflections as well as causing distributions with similar
 scattering vectors to differ. 
%\clearpage

\subsection{Results from Step 2: Projected point plots}
\label{sec:demo_task2}
\begin{figure}[htbp]
	\centering		
	\begin{subfigure}{.3\textwidth}
		\centering
		\includegraphics[width=1\linewidth]{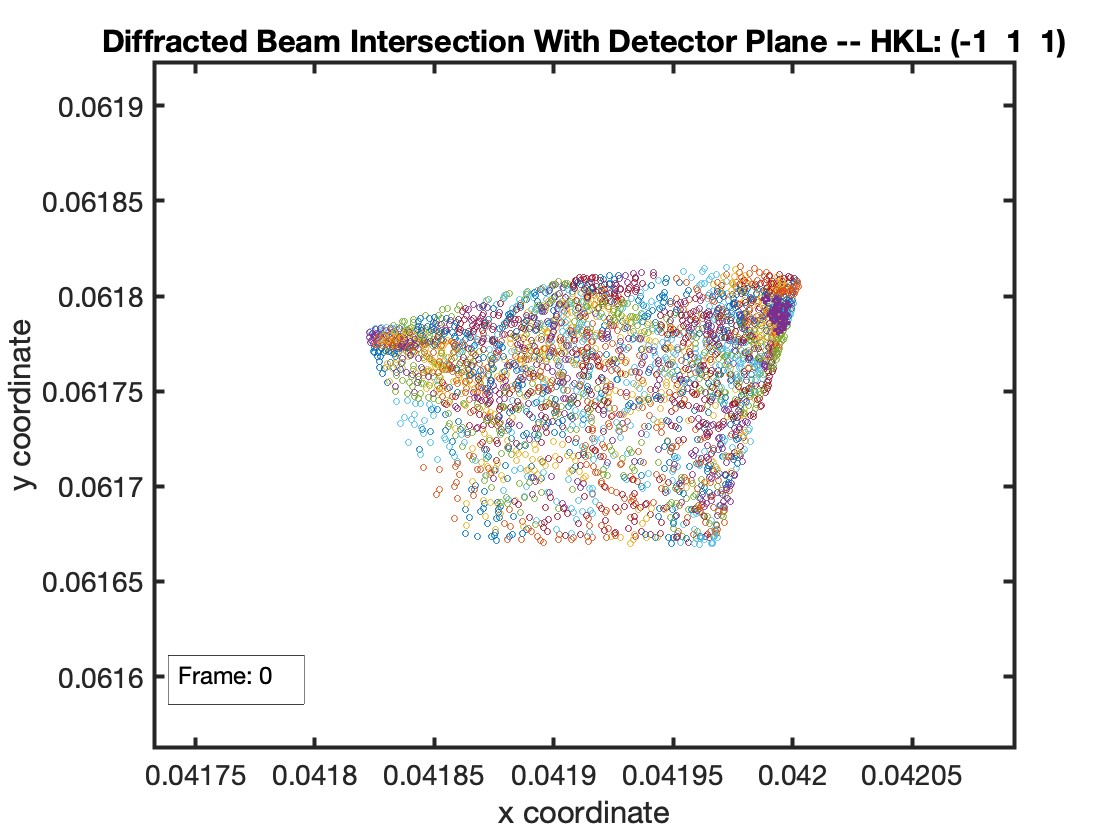}
		\caption{ }
		\label{fig:points_dectectorPlane_bar111_f0}
	\end{subfigure}%
	\quad
	\begin{subfigure}{.3\textwidth}
		\centering
		\includegraphics[width=1\linewidth]{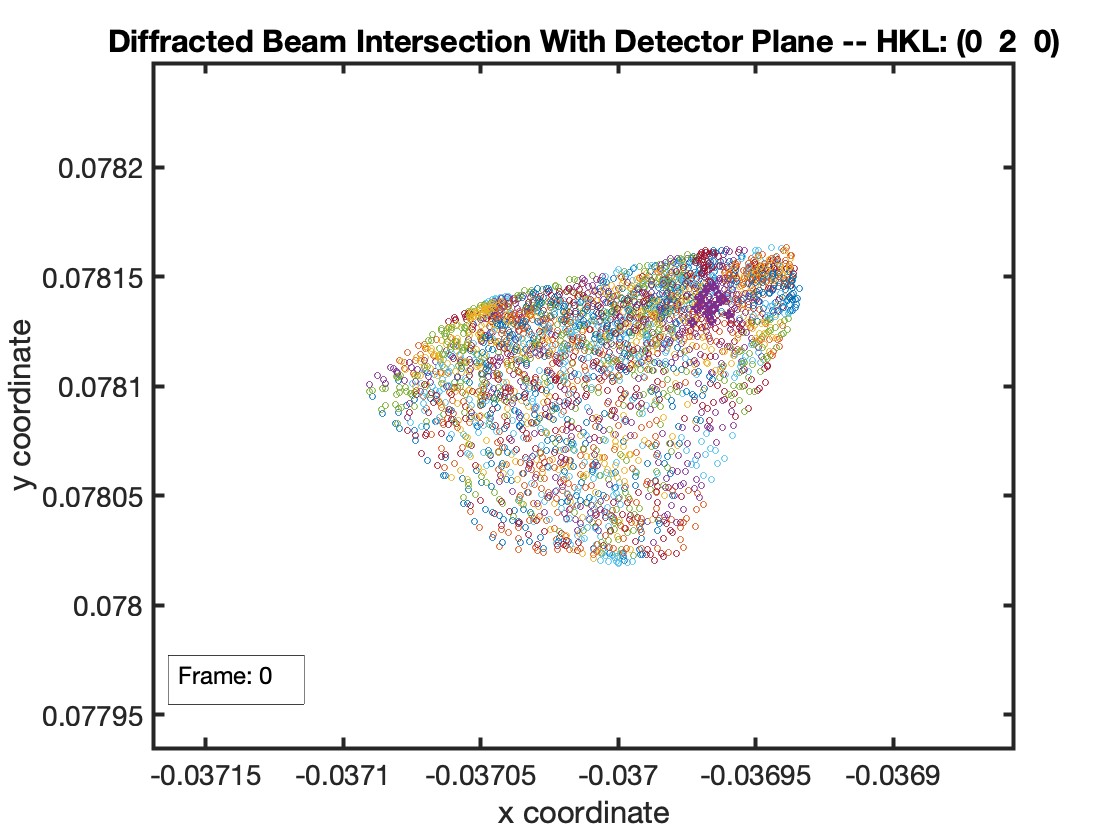}
		\caption{ }
		\label{fig:points_dectectorPlane_020_f0}
	\end{subfigure}	
	\quad
	\begin{subfigure}{.3\textwidth}
		\centering
		\includegraphics[width=1\linewidth]{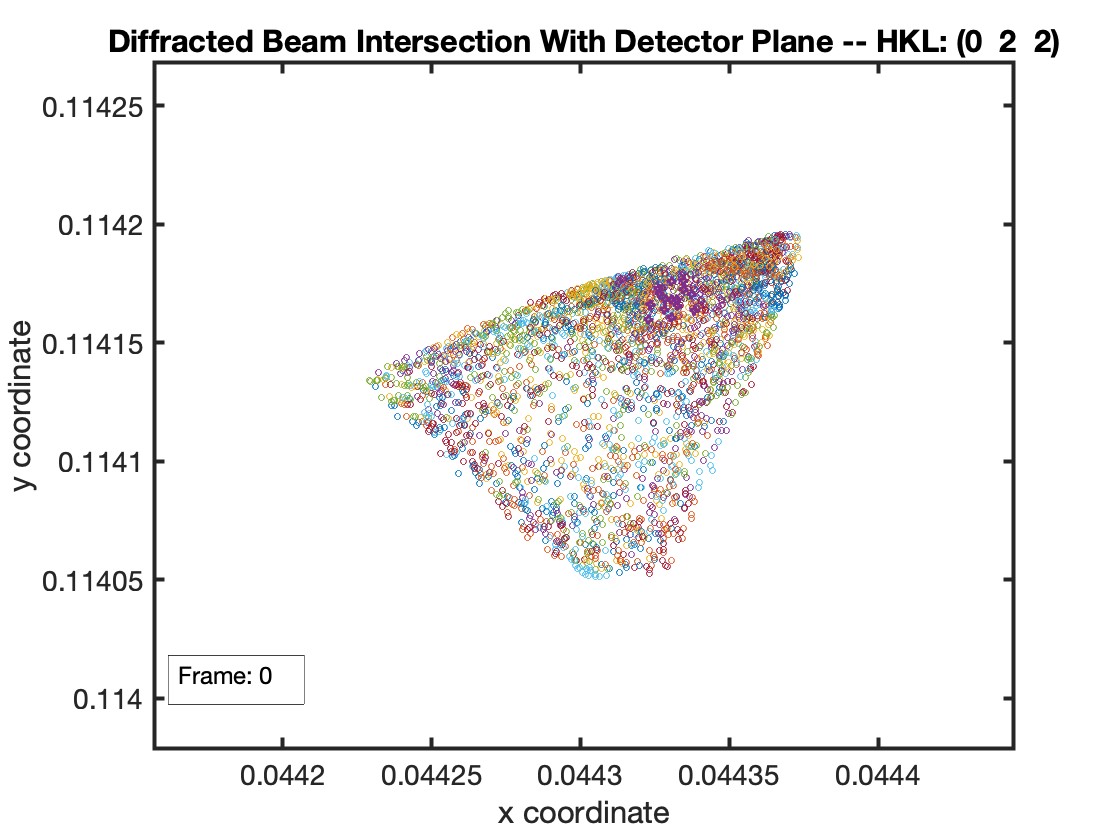}
		\caption{ }
		\label{fig:points_dectectorPlane_022_f0}
	\end{subfigure}
		\caption{Projected points on the detector plane under zero load for the most axial scattering vectors: (a) $\bar111$ reflection, (b) 020 reflection, and (c) 022 reflection. }
		\label{fig:points_on_detector_plane_unloaded_0deg}
\end{figure}

\begin{figure}[htbp]
	\centering		
	\begin{subfigure}{.3\textwidth}
		\centering
		\includegraphics[width=1\linewidth]{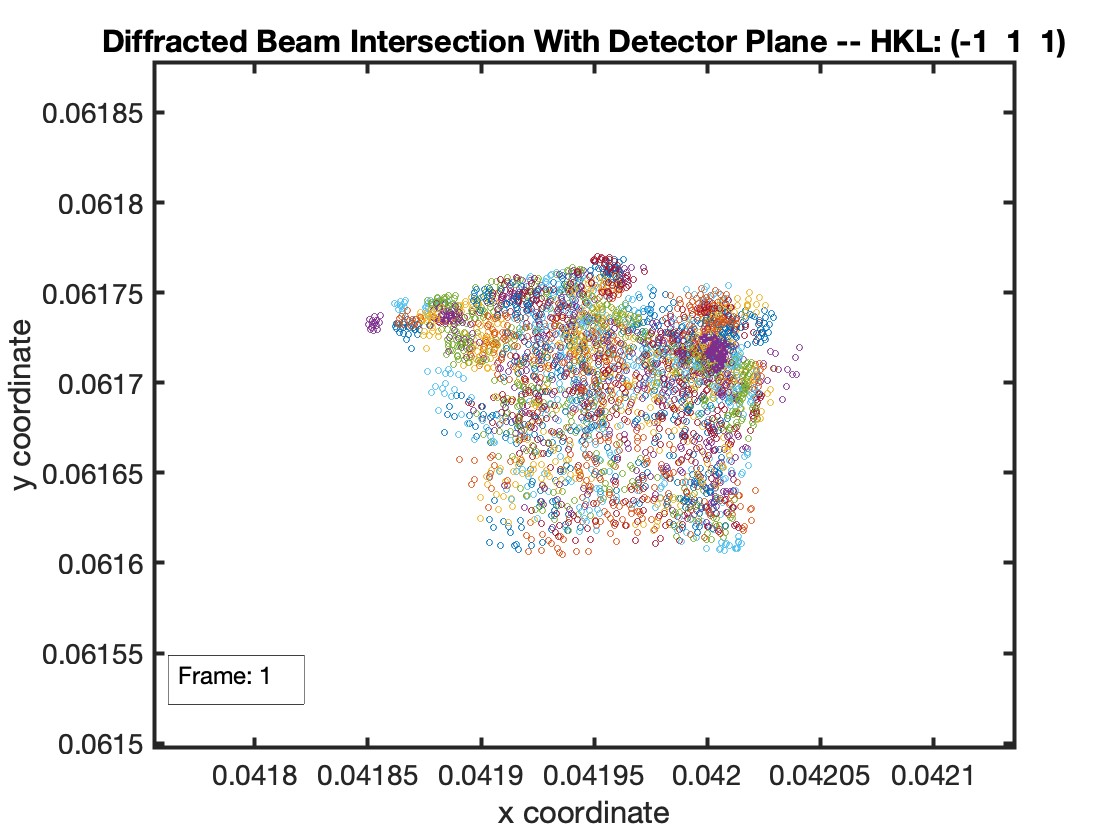}
		\caption{ }
		\label{fig:points_dectectorPlane_bar111l}
	\end{subfigure}%
	\quad
	\begin{subfigure}{.3\textwidth}
		\centering
		\includegraphics[width=1\linewidth]{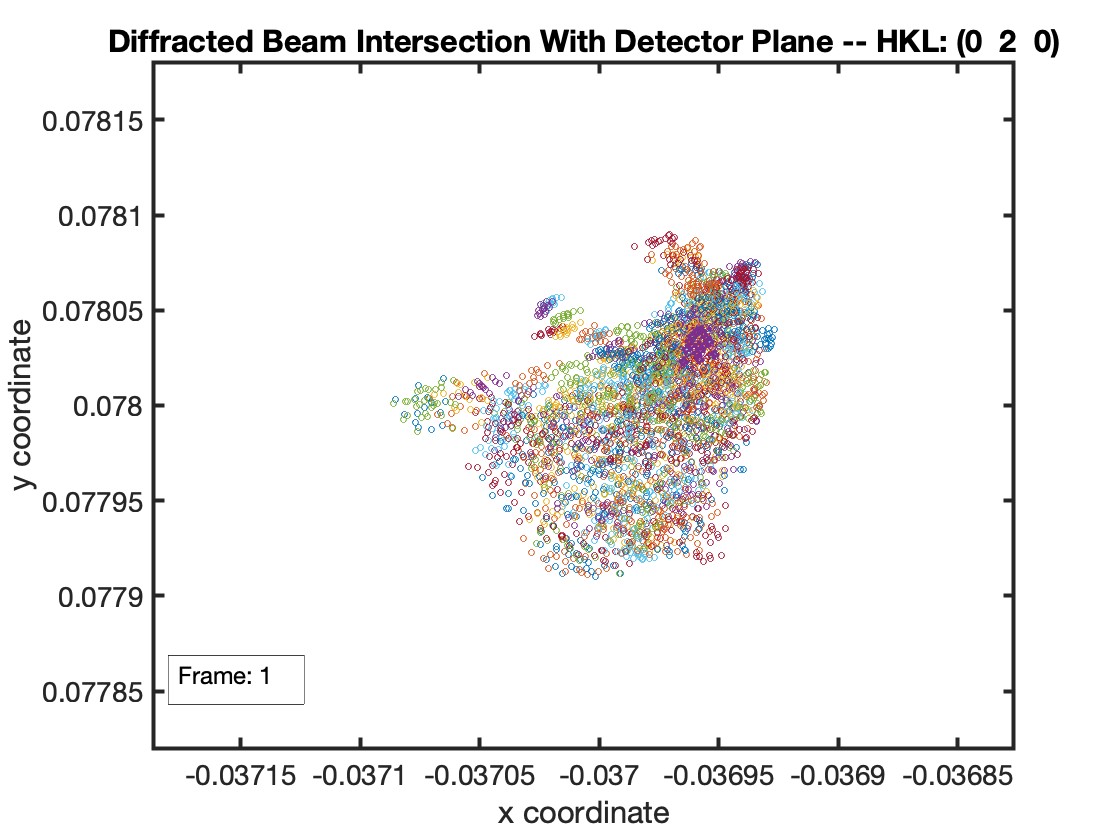}
		\caption{ }
		\label{fig:points_dectectorPlane_020l}
	\end{subfigure}	
	\quad
	\begin{subfigure}{.3\textwidth}
		\centering
		\includegraphics[width=1\linewidth]{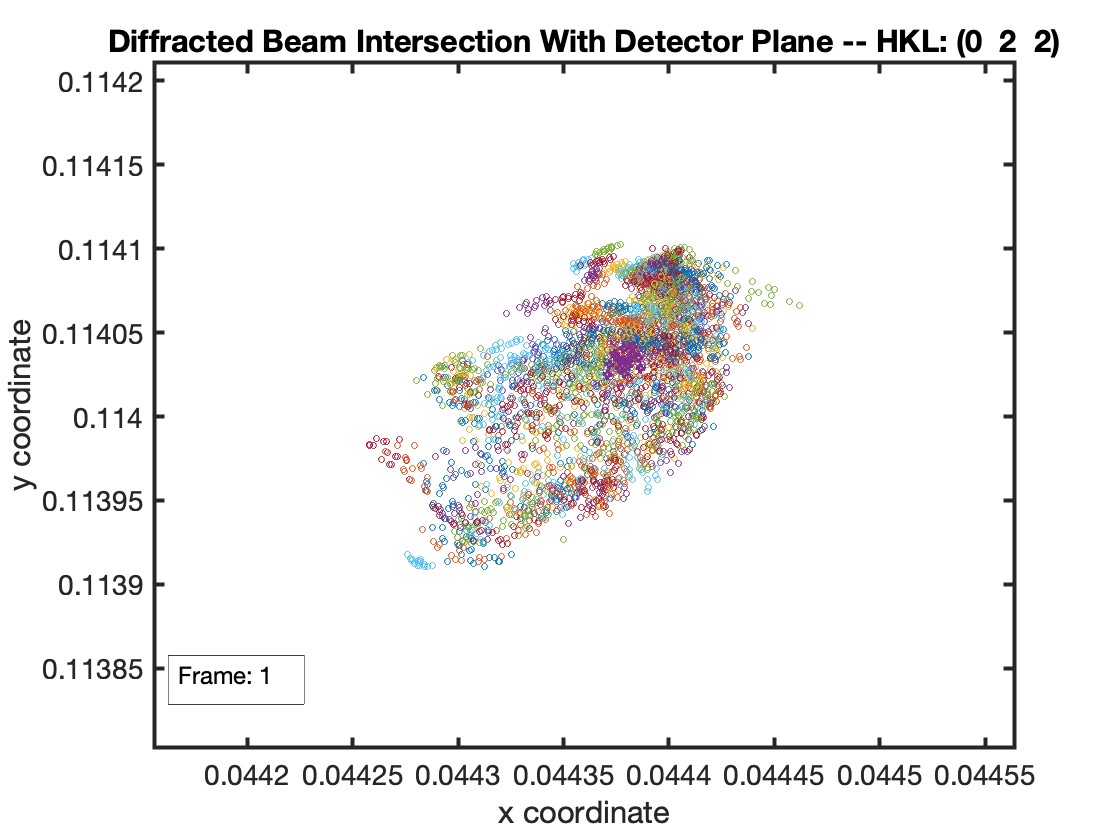}
		\caption{ }
		\label{fig:points_dectectorPlane_022l}
	\end{subfigure}
		\caption{Projected points on the detector plane under tensile load  for the most axial scattering vectors: (a) $\bar111$ reflection, (b) 020 reflection, and (c) 022 reflection.   }
		\label{fig:points_on_detector_plane_loaded_0deg}
\end{figure}

For each $hkl$, the virtual diffractometer projects a point onto the detector plane for every quadrature point of every finite element of the target grain.  
For Grain 80, 3735 points are projected points on detector plane (249 elements each with 15 quadrature points).
The projected point plots are shown in  \figref{fig:points_on_detector_plane_unloaded_0deg} and \figref{fig:points_on_detector_plane_loaded_0deg} for the 
unloaded and loaded states, respectively, for the 'near axial' scattering vectors.
The coordinates of points are given in detector coordinates ($x^d_i $); the orientation of the detector coordinate system is shown in \figref{fig:projection2detector} with $ \vctr{e}^d_3 = 1 $ in the laboratory coordinates.  
Thus, the view on the detector is in the direction opposite the direction of the incident beam.

\begin{figure}[htbp]
	\centering		
	\begin{subfigure}{.3\textwidth}
		\centering
		\includegraphics[width=1\linewidth]{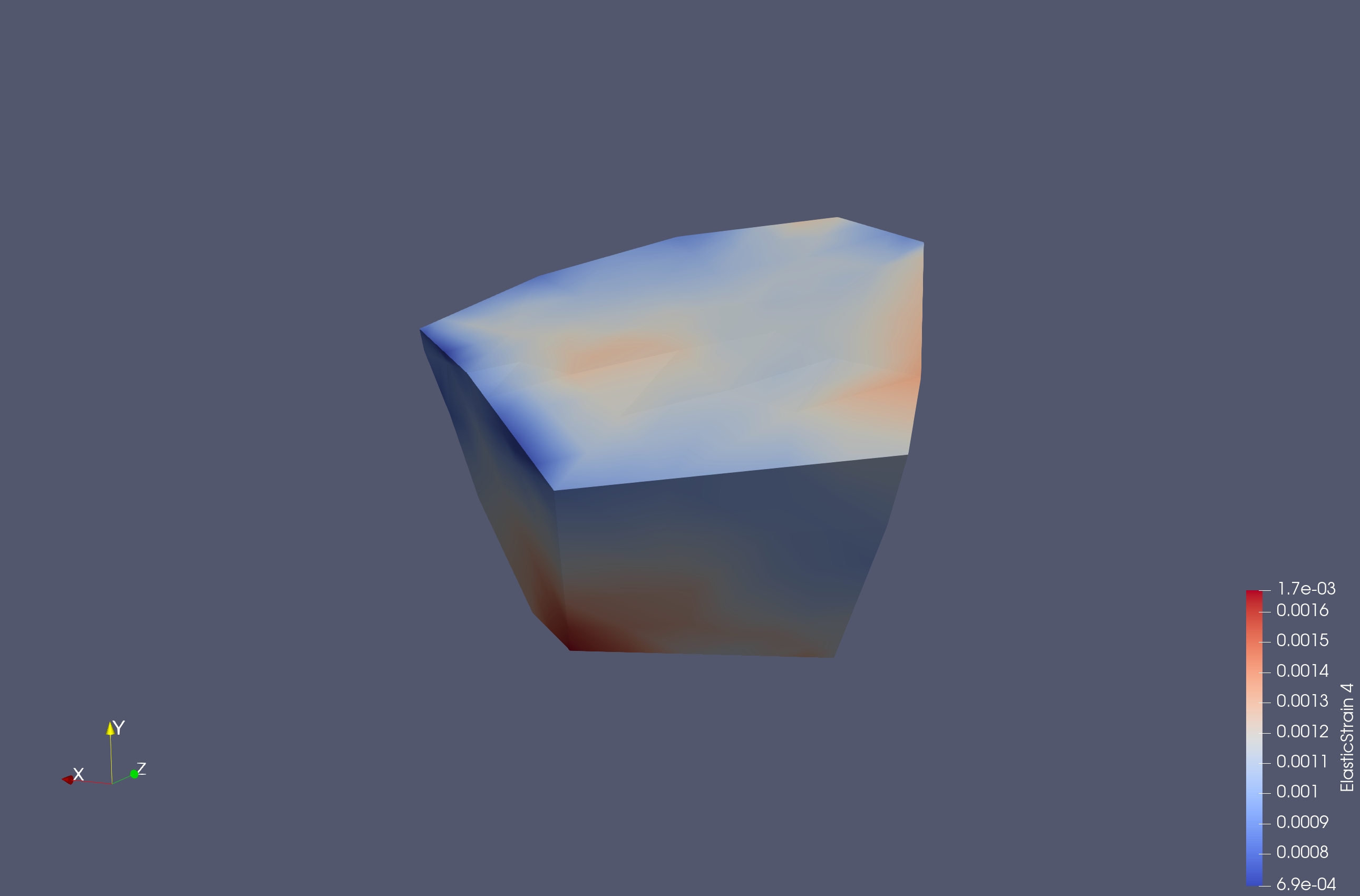}
		\caption{ }
		\label{fig:fig:omegarotation2p67}
	\end{subfigure}%
	\quad
	\begin{subfigure}{.3\textwidth}
		\centering
		\includegraphics[width=1\linewidth]{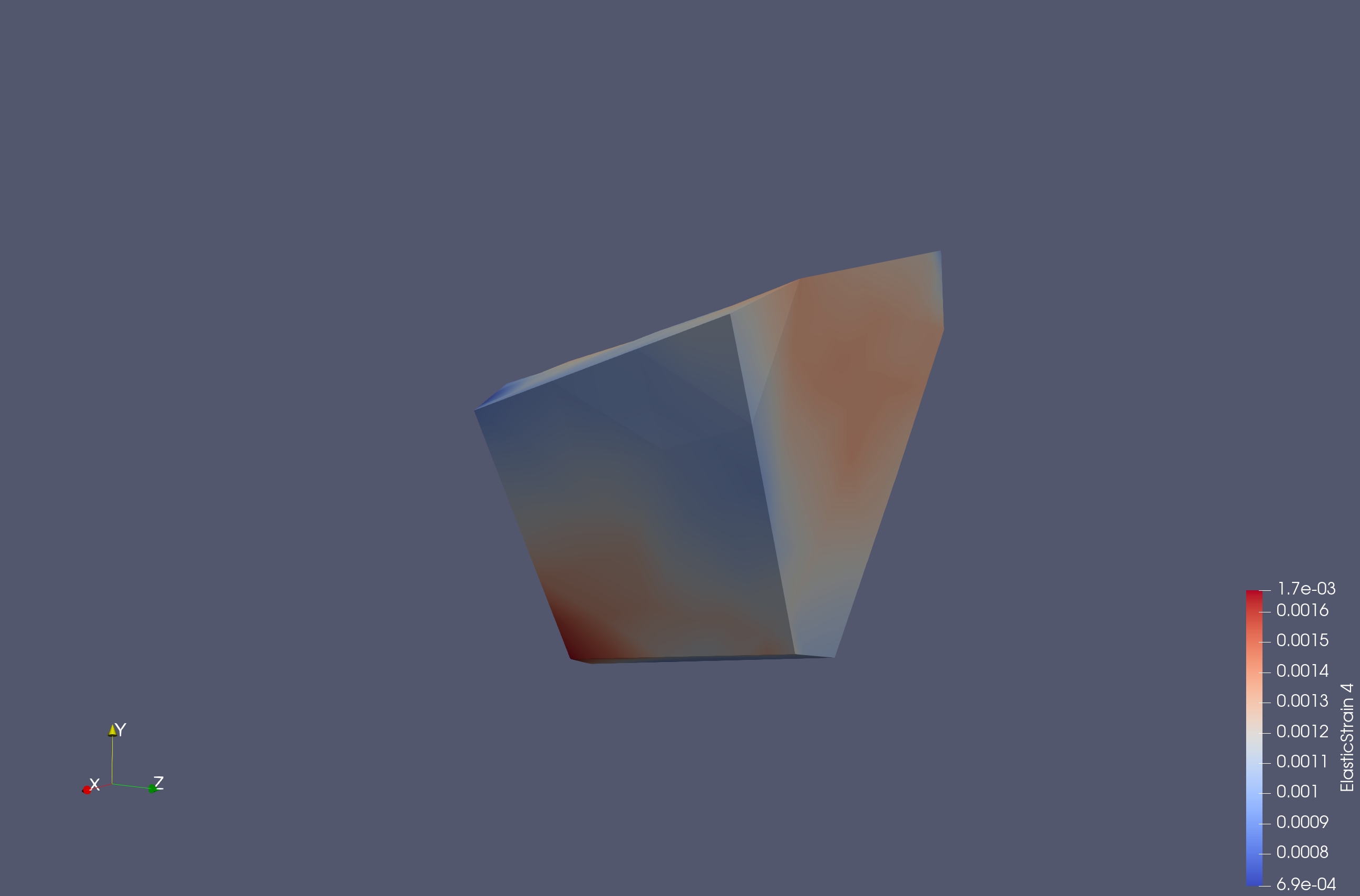}
		\caption{ }
		\label{fig:fig:omegarotation1p72}
	\end{subfigure}	
	\quad
	\begin{subfigure}{.3\textwidth}
		\centering
		\includegraphics[width=1\linewidth]{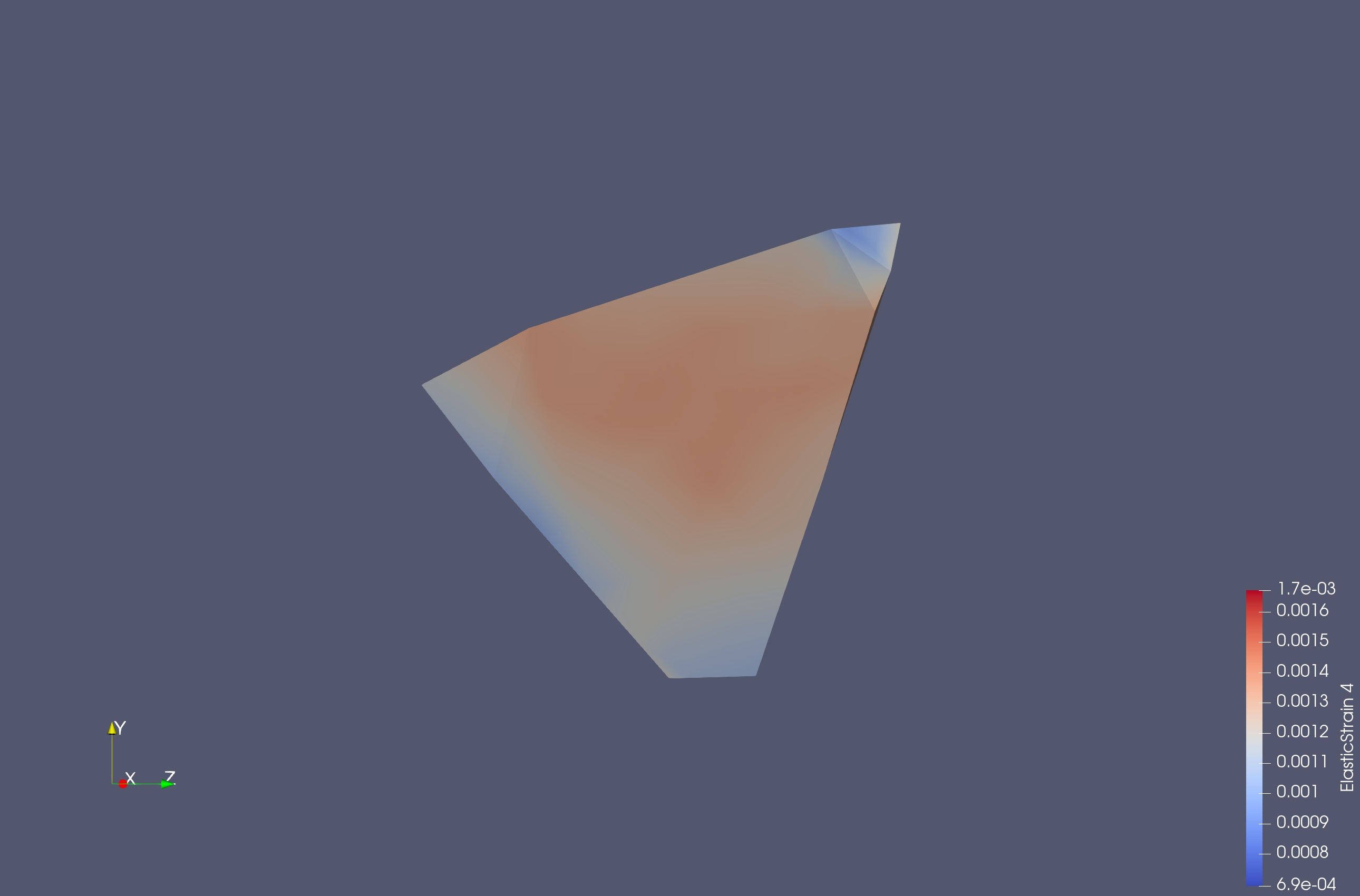}
		\caption{ }
		\label{fig:omegarotation1p44}
	\end{subfigure}
		\caption{Grain 80 rotated by $\omega$ from the view in \figref{fig:targetgrainzaxis} by approximately: (a) 2.67 rad; (b) 1.72 rad; and (c) 1.44 rad to correspond to rotations associated with the: (a) $\bar111$ reflection, (b) 020 reflection, and (c) 022 reflection shown in \figref{fig:points_on_detector_plane_unloaded_0deg}.  }
		\label{fig:grain80_omegarotations}
\end{figure}

In the plots, like-colored points are associated with the quadrature points of a single finite element.
The specific reflections for this case are the  $\bar111$,  $020$, and  $022$ planes.
The overall spatial pattern is defined in the unloaded case shown in \figref{fig:points_on_detector_plane_unloaded_0deg}  by the cross section that the grain presents to the incident beam.  The pattern is different for each reflection owing the the sample rotation, $\omega$. 
To make this point more evident, \figref{fig:grain80_omegarotations} shows Grain 80 rotated about the loading axis from the view in \figref{fig:targetgrainzaxis} by $\omega$ for each of the reflections shown in \figref{fig:points_on_detector_plane_unloaded_0deg}.
The outlines are not exactly one-to-one because the angles of rotation in \figref{fig:grain80_omegarotations} are not precisely the correct values of $\omega$ and the influence of projection onto the detector plane is not included. 
Comparing \figref{fig:points_on_detector_plane_unloaded_0deg} to \figref{fig:points_on_detector_plane_loaded_0deg}
on a reflection-by-reflection basis, the influence of the distortion of the lattice induced by the stress is evident.  
The outlines of the grain evident with the undeformed grain are not as clearly discerned once the grain is loaded.
There is an overall shift in the location of the points, which is manifest in the average strains 
listed in \tabref{tab:diffractionconditions}, but what is seen here are the changes in shape of a 
pattern that is due to strain heterogeneity. The diffracted beams for the collection of diffraction volumes (volumes associated with the quadrature points of the finite elements) are no longer parallel owing to changes in the unit cell as given in \eqnref{eqn:adeformed}.   \bigskip

\begin{figure}[htbp]
	\centering		
	\begin{subfigure}{.3\textwidth}
		\centering
		\includegraphics[width=1\linewidth]{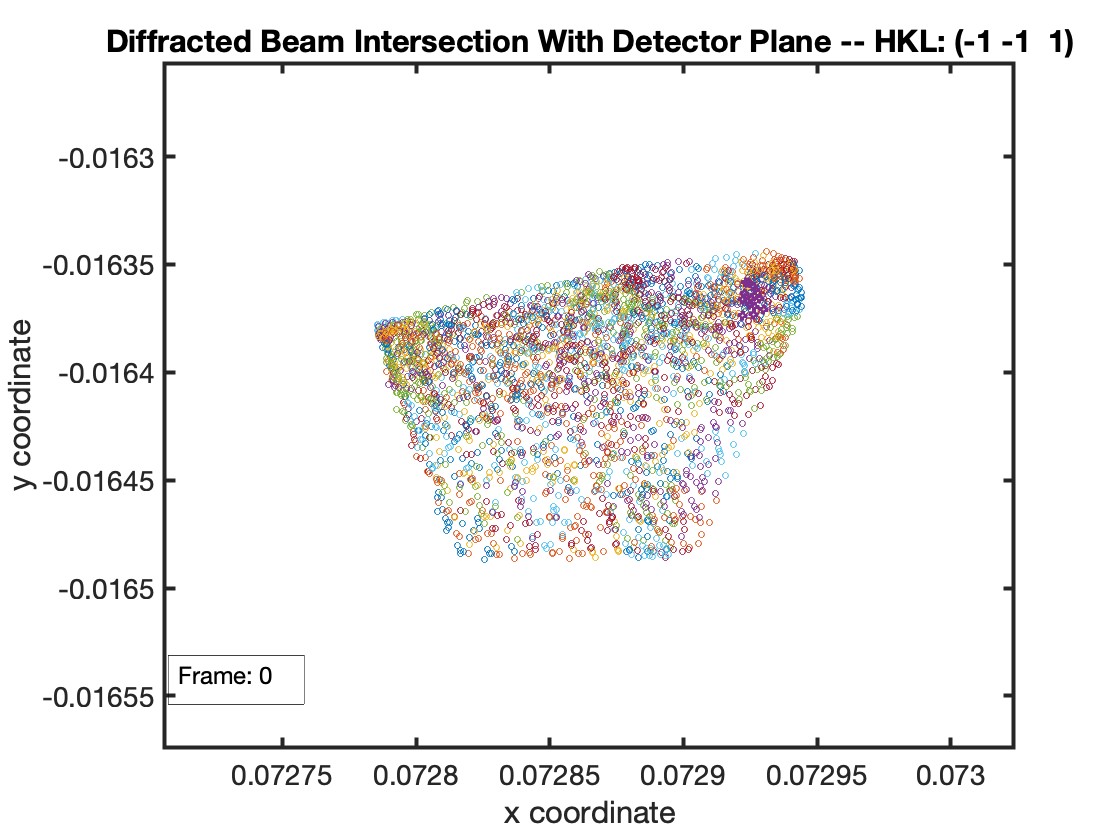}
		\caption{ }
		\label{fig:points_dectectorPlane_bar1bar11u}
	\end{subfigure}%
	\quad
	\begin{subfigure}{.3\textwidth}
		\centering
		\includegraphics[width=1\linewidth]{figure_03.jpg}
		\caption{ }
		\label{fig:points_dectectorPlane_200u}
	\end{subfigure}	
	\quad
	\begin{subfigure}{.3\textwidth}
		\centering
		\includegraphics[width=1\linewidth]{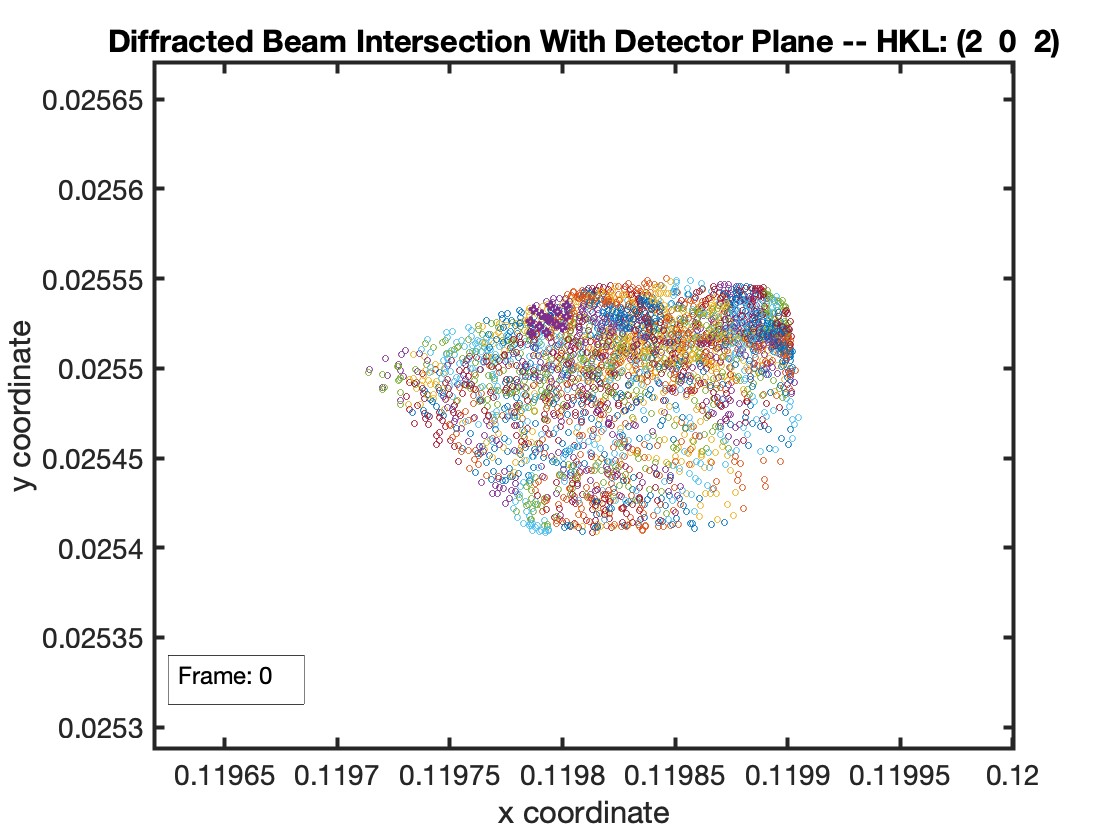}
		\caption{ }
		\label{fig:points_dectectorPlane_202u}
	\end{subfigure}
		\caption{Projected points on the detector plane under zero load for the most transverse scattering vectors: (a) $\bar1\bar11$ reflection, (b) 200 reflection, and (c) 202 reflection. }
		\label{fig:points_on_detector_plane_unloaded_90deg}
\end{figure}

\begin{figure}[htbp]
	\centering		
	\begin{subfigure}{.3\textwidth}
		\centering
		\includegraphics[width=1\linewidth]{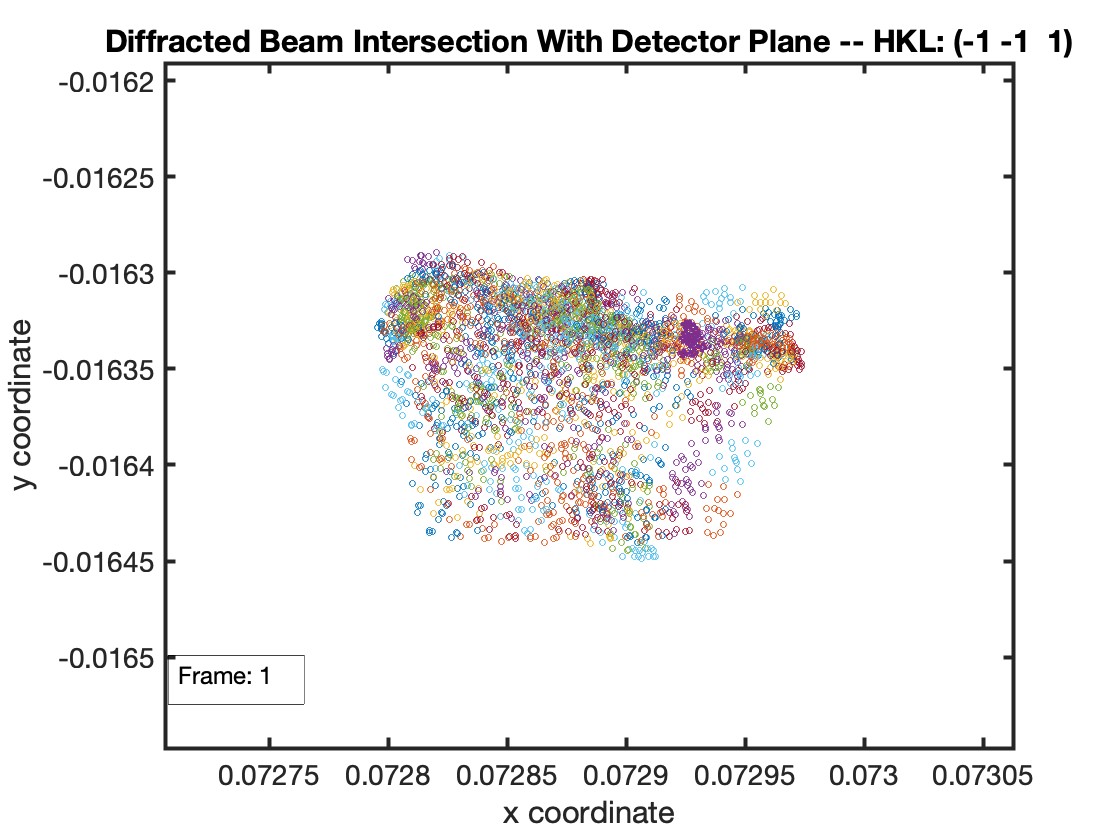}
		\caption{ }
		\label{fig:points_dectectorPlane_bar1bar11l}
	\end{subfigure}%
	\quad
	\begin{subfigure}{.3\textwidth}
		\centering
		\includegraphics[width=1\linewidth]{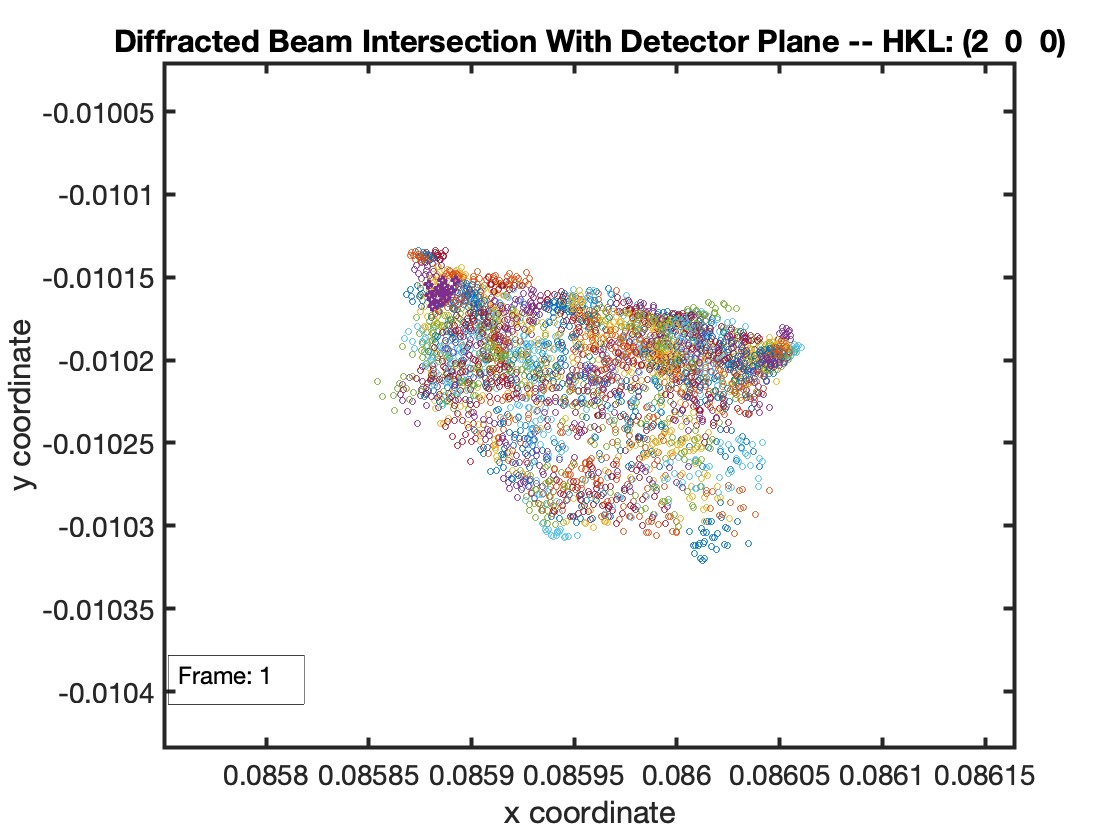}
		\caption{ }
		\label{fig:points_dectectorPlane_200l}
	\end{subfigure}	
	\quad
	\begin{subfigure}{.3\textwidth}
		\centering
		\includegraphics[width=1\linewidth]{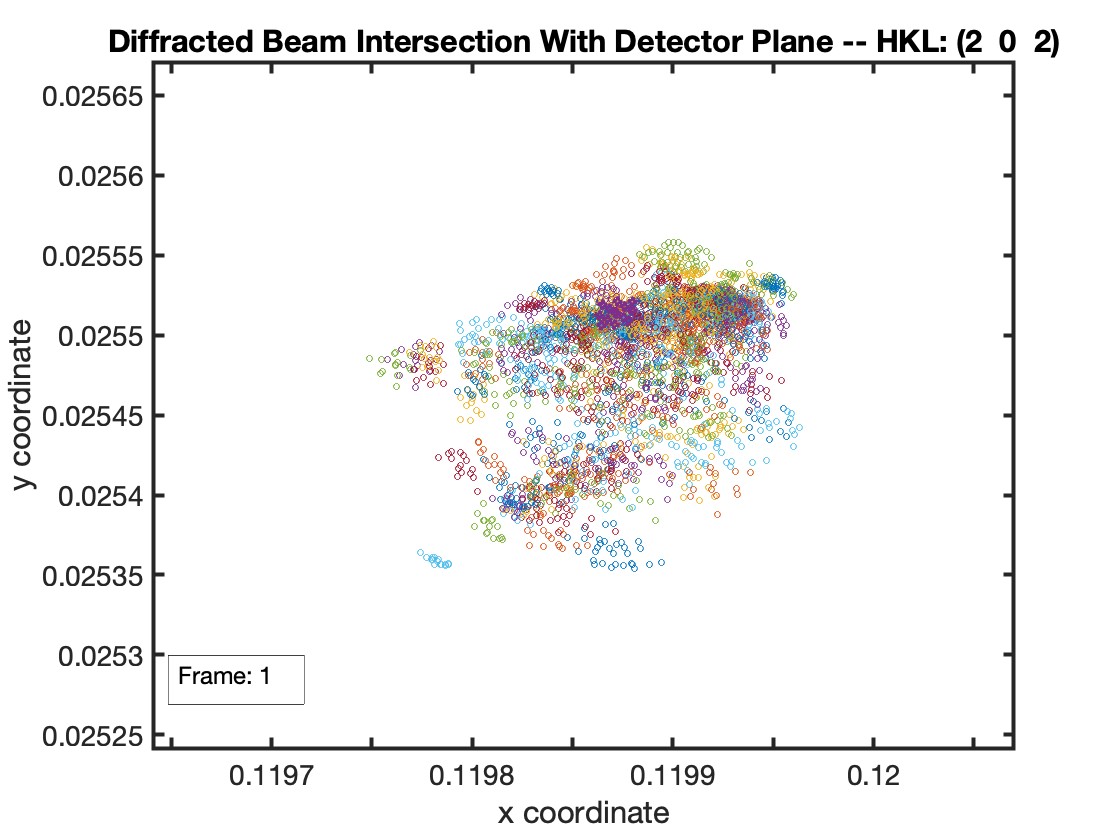}
		\caption{ }
		\label{fig:points_dectectorPlane_202l}
	\end{subfigure}
		\caption{Projected points on the detector plane under tensile load for the most transverse scattering vectors: (a) $\bar1\bar11$ reflection, (b) 200 reflection, and (c) 202 reflection. }
		\label{fig:points_on_detector_plane_loaded_90deg}
\end{figure}

The projected point plots are shown for the 'near transverse' scattering vectors in  \figref{fig:points_on_detector_plane_unloaded_90deg} and \figref{fig:points_on_detector_plane_loaded_90deg}.
Again, the grain shape is evident in the patterns associated with the undeformed state.
As with the patterns for the 'near axial' scattering vectors, there are slight differences
from reflection to reflection owing to differences is the directions of the diffracted beam.
We also see the effects of the lattice distortion on the patterns from the
undeformed to deformed states.
These changes are most pronounced for the  $ \{220\}$ family of planes.

%\clearpage

\subsection{Results from Step 3: Intensity distributions}
\label{sec:demo_task3}

The relative intensity distributions are an important intermediate goal for the diffractometer.  
They differ qualitatively from the projected point plots in that the field distribution of intensity 
much better conveys the actual spatial distribution that does the set of projected points. 
The points plot suffer from not including the point weights, not spreading the point contribution according
to the Gaussian spread function, and from the points possibly overlaying each other.
The intensity distributions rectify all these limitations.

The finite element mesh for the relative intensity distributions has 4096 (16x16) quadrilateral elements.  These overlay a detector area of 16 (4x4) pixels. The pixels are 0.2mm x 0.2mm.   The intensity distribution mesh is spatially uniform and registered with the pixels, so that there are 256 elements coinciding with each pixel.  
The intensity distribution mesh can be refined or coarsened as specified by the user, but is linked to the pixel size so that subsets of the total mesh always coincide exactly with the pixel positions.  
This facilitates integrating the intensity distributions over detector patches that are directly associated with one and only one pixel.
  
\begin{figure}[htbp]
	\centering		
	\begin{subfigure}{.3\textwidth}
		\centering
		\includegraphics[width=1\linewidth]{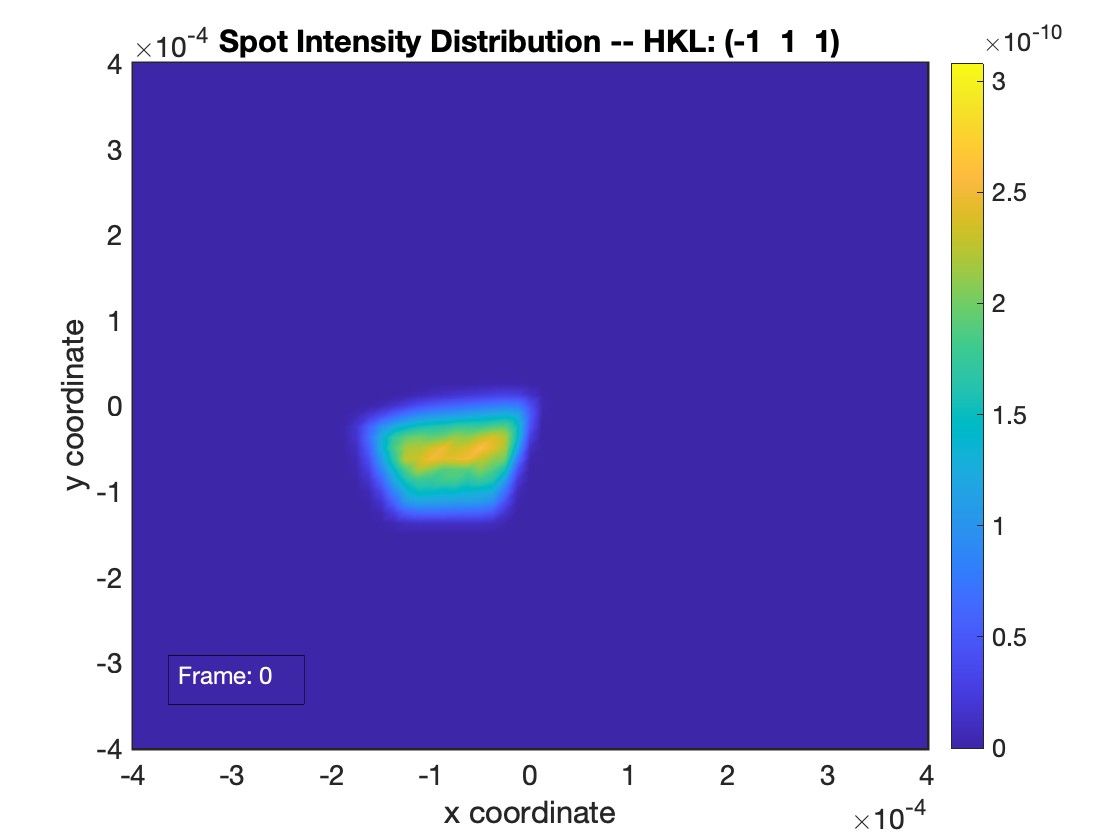}
		\caption{ }
		\label{fig:dist_dectectorPlane_bar111u}
	\end{subfigure}%
	\quad
	\begin{subfigure}{.3\textwidth}
		\centering
		\includegraphics[width=1\linewidth]{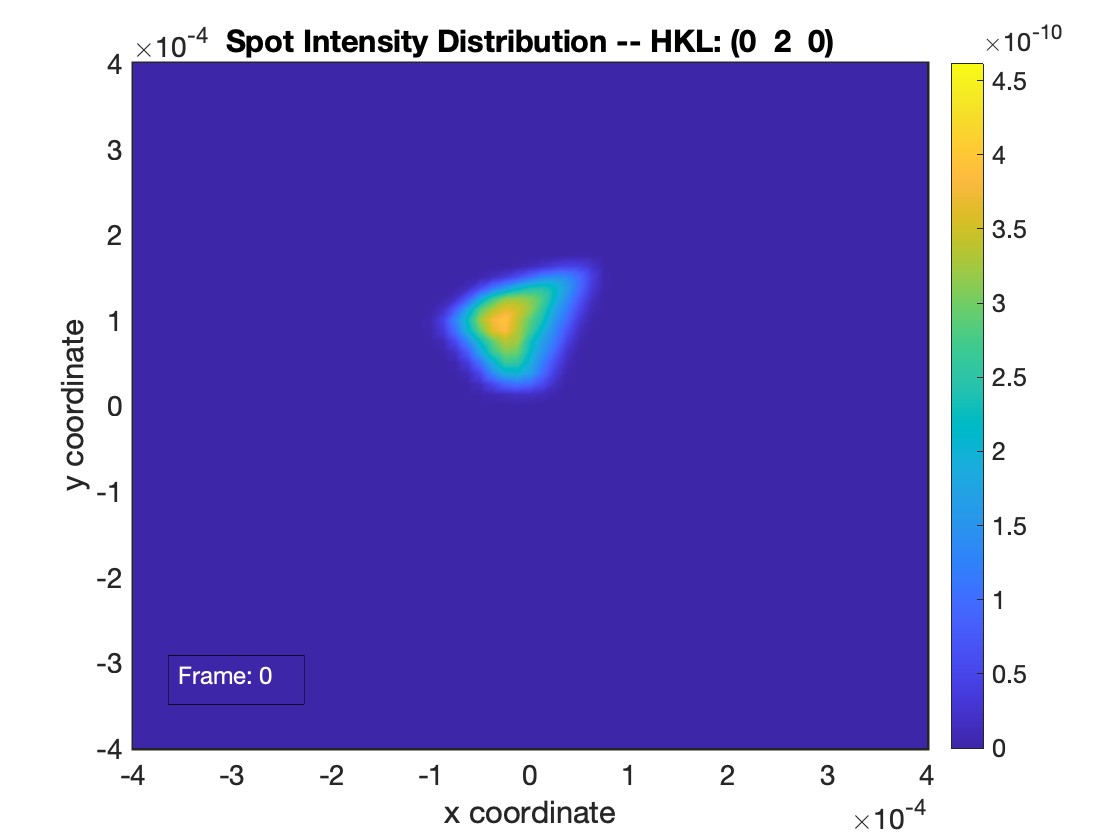}
		\caption{ }
		\label{fig:dist_dectectorPlane_020u}
	\end{subfigure}	
	\quad
	\begin{subfigure}{.3\textwidth}
		\centering
		\includegraphics[width=1\linewidth]{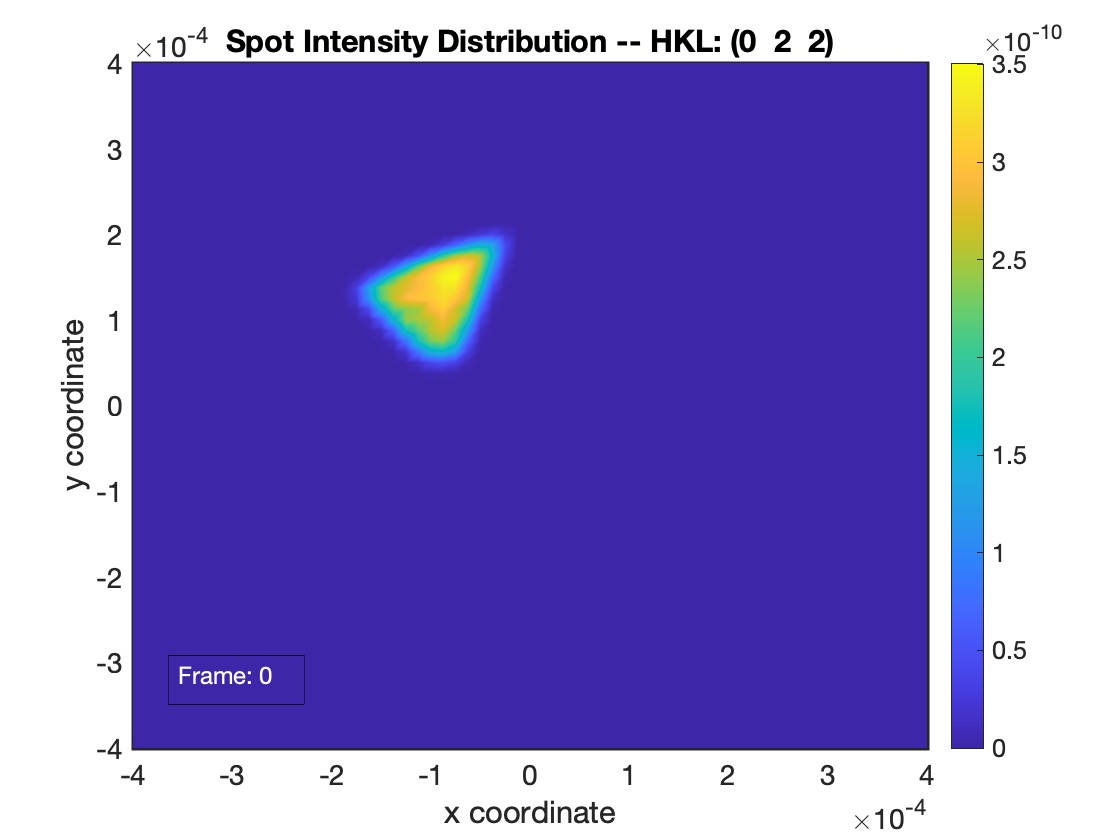}
		\caption{ }
		\label{fig:dist_dectectorPlane_022u}
	\end{subfigure}
		\caption{Intensity distribution over the detector plane under zero load for the most axial scattering vectors: (a) $\bar111$ reflection, (b) 020 reflection, and (c) 022 reflection.}
		\label{fig:dist_on_detector_plane_unloaded_0deg}
\end{figure}

 The relative intensity distributions for unloaded and loaded states are shown in  \figref{fig:dist_on_detector_plane_unloaded_0deg} and \figref{fig:dist_on_detector_plane_unloaded_0deg}
 for the 'near axial' scattering vectors and in 
 \figref{fig:dist_on_detector_plane_unloaded_90deg} and \figref{fig:dist_on_detector_plane_loaded_90deg}
 for the 'near transverse' scattering vectors.
Note the following:
the discretized regions for the unloaded and loaded cases are the same for each reflection
({\it e.g.} the two (a) subfigures have the $x-y$ domains as do the two (b)'s and the two (c)'s).   
This allows the user to better see the degree to which a spot shifts and spreads when loaded.  
Likewise the intensity scales are the same for the matching unloaded and loaded images.
The parameter that determines the amount of attenuation was set so that the attenuation was minimal.  This has little effect here as we are examining only a single grain rather than ones with different positions within the sample.

\begin{figure}[htbp]
	\centering		
	\begin{subfigure}{.3\textwidth}
		\centering
		\includegraphics[width=1\linewidth]{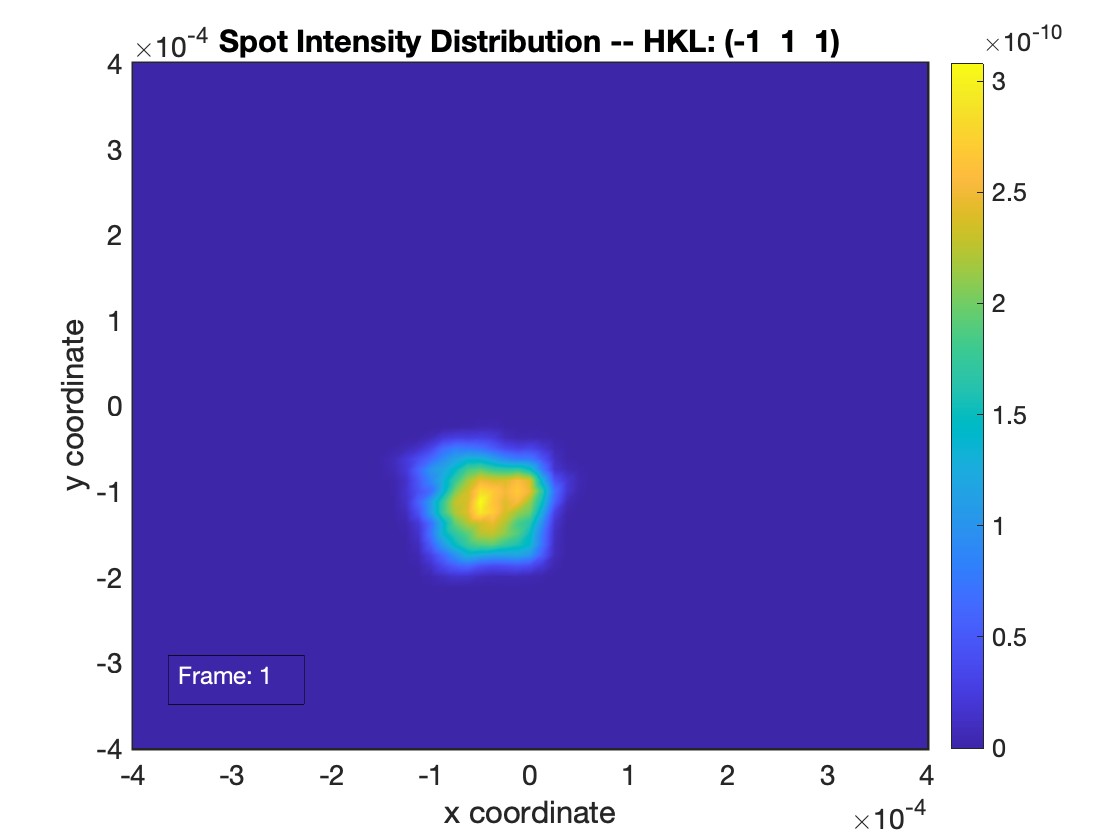}
		\caption{ }
		\label{fig:dist_dectectorPlane_bar111l}
	\end{subfigure}%
	\quad
	\begin{subfigure}{.3\textwidth}
		\centering
		\includegraphics[width=1\linewidth]{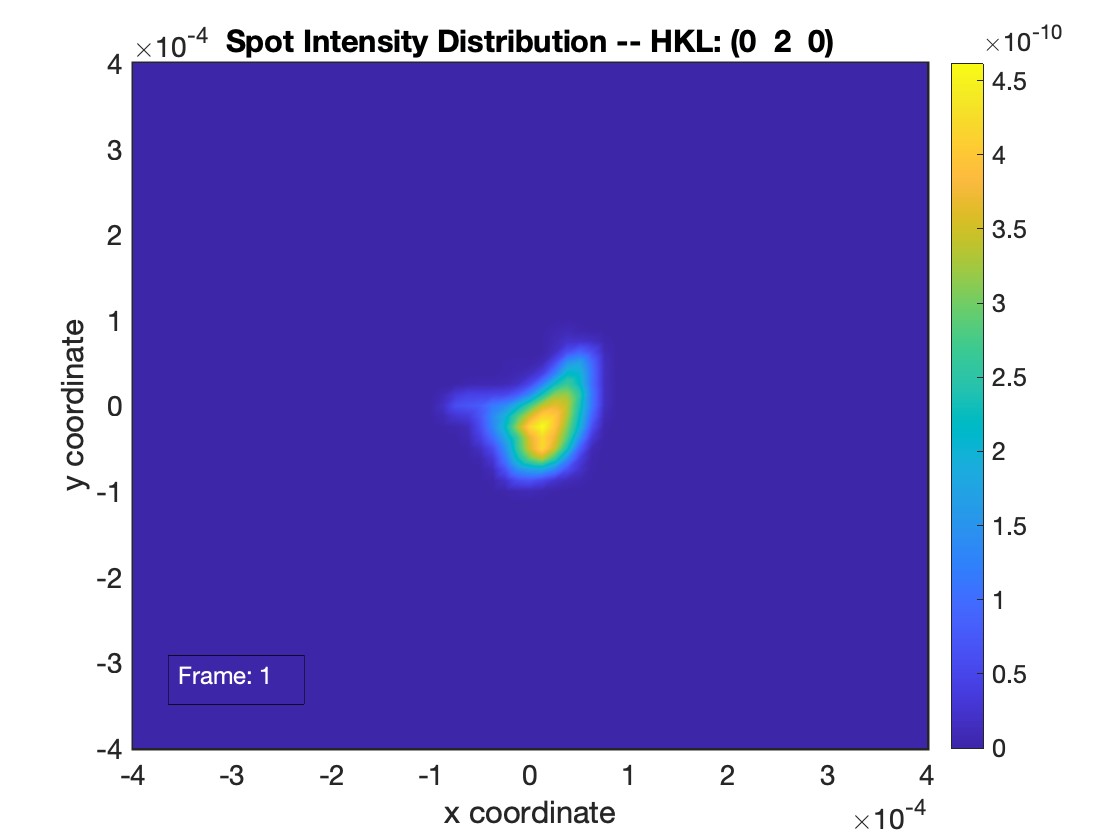}
		\caption{ }
		\label{fig:dist_dectectorPlane_020l}
	\end{subfigure}	
	\quad
	\begin{subfigure}{.3\textwidth}
		\centering
		\includegraphics[width=1\linewidth]{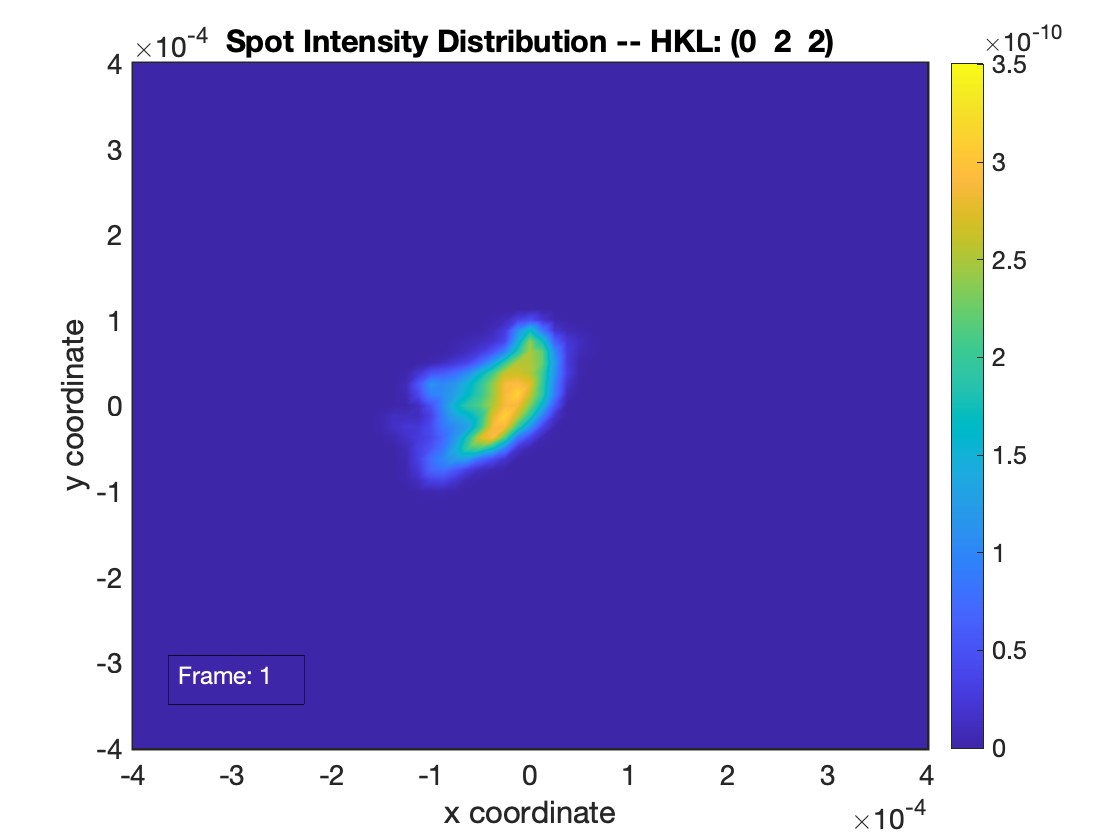}
		\caption{ }
		\label{fig:dist_dectectorPlane_022l}
	\end{subfigure}
		\caption{Intensity distribution over the detector plane under tensile load for the most axial scattering vectors: (a) $\bar111$ reflection, (b) 020 reflection, and (c) 022 reflection.}
		\label{fig:dist_on_detector_plane_loaded_0deg}
\end{figure}

\begin{figure}[htbp]
	\centering		
	\begin{subfigure}{.3\textwidth}
		\centering
		\includegraphics[width=1\linewidth]{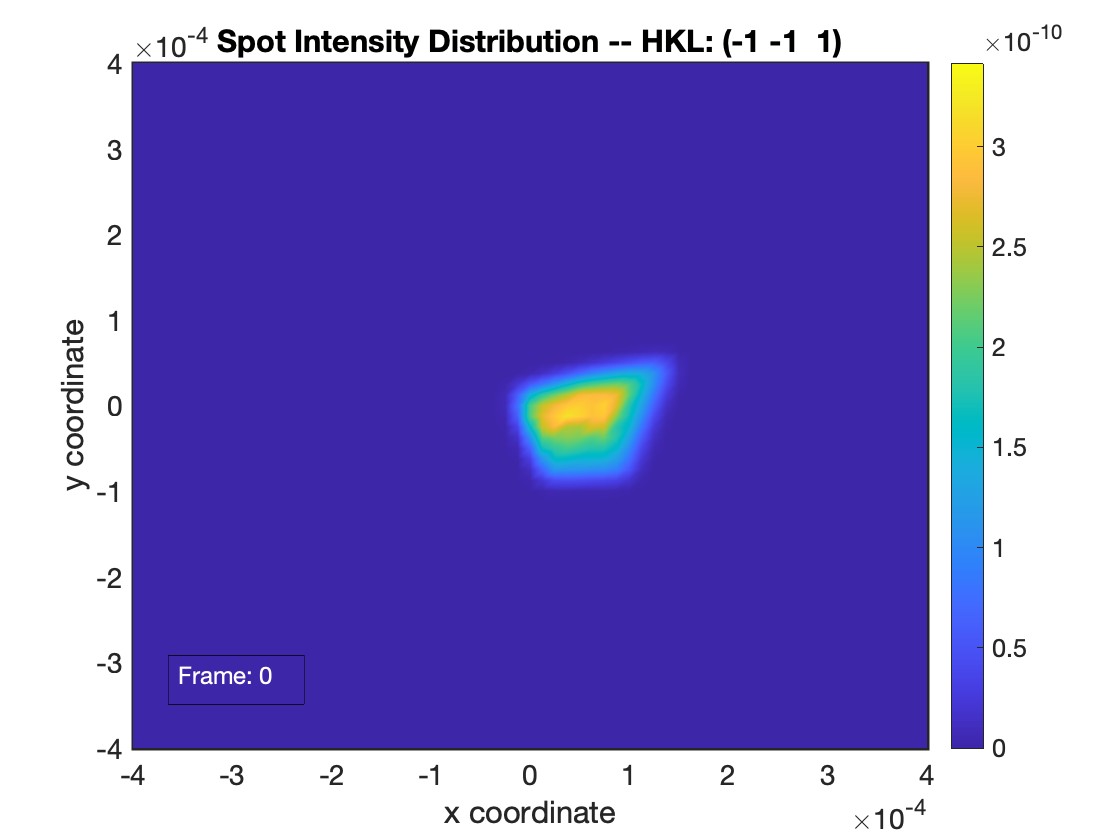}
		\caption{ }
		\label{fig:dist_dectectorPlane_bar1bar11u}
	\end{subfigure}%
	\quad
	\begin{subfigure}{.3\textwidth}
		\centering
		\includegraphics[width=1\linewidth]{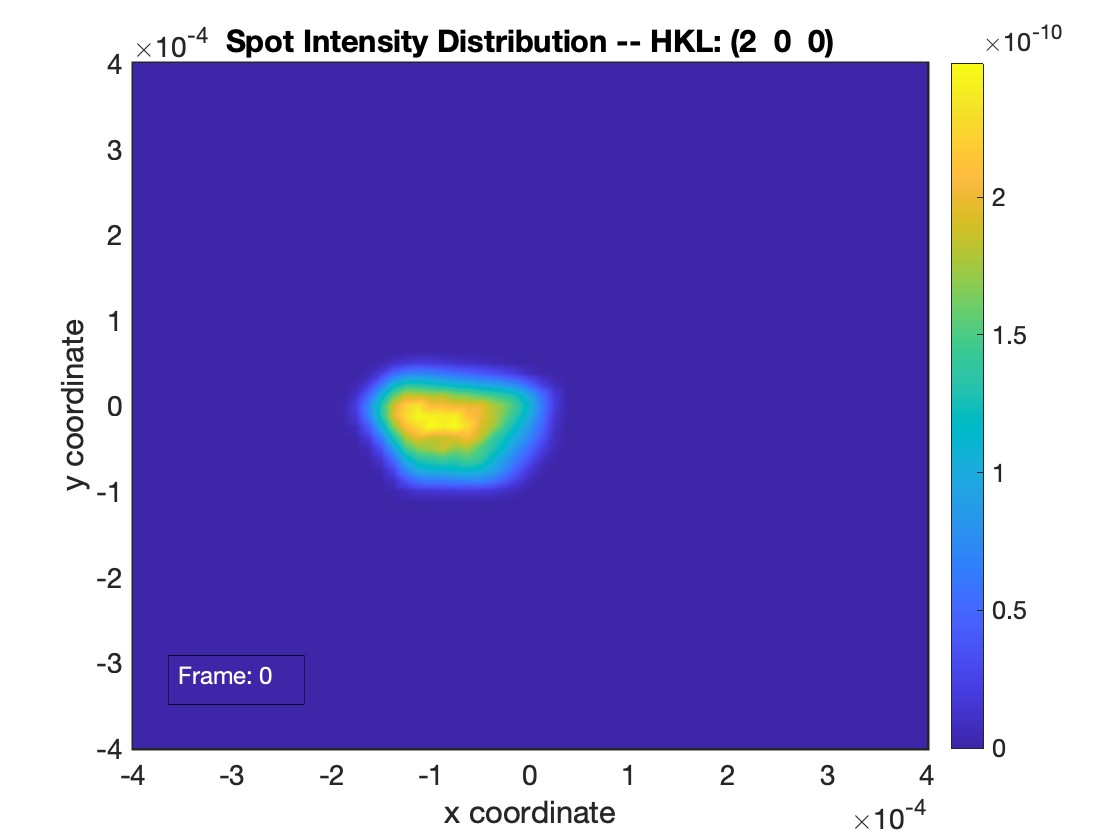}
		\caption{ }
		\label{fig:dist_dectectorPlane_200u}
	\end{subfigure}	
	\quad
	\begin{subfigure}{.3\textwidth}
		\centering
		\includegraphics[width=1\linewidth]{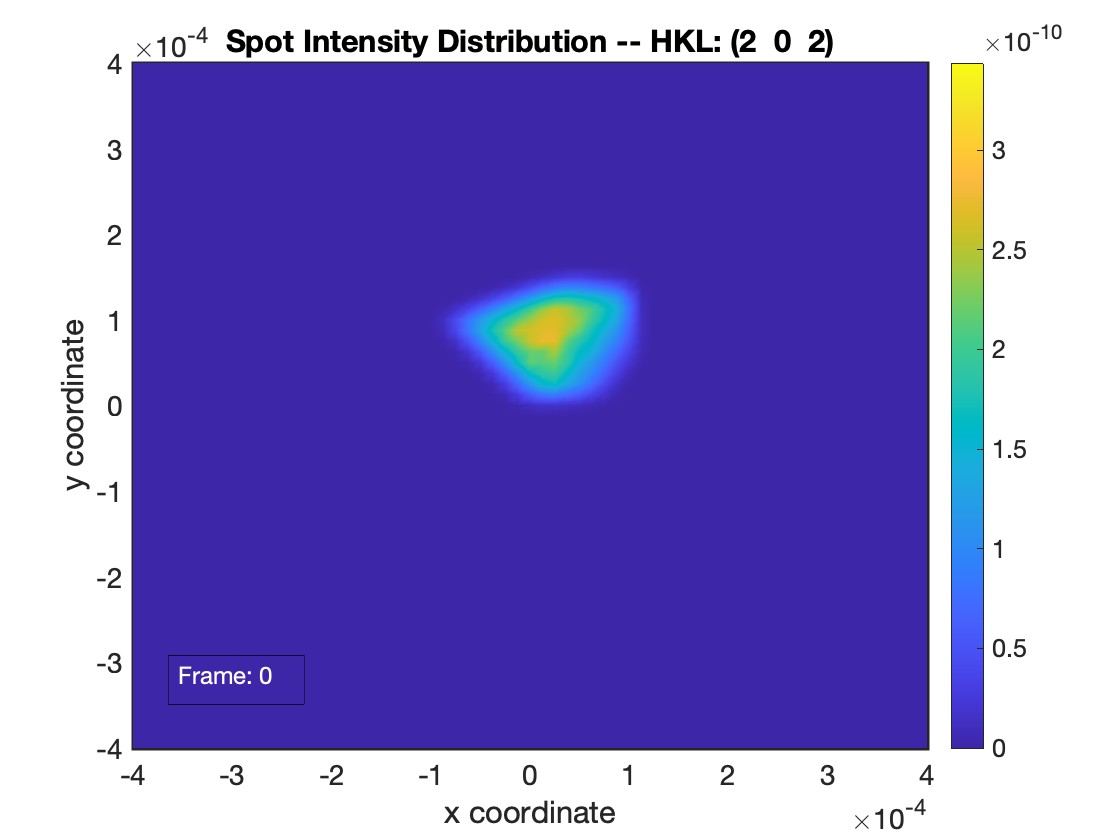}
		\caption{ }
		\label{fig:dist_dectectorPlane_202u}
	\end{subfigure}
		\caption{Intensity distribution over the detector plane under zero load for the most transverse scattering vectors: (a) $\bar1\bar11$ reflection, (b) 200 reflection, and (c) 202 reflection.}
		\label{fig:dist_on_detector_plane_unloaded_90deg}
\end{figure}

\begin{figure}[htbp]
	\centering		
	\begin{subfigure}{.3\textwidth}
		\centering
		\includegraphics[width=1\linewidth]{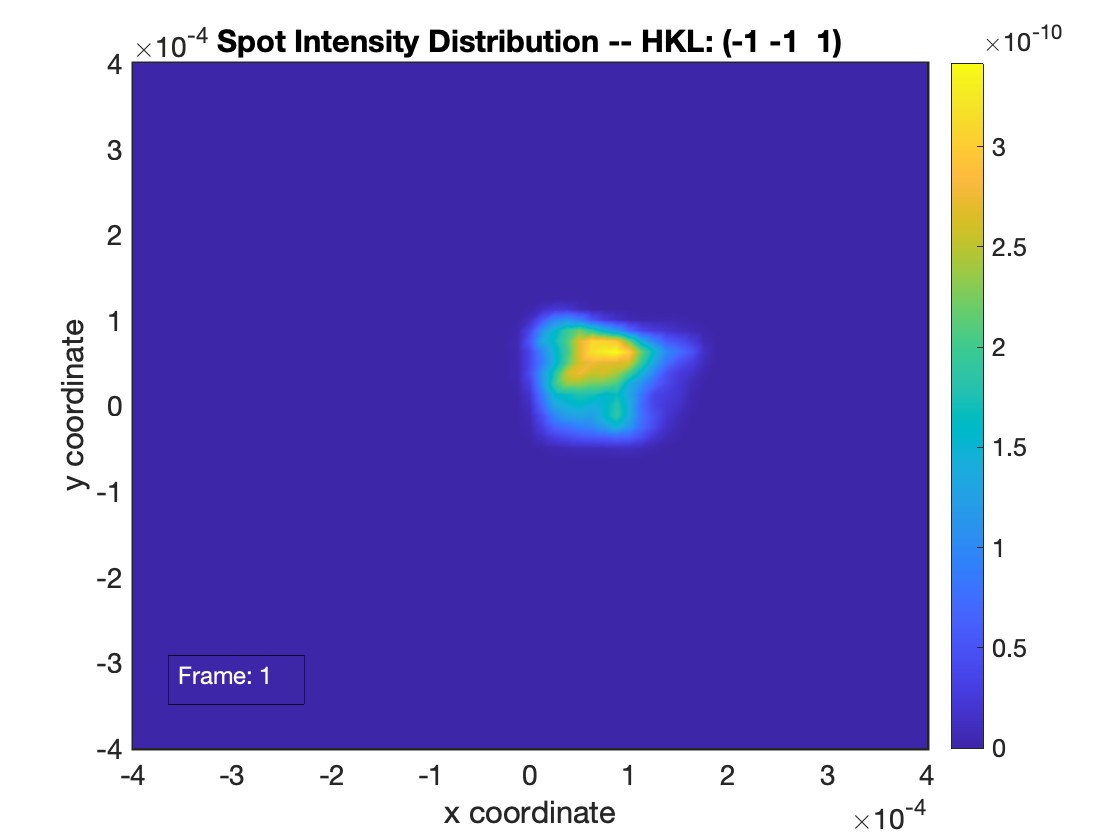}
		\caption{ }
		\label{fig:dist_dectectorPlane_bar1bar11l}
	\end{subfigure}%
	\quad
	\begin{subfigure}{.3\textwidth}
		\centering
		\includegraphics[width=1\linewidth]{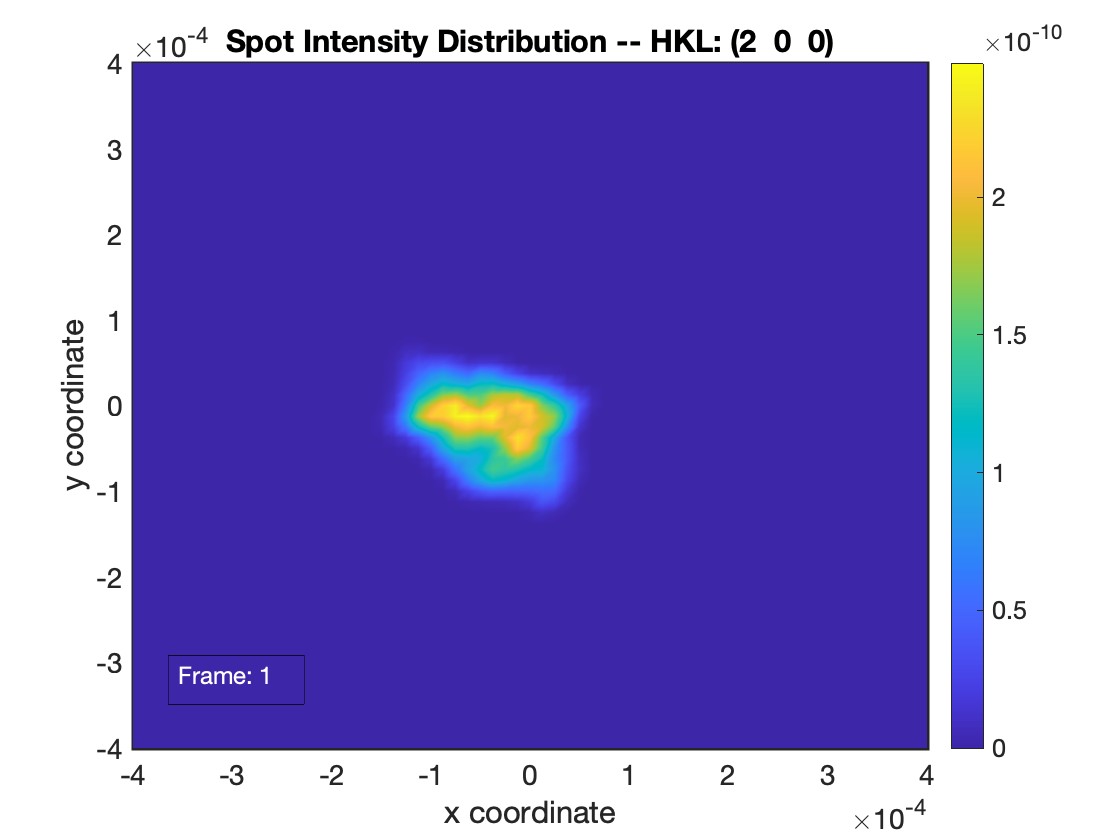}
		\caption{ }
		\label{fig:dist_dectectorPlane_200l}
	\end{subfigure}	
	\quad
	\begin{subfigure}{.3\textwidth}
		\centering
		\includegraphics[width=1\linewidth]{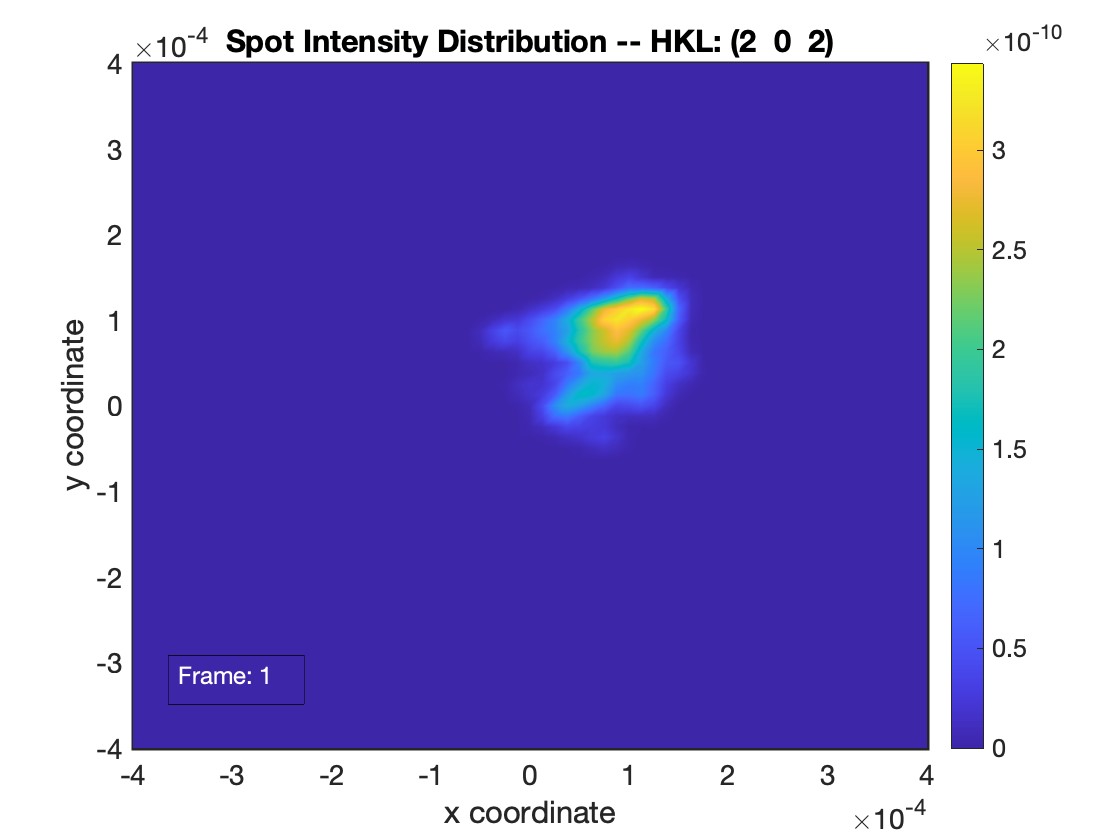}
		\caption{ }
		\label{fig:dist_dectectorPlane_202l}
	\end{subfigure}
		\caption{Intensity distribution over the detector plane under tensile load for the most transverse scattering vectors: (a) $\bar1\bar11$ reflection, (b) 200 reflection, and (c) 202 reflection.}
		\label{fig:dist_on_detector_plane_loaded_90deg}
\end{figure}

 For the axial scattering vectors, both the shifts and the intensity changes are quite apparent.  
 The shifts can be interpreted (and checked) more readily than the intensities using the actual spot positions available in \tabref{tab:diffractionconditions}.  
 Note that the spot shifts from unloaded to loaded states are consistent with the signs of the strains. 
 That is,  if the strain is tensile there is an increased in lattice plane ($d$) spacing  
 and a corresponding decrease in the Bragg ($\theta$) angle.  
This can be observed by combining the approximate spot center with the apparent movement of the spot between \figref{fig:dist_on_detector_plane_unloaded_0deg} and \figref{fig:dist_on_detector_plane_loaded_0deg} for the most axial reflections or between \figref{fig:dist_on_detector_plane_unloaded_90deg} and \figref{fig:dist_on_detector_plane_loaded_90deg} for the most transverse reflections.  
Because the intensity plots have been re-centered, it is not possible to confirm this by looking at the intensity distribution plots alone.

We do not attempt to make general statements regarding the relative intensity changes due lattice distortions 
under applied load.
Grain 80 is only one of the more than 400 grains in Layer 2.
Examination of the relative intensity across the full population of grains  should be made
before attempting to suggest trends in this regard.
Such an examination is possible using the virtual diffractometer, but has not been attempted here.

\subsection{Results from Step 4: Detector images}
\label{sec:demo_task4}
Synthetic detector images are the end goal for the virtual diffractometer. 
In contrast to the intensity distributions of Step 3, the detector images reflect the finite resolution of inherent with pixels from which a detector is built.  From a experimental planning perspective, having both the intensity distribution plots and the detector images facilitates evaluating whether or not the choice and positioning of the detector is optimal for the experimental objectives.  Are critical aspects of the diffraction patterns captured on the detector?

A pixel-based mesh is constructed for computing detector images.  
This mesh is comprised of quadrilateral elements with piecewise constant interpolation (one value of the detector intensity over the element). 
Elements are the same size as pixels.  
In this simulation, parameters for the detector are the same as those employed by Wong {\it et al.}~\cite{won_par_mil_daw_13} with a pixel size of 0.2mm.   
The intensity distribution mesh was constructed first to overlay a patch equal in overall size to 16 pixels. 
For the detector image, the mesh has 16 elements that coincide with the 16 pixels.  The intensity distribution from Step 3 is integrated over those  elements associated with a pixel (and thus an element of the detector image mesh) to give the single value displayed by that pixel. 
Because the point spread function enlarges the detector spot, the detector mesh is doubled in size about its center.  This is why all of the detector image plots span twice the are as the intensity distribution plots and cover an area of 64 (8x8) pixels.

\begin{figure}[htbp]
	\centering		
	\begin{subfigure}{.3\textwidth}
		\centering
		\includegraphics[width=1\linewidth]{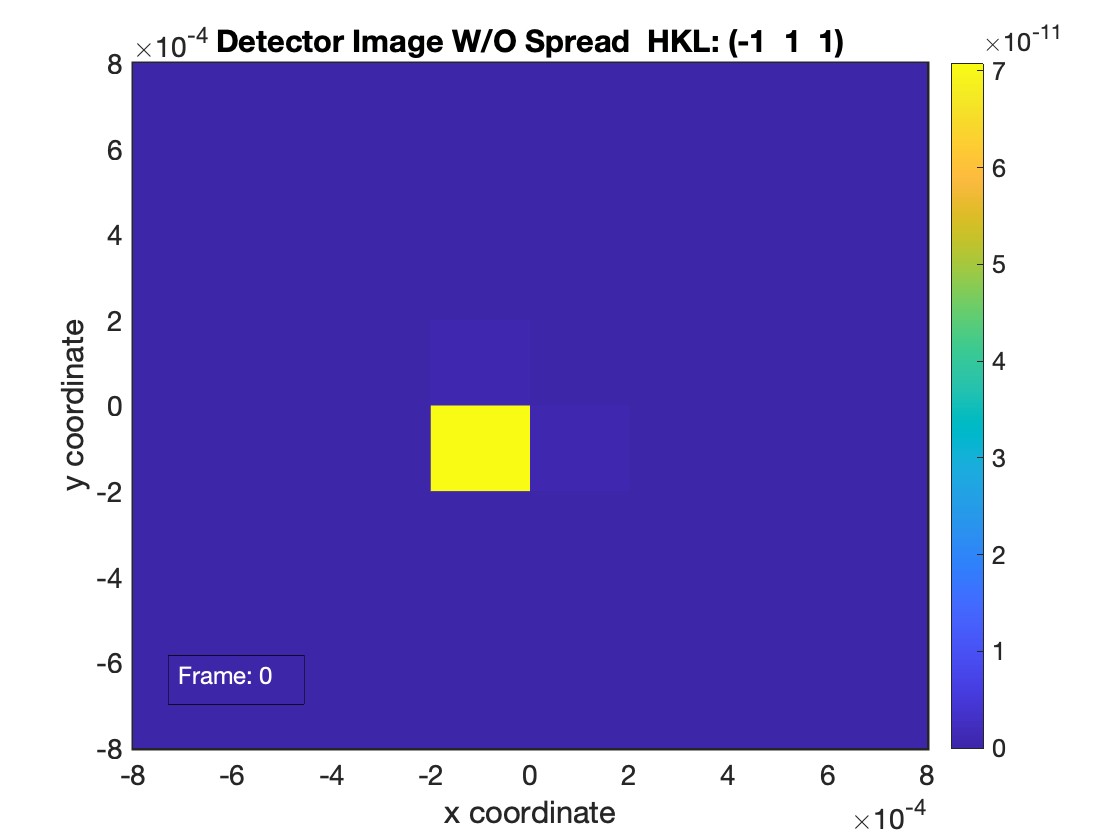}
		\caption{ }
		\label{fig:dectectorimage_bar111u}
	\end{subfigure}%
	\quad
	\begin{subfigure}{.3\textwidth}
		\centering
		\includegraphics[width=1\linewidth]{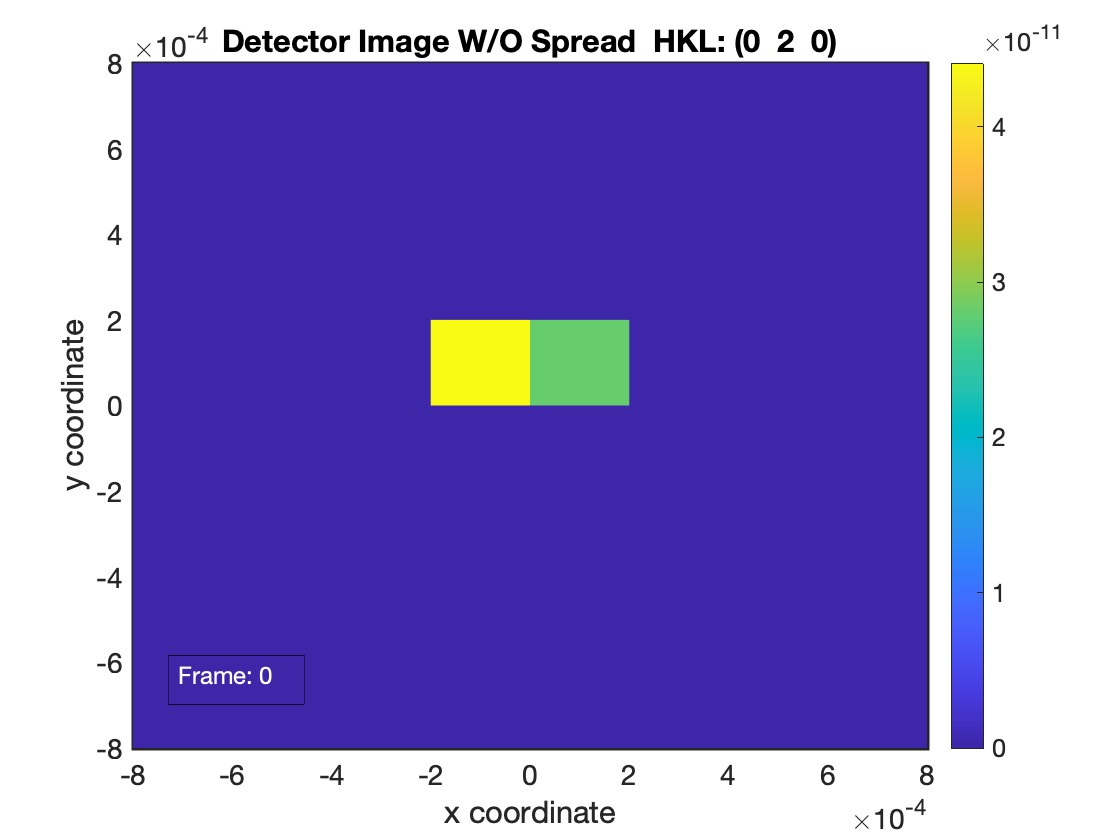}
		\caption{ }
		\label{fig:dectectorimage_020u}
	\end{subfigure}	
	\quad
	\begin{subfigure}{.3\textwidth}
		\centering
		\includegraphics[width=1\linewidth]{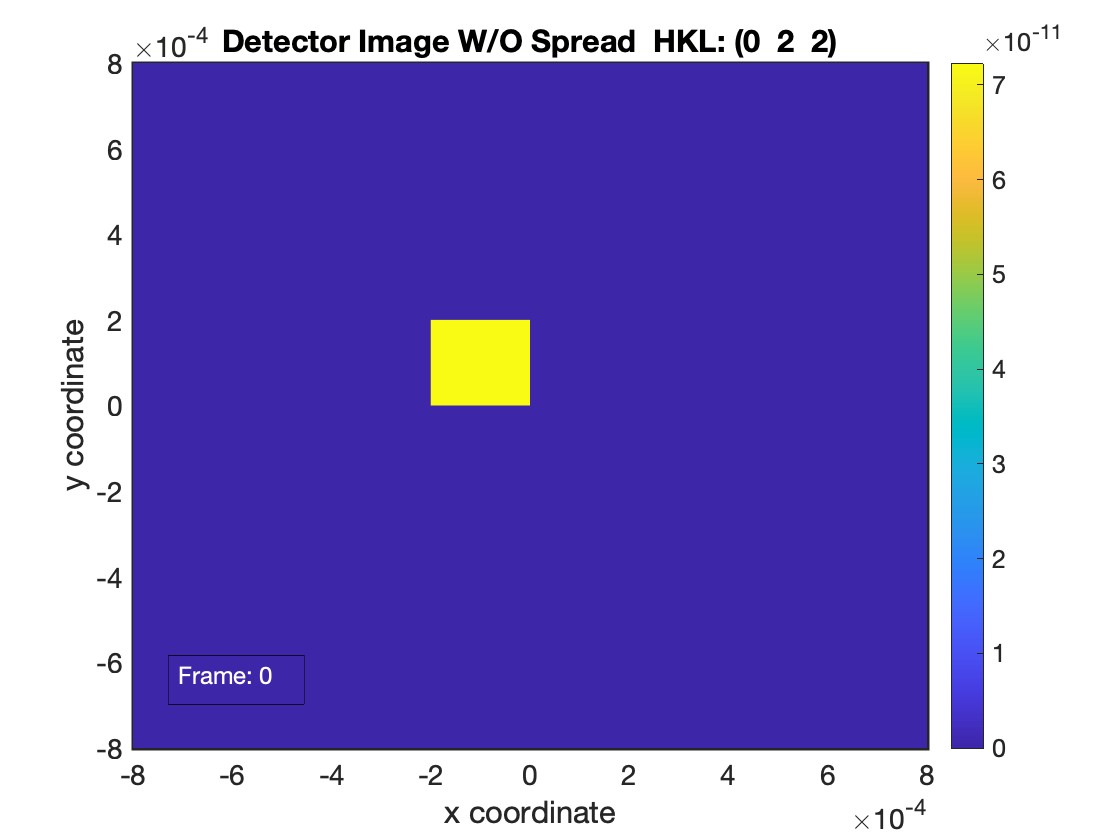}
		\caption{ }
		\label{fig:dectectorimage_022u}
	\end{subfigure}
		\caption{Detector images under zero load for the most axial scattering vectors: (a) $\bar111$ reflection, (b) 020 reflection, and (c) 022 reflection.  The pixel size is 0.2mm.}
		\label{fig:detector_image_unloaded_0deg}
\end{figure}

\begin{figure}[htbp]
	\centering		
	\begin{subfigure}{.3\textwidth}
		\centering
		\includegraphics[width=1\linewidth]{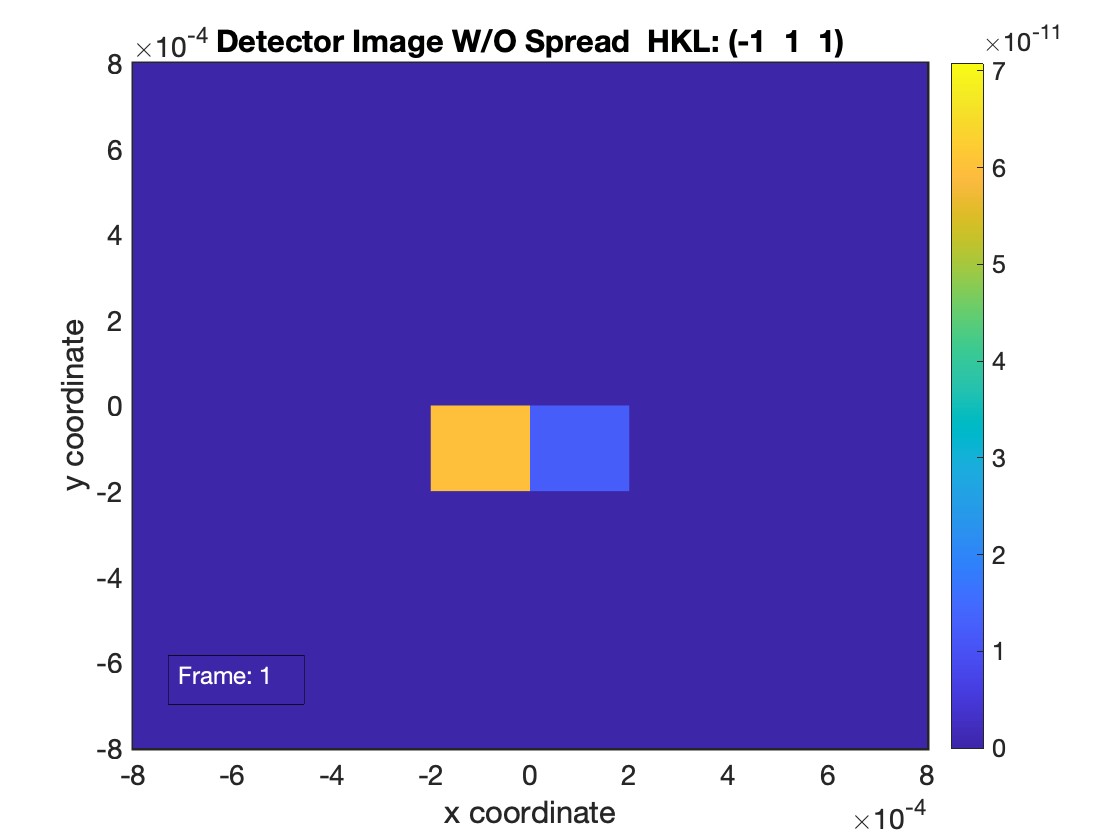}
		\caption{ }
		\label{fig:detectorimage_bar111l}
	\end{subfigure}%
	\quad
	\begin{subfigure}{.3\textwidth}
		\centering
		\includegraphics[width=1\linewidth]{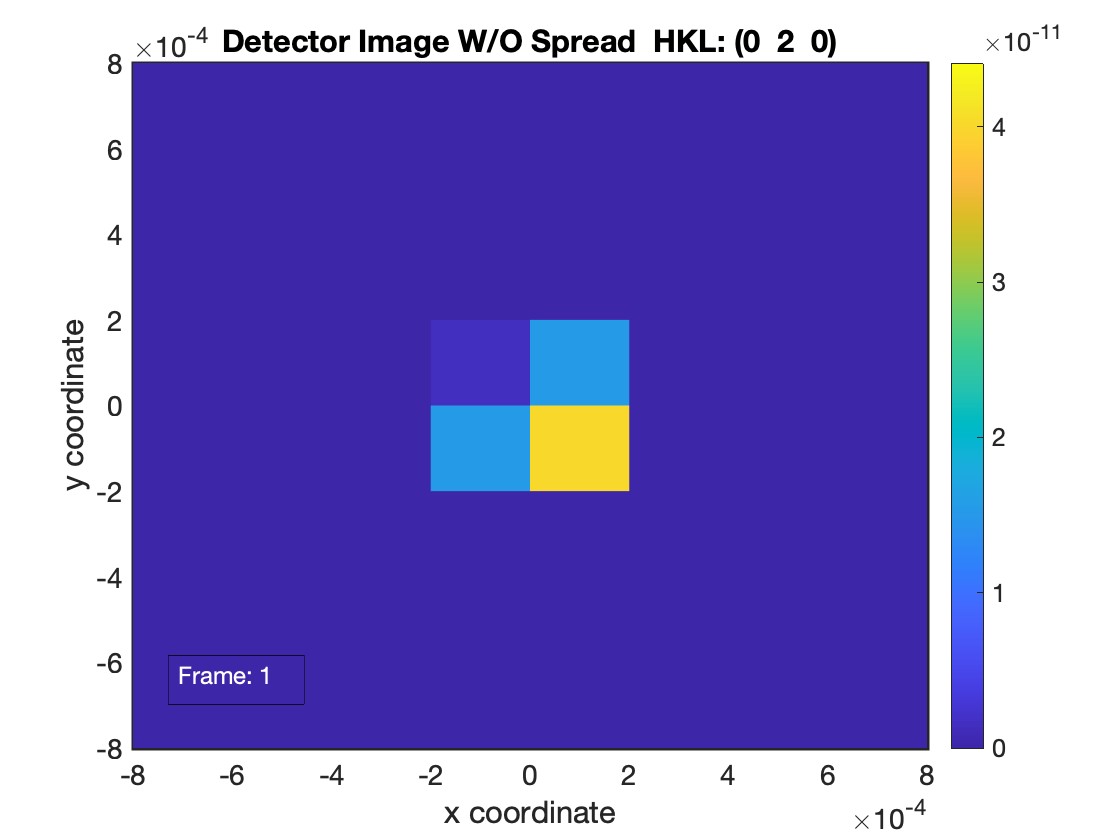}
		\caption{ }
		\label{fig:dectectorimage_020l}
	\end{subfigure}	
	\quad
	\begin{subfigure}{.3\textwidth}
		\centering
		\includegraphics[width=1\linewidth]{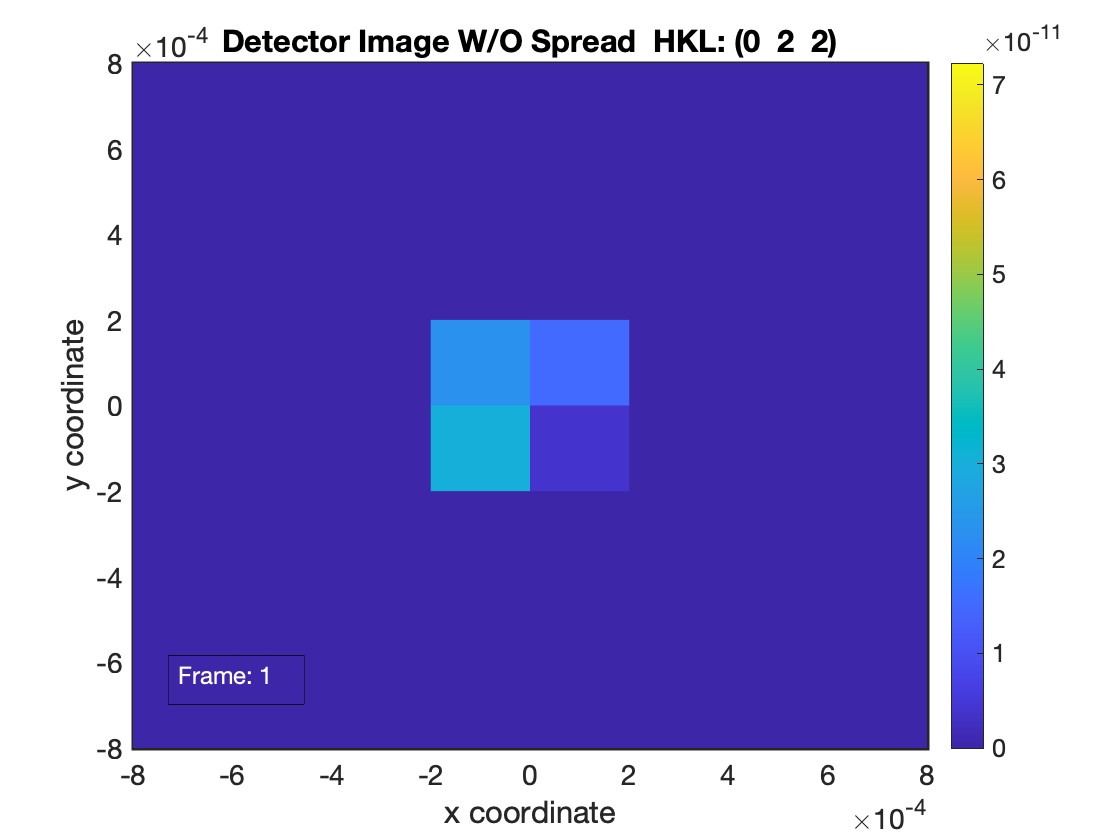}
		\caption{ }
		\label{fig:dectectorimage_022l}
	\end{subfigure}
		\caption{Detector images under tensile load for the most axial scattering vectors: (a) $\bar111$ reflection, (b) 020 reflection, and (c) 022 reflection.  The pixel size is 0.2mm.}
		\label{fig:detector_image_loaded_0deg}
\end{figure}

\begin{figure}[htbp]
	\centering		
	\begin{subfigure}{.3\textwidth}
		\centering
		\includegraphics[width=1\linewidth]{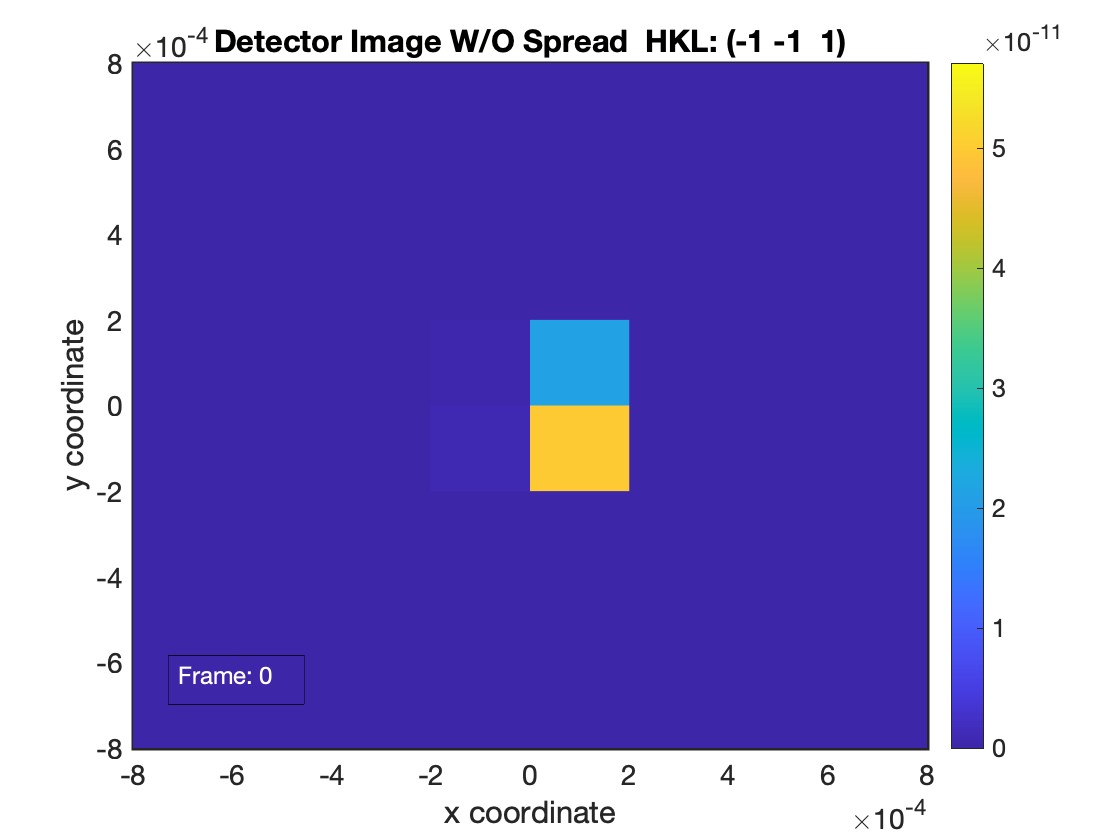}
		\caption{ }
		\label{fig:dectectorimage_bar1bar11u}
	\end{subfigure}%
	\quad
	\begin{subfigure}{.3\textwidth}
		\centering
		\includegraphics[width=1\linewidth]{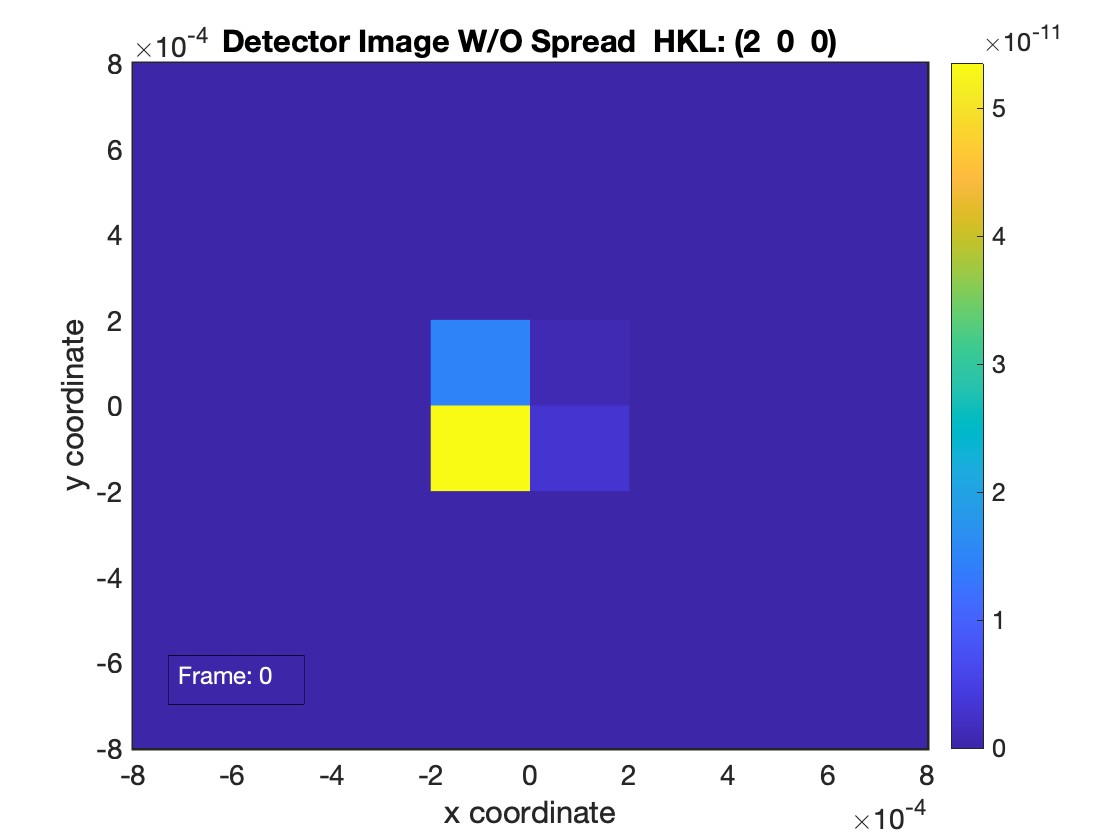}
		\caption{ }
		\label{fig:dectectorimage_200u}
	\end{subfigure}	
	\quad
	\begin{subfigure}{.3\textwidth}
		\centering
		\includegraphics[width=1\linewidth]{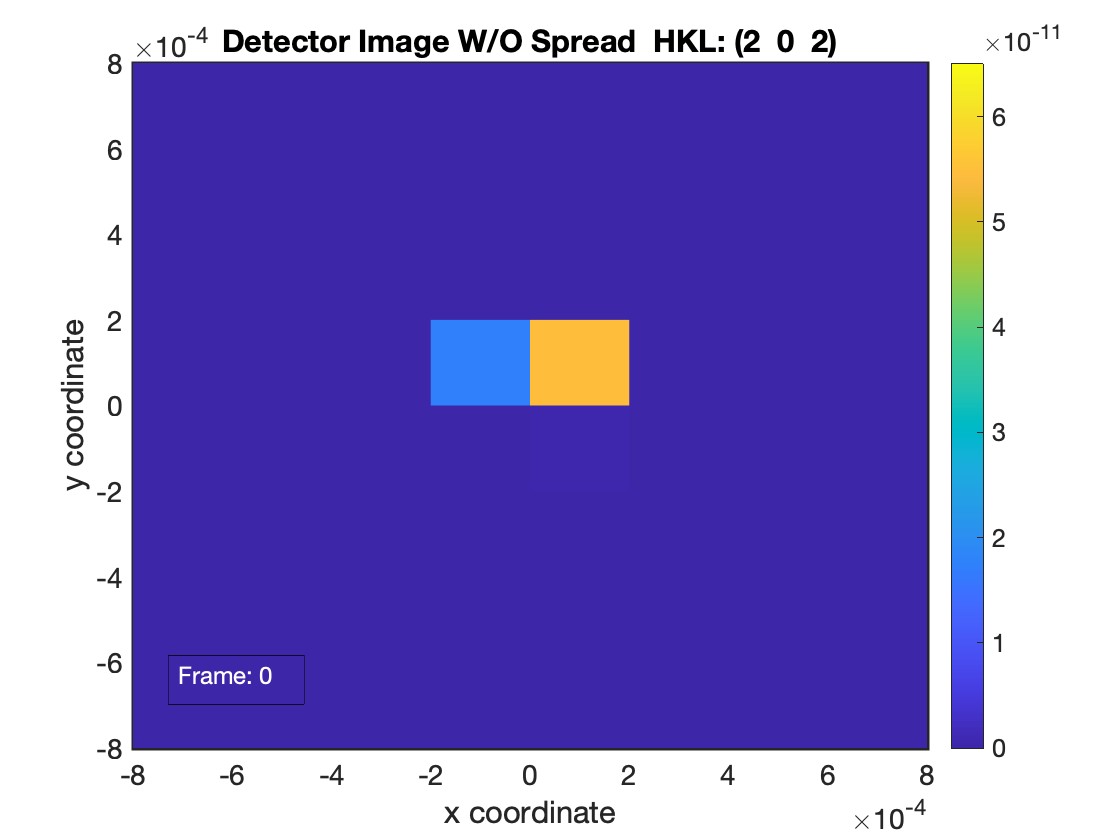}
		\caption{ }
		\label{fig:dectectorimage_202u}
	\end{subfigure}
		\caption{Detector images under zero load for the most transverse scattering vectors: (a) $\bar1\bar11$ reflection, (b) 200 reflection, and (c) 202 reflection.  The pixel size is 0.2mm.}
		\label{fig:detector_image_unloaded_90deg}
\end{figure}

\begin{figure}[htbp]
	\centering		
	\begin{subfigure}{.3\textwidth}
		\centering
		\includegraphics[width=1\linewidth]{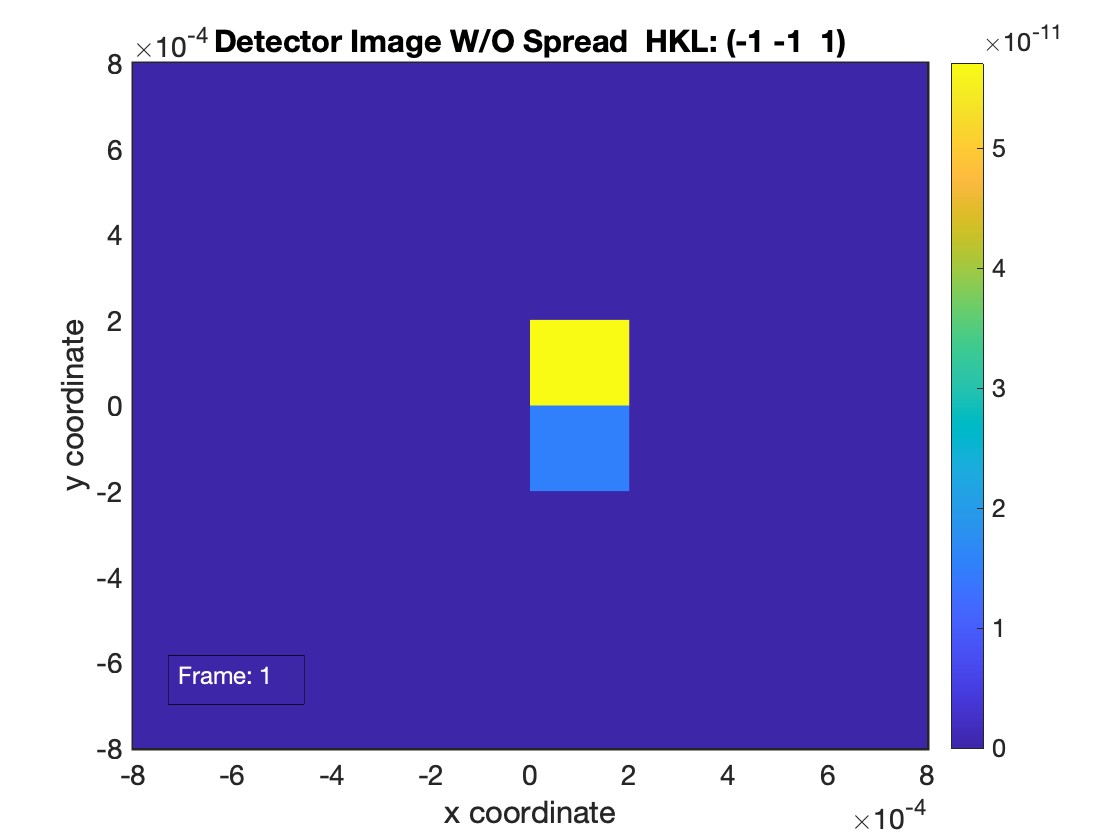}
		\caption{ }
		\label{fig:dectectorimage_bar1bar11l}
	\end{subfigure}%
	\quad
	\begin{subfigure}{.3\textwidth}
		\centering
		\includegraphics[width=1\linewidth]{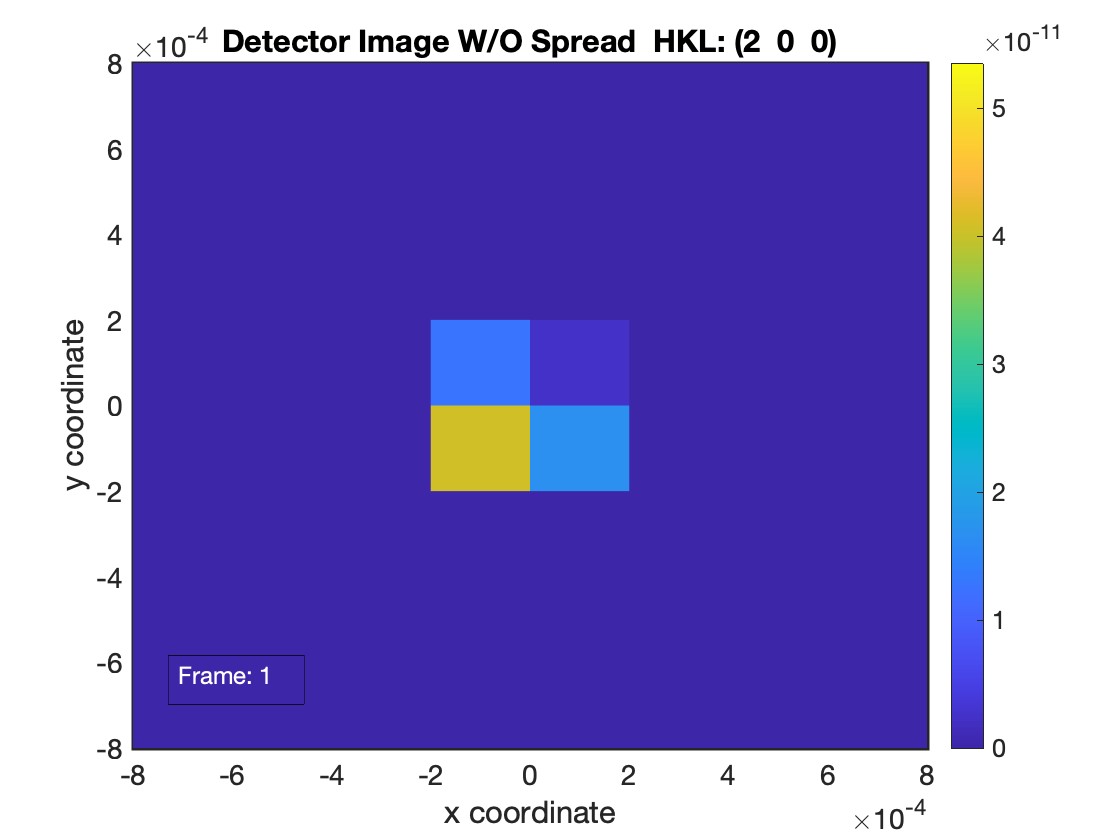}
		\caption{ }
		\label{fig:dectectorimage_200l}
	\end{subfigure}	
	\quad
	\begin{subfigure}{.3\textwidth}
		\centering
		\includegraphics[width=1\linewidth]{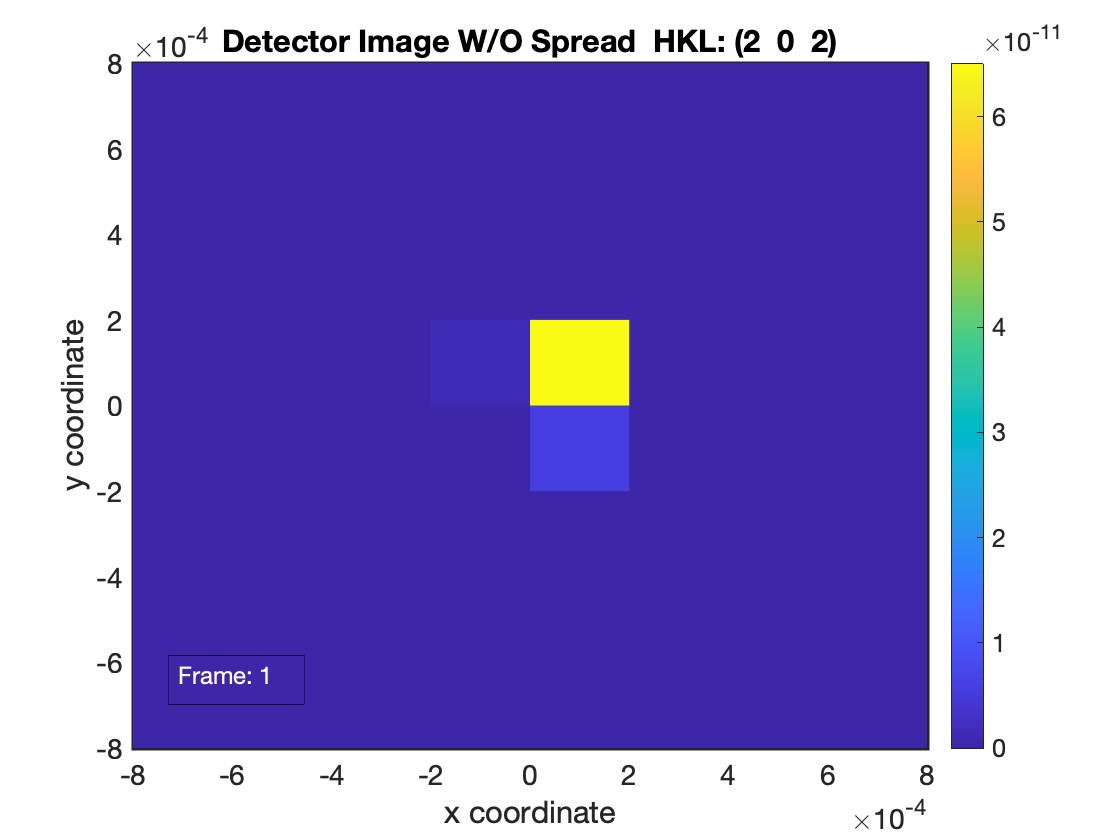}
		\caption{ }
		\label{fig:dectectorimage_202l}
	\end{subfigure}
		\caption{Detector images under tensile load for the most transverse scattering vectors: (a) $\bar1\bar11$ reflection, (b) 200 reflection, and (c) 202 reflection.  The pixel size is 0.2mm.}
		\label{fig:detector_image_loaded_90deg}
\end{figure}

First, detector images prior to the application of the point spread function are shown in 
    \figref{fig:detector_image_unloaded_0deg} and \figref{fig:detector_image_loaded_0deg} 
    for the most axial reflections or between \figref{fig:detector_image_unloaded_90deg} and \figref{fig:detector_image_loaded_90deg}
    for the most transverse reflections. 
    To accommodate the spreading of the spot, the patch size is four times (twice as many pixels in each detector direction) larger than in used for the intensity distribution plots (Figures~\ref{fig:dist_on_detector_plane_unloaded_0deg}-\ref{fig:dist_on_detector_plane_loaded_90deg}).
As is evident from comparison of the two set of images, the detector images are very coarse with this combination of detector position and pixel size. 
The spots shown in the relative intensity distributions are comparable in area to a single pixel. Because the spots are not centered on one pixel, a few pixels typically are activated even though the net area of the intensity distribution spot is comparable to a single pixel.  
As a consequence, much of the detail apparent in the intensity distribution plots is lost in the process of binning the distributions to construct the detector image with the specified pixel size of 0.2mm.  
Some features are still evident, however, including an overall shift in the spot stemming from the change in lattice spacing with strain.  Note also that the scales on the detector images overall are lower than those on the corresponding intensity distribution images.  This is due to the averaging over the intensity distributions done to create the pixel intensity values.

Next, a point spread function is applied to the detector image. 
The point spread function here is the same as reported by Wong {\it et al.}~\cite{won_par_mil_daw_13} together with the fitting parameters determined from line spread data reported by Lee~\cite{Lee2008}.
Values of the point spread template given in \eqnref{eq:ps_template} are:
\begin{equation}
\begin{bmatrix}
 &  & 0.0116 &  & \\
 & 0.0400 & 0.0868 &0.0400  & \\
0.0116 & 0.0868& 0.4462 & 0.0868 &  0.0116 \\
 & 0.0400  & 0.0868 & 0.0400  & \\
 &  &  0.0116  &  & \\
\end{bmatrix}
\label{eq:ps_template_demo}
\end{equation}
Detector images following the application of the point spread function are shown in 
    \figref{fig:detector_image_unloaded_0deg} and \figref{fig:detector_image_loaded_0deg} 
    for the most axial reflections or between \figref{fig:detector_image_unloaded_90deg} and \figref{fig:detector_image_loaded_90deg}
    for the most transverse reflections. 
 Probably the most apparent take-away is that the detector image with point spread is highly influenced by the detector resolution and its characteristic spread function.
\begin{figure}[htbp]
	\centering		
	\begin{subfigure}{.3\textwidth}
		\centering
		\includegraphics[width=1\linewidth]{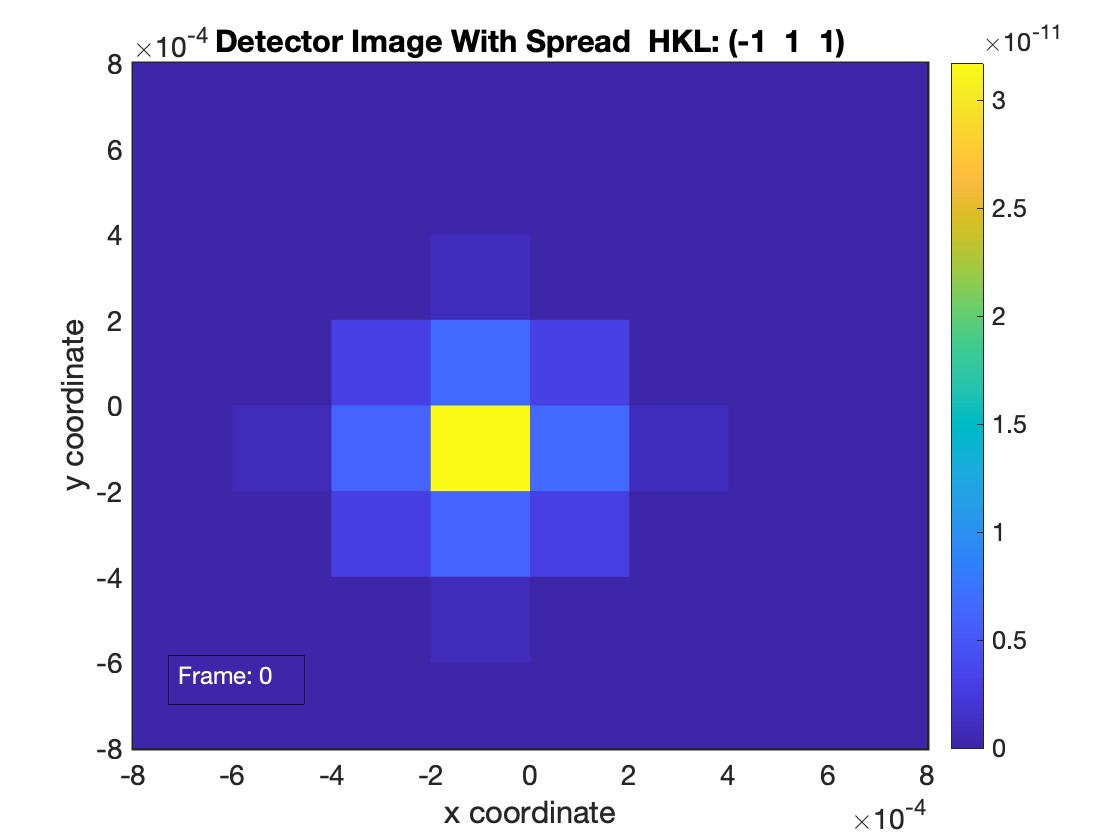}
		\caption{ }
		\label{fig:det+psimage_bar111u}
	\end{subfigure}%
	\quad
	\begin{subfigure}{.3\textwidth}
		\centering
		\includegraphics[width=1\linewidth]{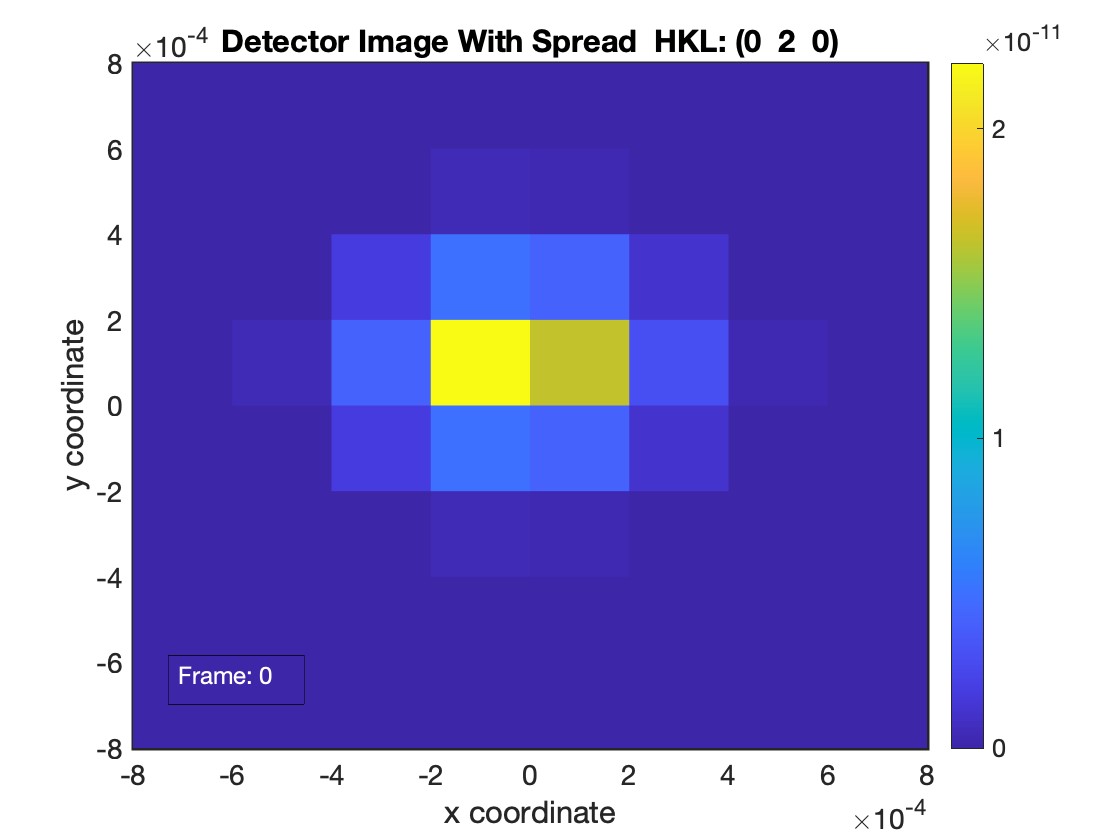}
		\caption{ }
		\label{fig:det+psimage_020u}
	\end{subfigure}	
	\quad
	\begin{subfigure}{.3\textwidth}
		\centering
		\includegraphics[width=1\linewidth]{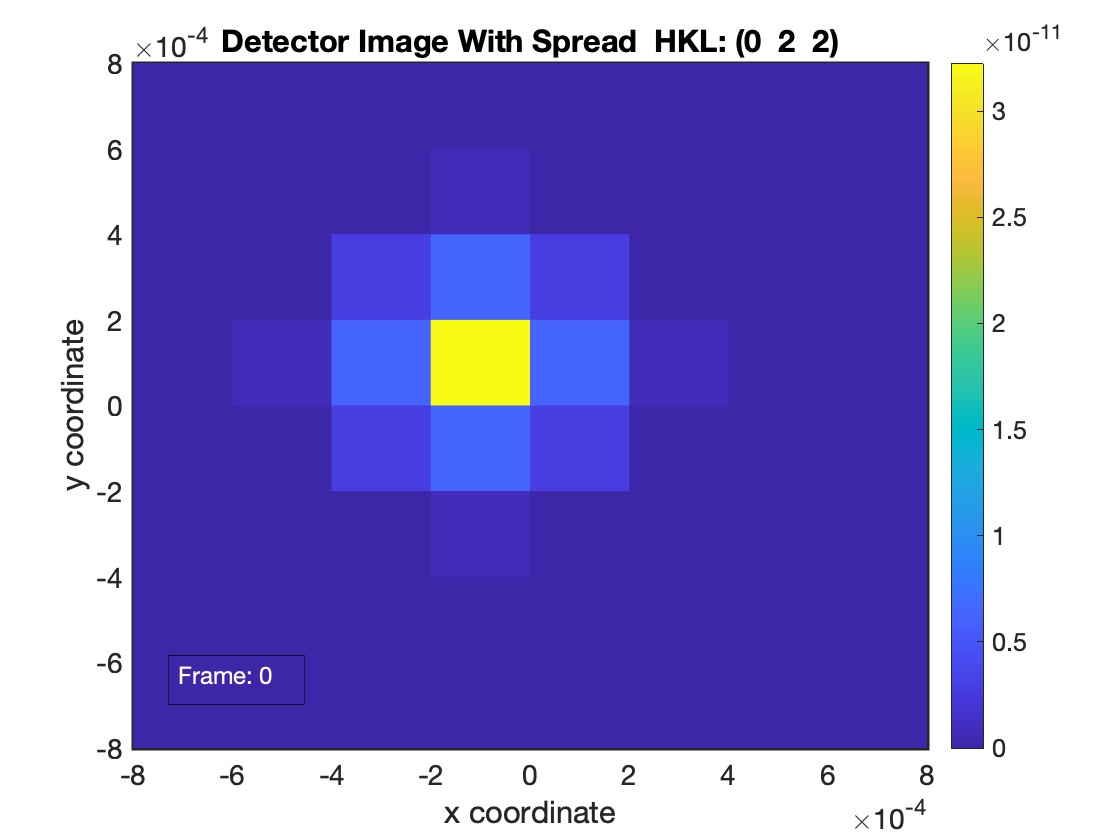}
		\caption{ }
		\label{fig:detpsimage_022u}
	\end{subfigure}
		\caption{Detector images under zero load for the most axial scattering vectors: (a) $\bar111$ reflection, (b) 020 reflection, and (c) 022 reflection.  The pixel size is 0.2mm.}
		\label{fig:det+ps_image_unloaded_0deg}
\end{figure}
\begin{figure}[htbp]
	\centering		
	\begin{subfigure}{.3\textwidth}
		\centering
		\includegraphics[width=1\linewidth]{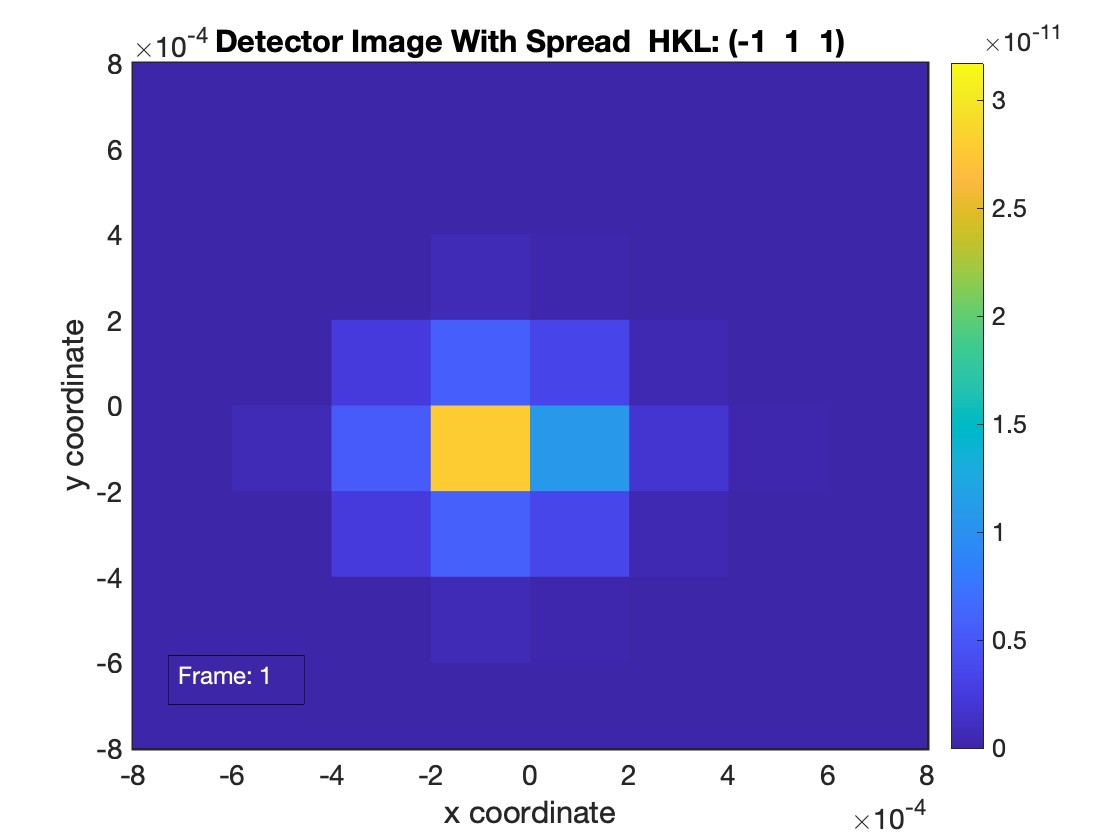}
		\caption{ }
		\label{fig:det+psimage_bar111l}
	\end{subfigure}%
	\quad
	\begin{subfigure}{.3\textwidth}
		\centering
		\includegraphics[width=1\linewidth]{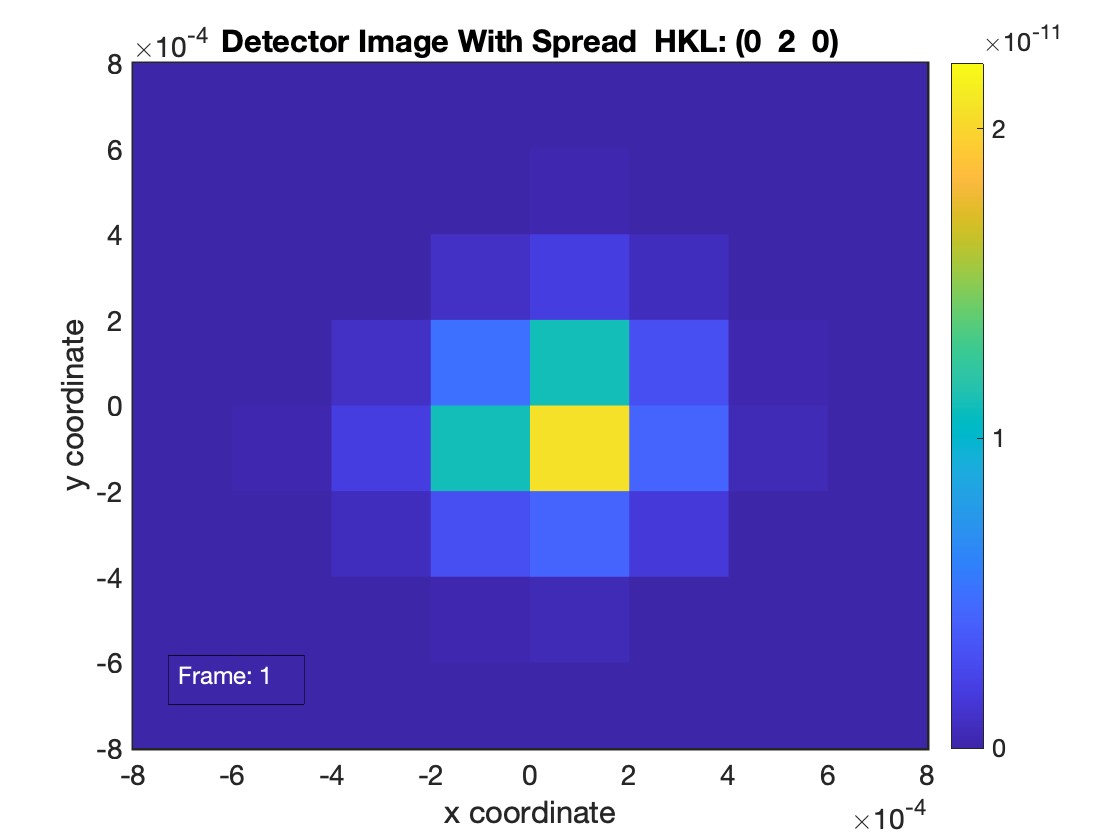}
		\caption{ }
		\label{fig:det+psimage_020l}
	\end{subfigure}	
	\quad
	\begin{subfigure}{.3\textwidth}
		\centering
		\includegraphics[width=1\linewidth]{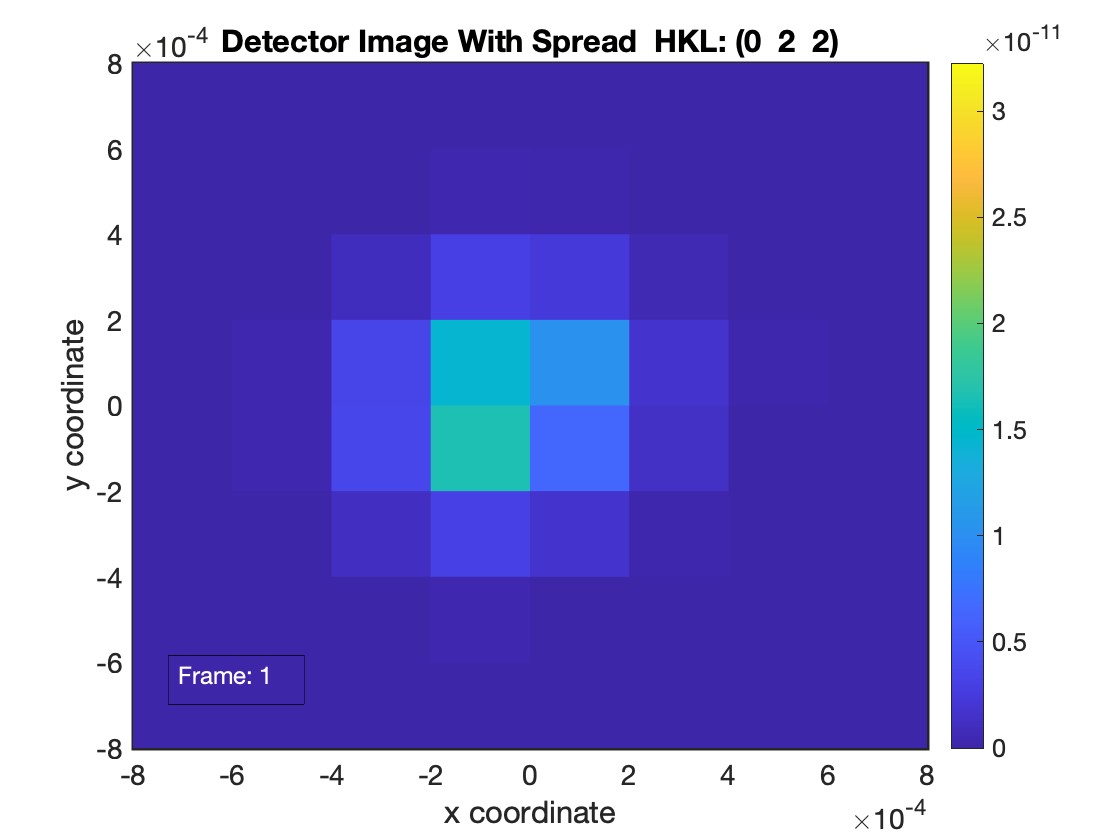}
		\caption{ }
		\label{fig:det+psimage_022l}
	\end{subfigure}
		\caption{Detector images under tensile load for the most axial scattering vectors: (a) $\bar111$ reflection, (b) 020 reflection, and (c) 022 reflection.  The pixel size is 0.2mm.}
		\label{fig:det+ps_image_loaded_0deg}
\end{figure}
\begin{figure}[htbp]
	\centering		
	\begin{subfigure}{.3\textwidth}
		\centering
		\includegraphics[width=1\linewidth]{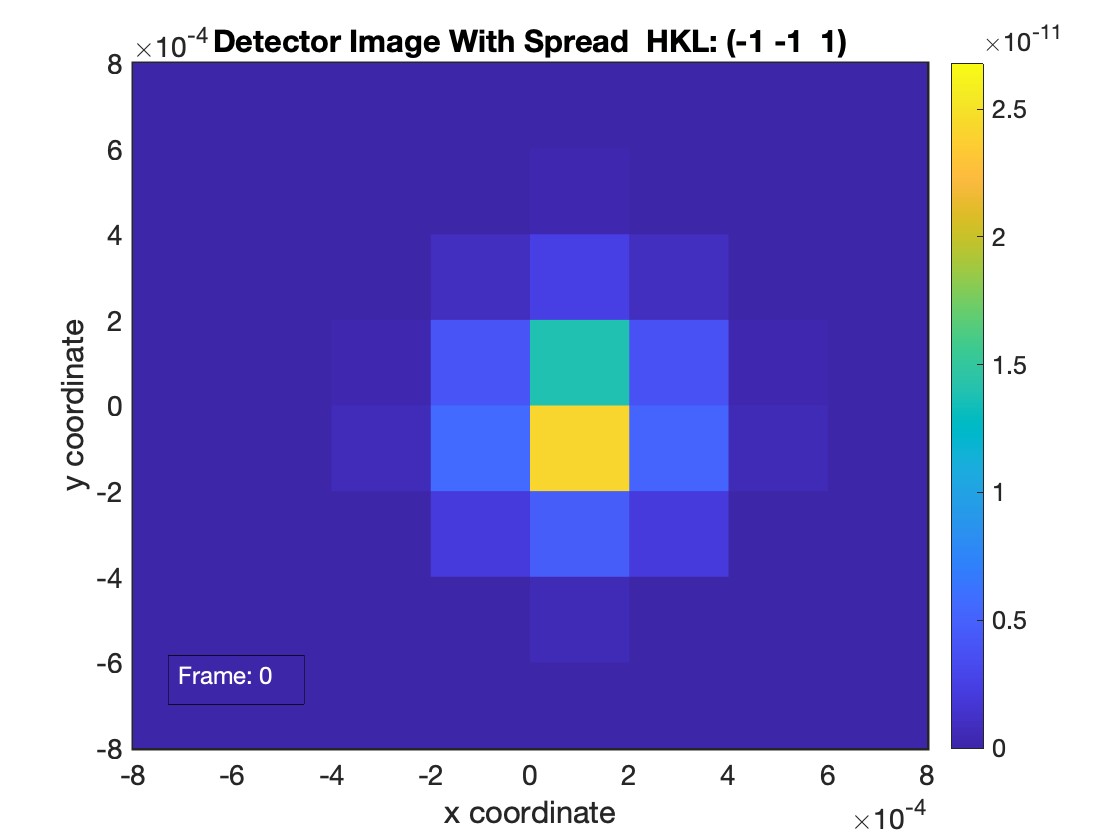}
		\caption{ }
		\label{fig:det+psimage_bar1bar11u}
	\end{subfigure}%
	\quad
	\begin{subfigure}{.3\textwidth}
		\centering
		\includegraphics[width=1\linewidth]{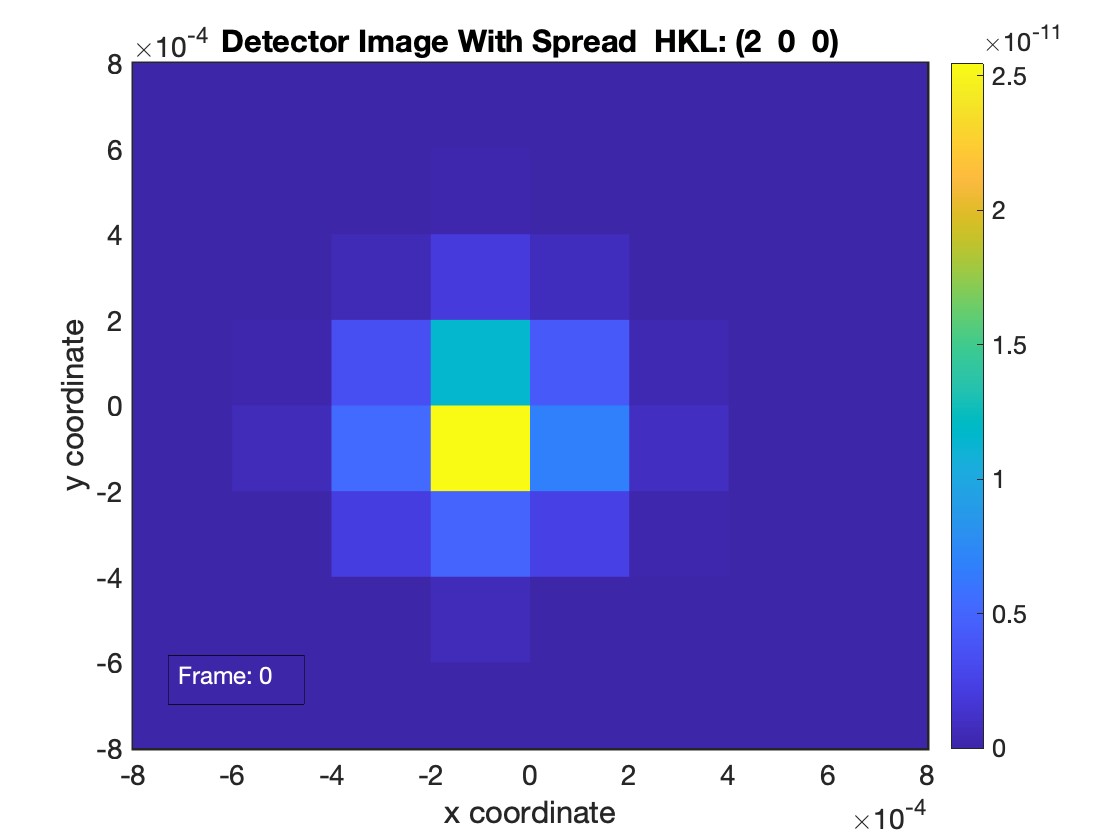}
		\caption{ }
		\label{fig:det+psimage_200u}
	\end{subfigure}	
	\quad
	\begin{subfigure}{.3\textwidth}
		\centering
		\includegraphics[width=1\linewidth]{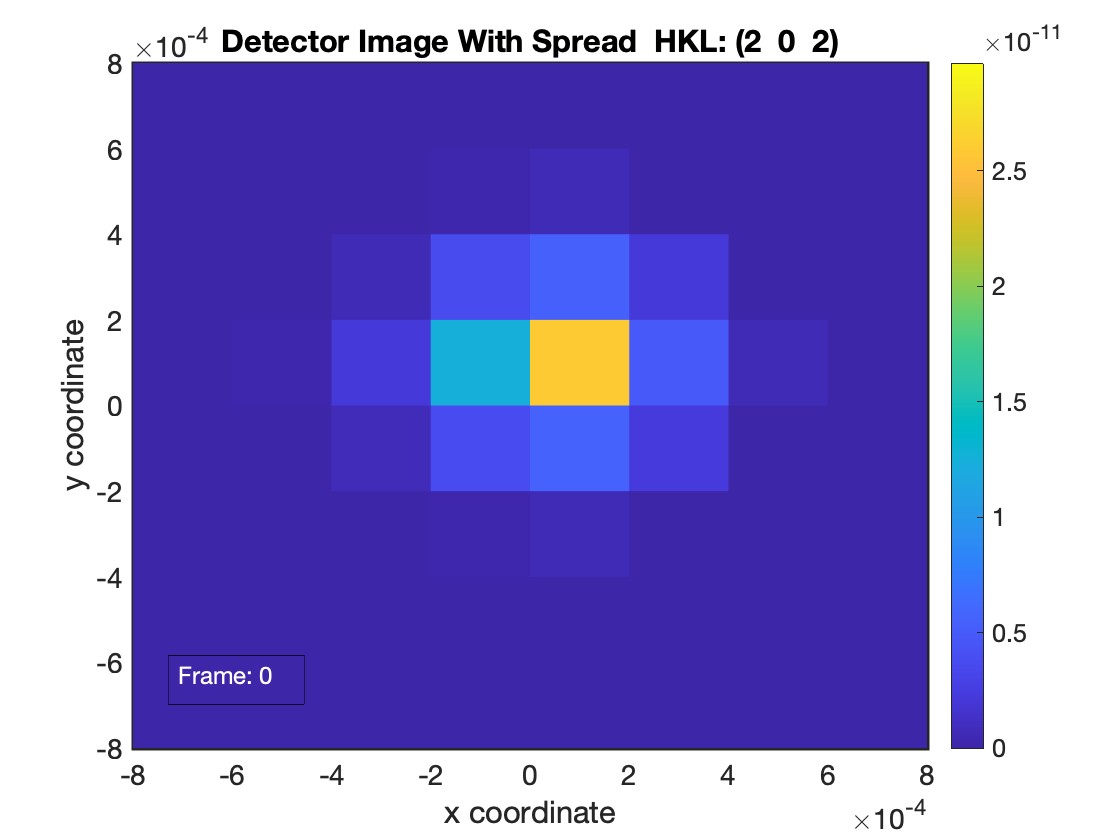}
		\caption{ }
		\label{fig:det+psimage_202u}
	\end{subfigure}
		\caption{Detector images under zero load for the most transverse scattering vectors: (a) $\bar1\bar11$ reflection, (b) 200 reflection, and (c) 202 reflection.  The pixel size is 0.2mm.}
		\label{fig:det+ps_image_unloaded_90deg}
\end{figure}

\begin{figure}[htbp]
	\centering		
	\begin{subfigure}{.3\textwidth}
		\centering
		\includegraphics[width=1\linewidth]{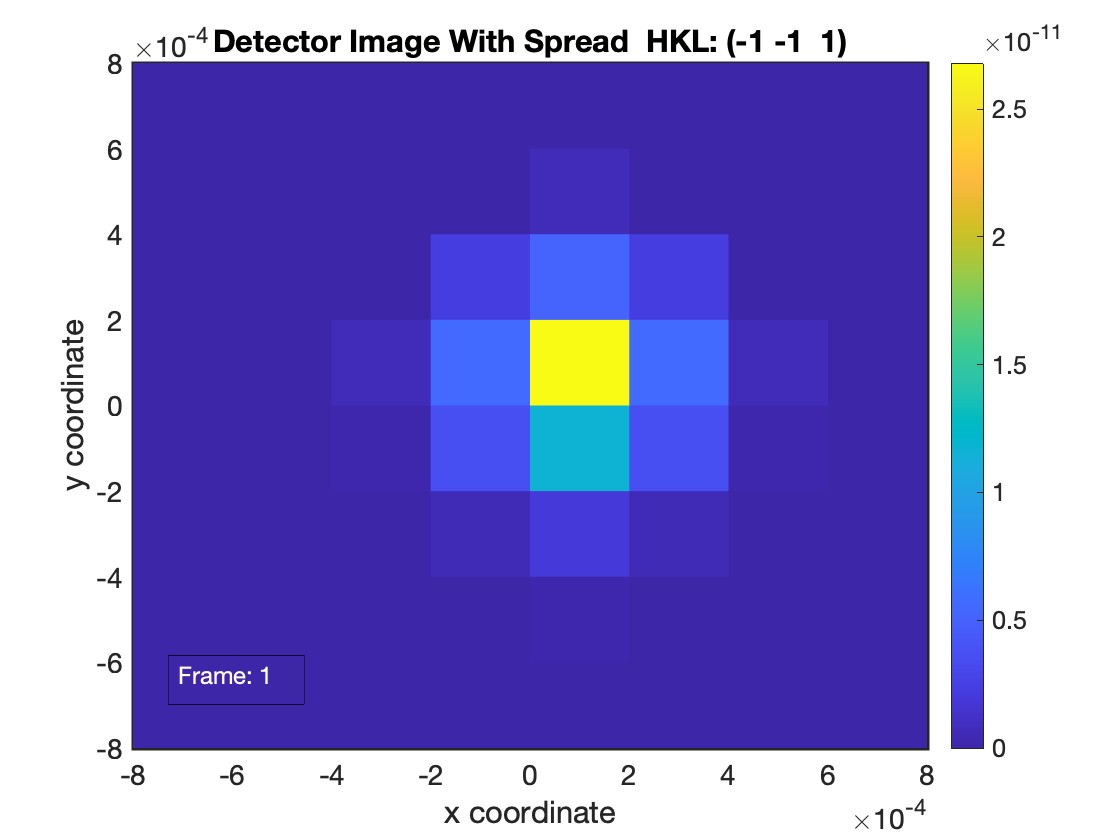}
		\caption{ }
		\label{fig:det+psimage_bar1bar11l}
	\end{subfigure}%
	\quad
	\begin{subfigure}{.3\textwidth}
		\centering
		\includegraphics[width=1\linewidth]{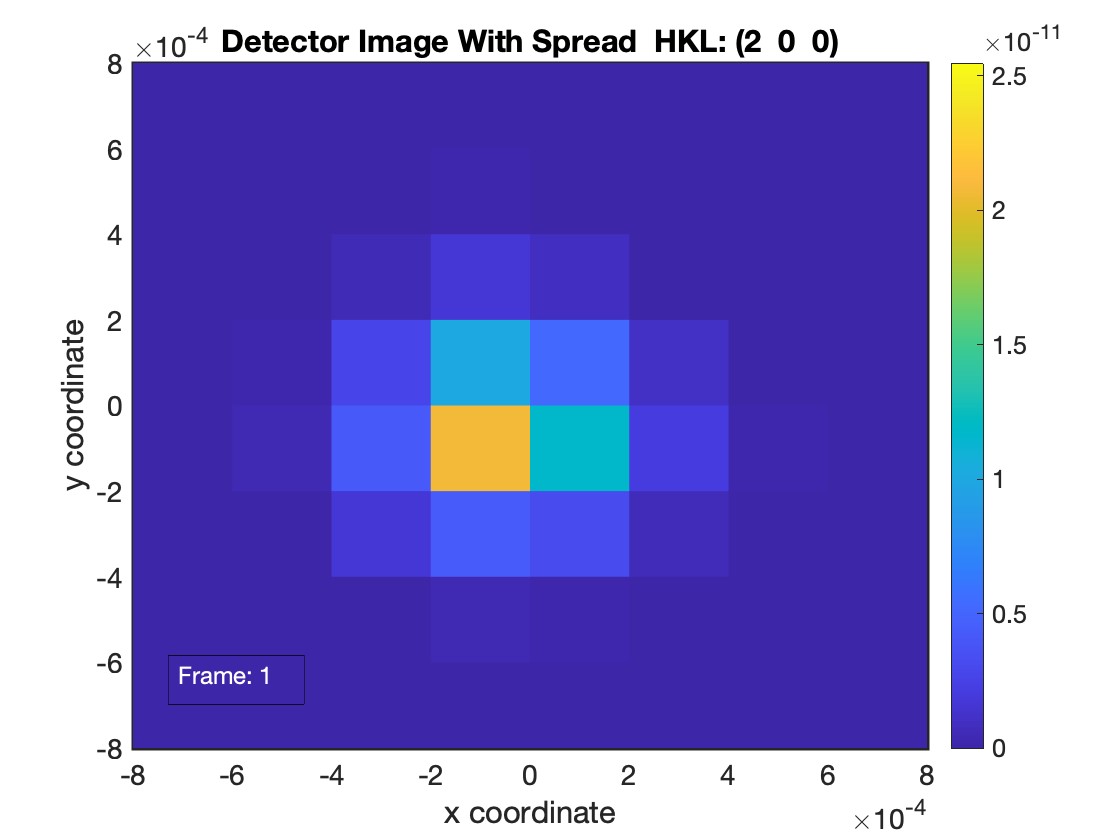}
		\caption{ }
		\label{fig:det+psimage_200l}
	\end{subfigure}	
	\quad
	\begin{subfigure}{.3\textwidth}
		\centering
		\includegraphics[width=1\linewidth]{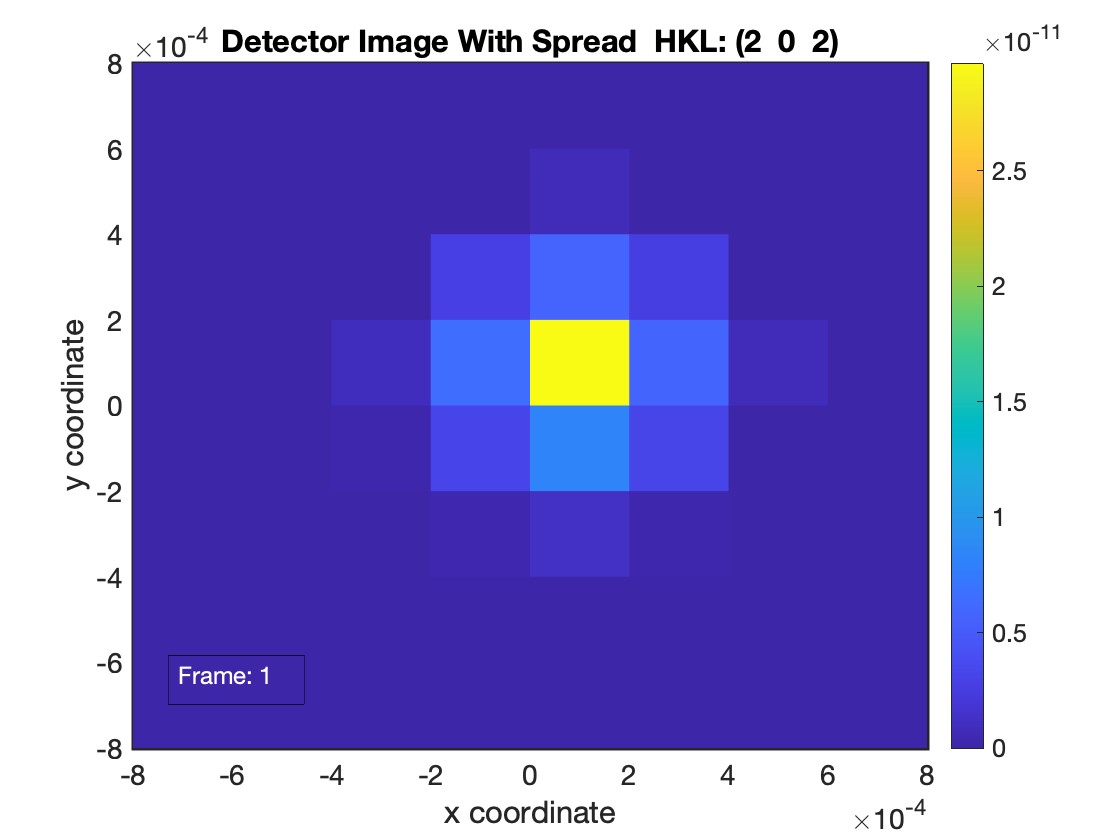}
		\caption{ }
		\label{fig:det+psimage_202l}
	\end{subfigure}
		\caption{Detector images under tensile load for the most transverse scattering vectors: (a) $\bar1\bar11$ reflection, (b) 200 reflection, and (c) 202 reflection.  The pixel size is 0.2mm.}
		\label{fig:det+ps_image_loaded_90deg}
\end{figure}
\section{Summary}
\label{sec:summary}

A computational tool for creating synthetic diffraction images (spots) associated with HEDM experiments is documented. 
Attention is focused on polycrystalline samples that are subjected to {\it in situ} loading.
The tool, referred to as a virtual diffractometer, computes images for diffraction volumes that are
sub-crystal in scale.  
Under these conditions, the diffraction images consist of isolated peaks (or spots) derived from beams diffracted from individual crystals.  
The virtual diffractometer is configured to compute the spots created by diffraction within a single, target crystal. 
The user designates the families of crystallographic planes that spawn diffracted beams 
and the code identifies the specific planes that are closet to desired scattering vectors.
From the diffraction conditions computed for Laue diffraction, morphology of the target crystal, and details
of the experimental configuration, the code produces virtual diffraction spots.
These spots embody information regarding the its detector coordinates and a spatial map of relative intensity
of the diffracted beam.  

In this article, the underlying methodology used for the virtual diffractometer is summarized and an example case illustrates the its capabilities.  Principal features include:
\begin{itemize}
\item The code is written in MatLab and is available freely from GitHub.
\item  The code takes in the following input data: 
\begin{enumerate}
\item definition of a virtual sample in the form of a meshed tessellation of the grains; and,
\item data arrays that provide the lattice orientations and elastic strains coinciding with the mesh at loaded and unloaded states.
\end{enumerate}
\item The code requires data related to the material, the x-ray beam, and the detector.  These can be re-defined as needed. The data used includes:
\begin{enumerate}
\item the lattice type and relevant lattice constants, families of $hkl$'s  to be considered, and attenuation characteristics;
\item the incident beam energy and direction and the spatial variance of the diffracted beam;
\item the plane of the detector and the resolution of the finite element mesh used to represent the intensity field in the vicinity of a spot; and,
\item the detector resolution (pixel size).
\end{enumerate}
\item The code computes spot images for a combination of target grain and scattering vector direction designated by the user.  The output includes the following items for each reflection and load point:
\begin{enumerate}
\item the average strains associated with the scattering vectors (loaded points only);
\item plots showing distributions of the rotation angle $\omega$ at which the diffraction conditions are satisfied (loaded points only);
\item plots showing points at which diffracted beams intersect the detector plane;
\item images of the spot intensity distributions; and,
\item  pixelated detector images including the effects of point spread.
\end{enumerate}
\item A demonstration example is provided for a stainless steel sample.
\end{itemize}

\section*{Acknowledgements}
The research reported here was supported by the ONR under grant \# N00014-16-1-3126, Dr. William Mullins Program Manager.  The authors thank Wiley Kirks for his assistance in providing independent checks of computations performed in Step 1 to determine the diffraction conditions of elemental volumes.

% Authors must disclose all relationships or interests that 
% could have direct or potential influence or impart bias on 
% the work: 
%
 \section*{Conflict of interest}
 The authors declare that they have no conflict of interest.

% BibTeX users please use one of
%\bibliographystyle{spbasic}      % basic style, author-year citations
%\bibliographystyle{spmpsci}      % mathematics and physical sciences
%\bibliographystyle{spphys}       % APS-like style for physics
\bibliographystyle{unsrt}
\bibliography{VirtualDiffractometer.bib}  

\begin{thebibliography}{10}

\bibitem{poulsen_book}
H~Poulsen.
\newblock {\em {Three-Dimensional X-Ray Diffraction Microscopy}}.
\newblock {Springer}, Heidelberg, U.K., 2004.

\bibitem{suter_jemt_2008}
R.~M. Suter, C~M Heffernan, S~F Li, D~Hennessy, and C~Xiao.
\newblock {Probing Microstructure Dynamics With X-Ray Diffraction Microscopy}.
\newblock {\em Journal of Engineering Materials and Technology}, 130:021007,
  2008.

\bibitem{lienert_jom_2010}
U~Lienert, S~F Li, C~M Heffernan, J~Lind, R.~M. Suter, J.~V. Bernier, N.~R.
  Barton, M~Brandes, M~J Mills, M~P Miller, C~Wejdemann, and W~Pantleon.
\newblock {High Energy Diffraction Microscopy at the Advanced Photon Source}.
\newblock {\em JOM}, 63(7):70--77, 2011.

\bibitem{lienert_titanium_2009}
U~Lienert, M~C Brandes, J.~V. Bernier, J~Weiss, S~D Shastri, M~J Mills, and M~P
  Miller.
\newblock {In situ single-grain peak profile measurements on Ti-7Al during
  tensile deformation}.
\newblock {\em Materials Science and Engineering A}, 524(1-2):46--54, 2009.

\bibitem{Oddershede2010a}
Jette Oddershede, S{\o}ren Schmidt, Henning~Friis Poulsen, Henning~Osholm
  S{\o}rensen, Jonathan Wright, and Walter Reimers.
\newblock Determining grain resolved stresses in polycrystalline materials
  using three-dimensional x-ray diffraction.
\newblock {\em Journal of Applied Crystallography}, 43(3):539--549, 2010.

\bibitem{Bernier2011}
J~V Bernier, N~R Barton, U~Lienert, and M~P Miller.
\newblock Far-field high-energy diffraction microscopy: a tool for
  intergranular orientation and strain analysis.
\newblock {\em The Journal of Strain Analysis for Engineering Design},
  46(7):527--547, 2011.

\bibitem{20202708900950}
Matthew~P. Miller, Darren~C. Pagan, Armand~J. Beaudoin, Kelly~E. Nygren, and
  Dalton~J. Shadle.
\newblock Understanding micromechanical material behavior using synchrotron
  x-rays and in situ loading.
\newblock {\em Metallurgical and Materials Transactions A: Physical Metallurgy
  and Materials Science}, 51(9):4360 -- 4376, 2020.

\bibitem{obs_won_daw_mil_14}
M.~Obstalecki, S.-L. Wong, P.~Dawson, and M.~Miller.
\newblock Quantitative analysis of crystal scale deformation heterogeneity
  during cyclic plasticity using high-energy \protect{X-ray} diffraction and
  finite-element simulation.
\newblock {\em Acta Materialia}, 75:259--272, 2014.

\bibitem{bertin2018computation}
N.~Bertin and W.~Cai.
\newblock Computation of virtual x-ray diffraction patterns from discrete
  dislocation structures.
\newblock {\em Computational Materials Science}, 146:268--277, 2018.

\bibitem{Pagan2020}
D.~C. Pagan, K.~K. Jones, J.~V. Bernier, and T.~Q. Phan.
\newblock A finite energy bandwidth-based diffraction simulation framework for
  thermal processing applications.
\newblock {\em Journal of Materials}, 72(12):4539 -- 4550, 2020.

\bibitem{Ribart:vl5004}
C.~Ribart, A.~King, W.~Ludwig, J.~P.~C. Bertoldo, and H.~Proudhon.
\newblock {{\it In situ} synchrotron X-ray multimodal experiment to study
  polycrystal plasticity}.
\newblock {\em Journal of Synchrotron Radiation}, 30(2):379--389, Mar 2023.

\bibitem{won_par_mil_daw_13}
S.~L. Wong, J.-S. Park, M.~P. Miller, and P.~R. Dawson.
\newblock A framework for generating synthetic diffraction images from
  deforming polycrystals using crystal-based finite element formulations.
\newblock {\em Computational Materials Science}, 77:456--466, 2013.

\bibitem{Cullity2001}
B.~D. Cullity and S.~R. Stock.
\newblock {\em Elements of X-Ray Diffraction}.
\newblock Prentice Hall, 2001.

\bibitem{Dawson_FEpX_2015}
P.~R. {Dawson} and D.~E. {Boyce}.
\newblock {FEpX -- Finite Element Polycrystals: Theory, Finite Element
  Formulation, Numerical Implementation and Illustrative Examples}.
\newblock {\em ArXiv e-prints}, April 2015.

\bibitem{Krawitz:2001}
A.~D. Krawitz.
\newblock {\em Introduction to Diffraction in Materials Science and
  Engineering}.
\newblock Wiley-Interscience, New York, 2001.

\bibitem{Ledbetter01a}
H.~Ledbetter.
\newblock Monocrystal-polycrystal elastic constants of stainless steels.
\newblock In M.~Levy, editor, {\em Handbook of Elastic Properties of Solids,
  Liquids, and Gases}, volume~3, chapter~17, pages 291--297. Academic Press,
  2001.

\bibitem{Lee2008}
John~H. Lee, C.~Can Ayd{\i}ner, Jonathan Almer, Joel Bernier, Karena~W.
  Chapman, Peter~J. Chupas, Dean Haeffner, Ken Kump, Peter~L. Lee, Ulrich
  Lienert, Antonino Miceli, and German Vera.
\newblock {Synchrotron applications of an amorphous silicon flat-panel
  detector}.
\newblock {\em Journal of Synchrotron Radiation}, 15(5):477--488, Sep 2008.

\end{thebibliography}

\end{document}